\documentclass[12pt]{iopart}
\pdfoutput=1
\usepackage{iopams}
\usepackage{graphicx}
\usepackage{multirow}
\begin{document}

\title{The adaptive nature of liquidity taking in limit order books}
\author{Damian~Eduardo~Taranto$^1$, Giacomo~Bormetti$^{1,2}$, and Fabrizio~Lillo$^{1,2,3,4}$}

\address{$^1$ Scuola Normale Superiore, Piazza dei Cavalieri 7, Pisa, 56126, Italy}
\address{$^2$ QUANTLab, via Pietrasantina 123, Pisa, 56122, Italy}
\address{$^3$ Dipartimento di Fisica e Chimica, Universit{\`a} degli Studi di Palermo, Viale delle Scienze Ed. 18, Palermo, 90128, Italy}
\address{$^4$ Santa Fe Institute, 1399 Hyde Park Road, Santa Fe, NM 87501, USA}
\eads{\mailto{damian.taranto@sns.it}, \mailto{giacomo.bormetti@sns.it}, and \mailto{fabrizio.lillo@unipa.it}}

\begin{abstract}
In financial markets, the order flow, defined as the process assuming value one for buy market orders and minus one for sell market orders, displays a very slowly decaying autocorrelation function. Since orders impact prices, reconciling the persistence of the order flow with market efficiency is a subtle issue. A possible solution is provided by asymmetric liquidity, which states that the impact of a buy or sell order is inversely related to the probability of its occurrence. We empirically find that when the order flow predictability increases in one direction, the liquidity in the opposite side decreases, but the probability that a trade moves the price decreases significantly. While the last mechanism is able to counterbalance the persistence of order flow and restore efficiency and diffusivity, the first acts in opposite direction. We introduce a statistical order book model where the persistence of the order flow is mitigated by adjusting the market order volume to the predictability of the order flow. The model reproduces the diffusive behaviour of prices at all time scales without fine-tuning the values of parameters, as well as the behaviour of most order book quantities as a function of the local predictability of order flow.
\end{abstract}

\submitto{JSTAT}


\maketitle

\section{Introduction}
A well established property of financial markets is that the order flow, defined as the process assuming value one for buyer initiated trades and minus one for seller initiated trades, displays a very slowly decaying autocorrelation function~\cite{lillo2004long,bouchaud2004fluctuations}. Since a buyer initiated trade moves on average the price up and a seller initiated trade moves it down, one would naively expect that a correlated flow induces a correlated return time series. However this latter correlation is not observed in real data because it would allow to easily predict price movements, and therefore would provide arbitrage opportunity. Reconciling correlated order flow with uncorrelated price returns is therefore a subtle issue, which is subject of current research (see also~\cite{bouchaud2009markets} for a recent review). The autocorrelation of order flow is strictly connected with the fact that large trading volumes are typically fragmented in small trades and executed incrementally (see~\cite{toth2011order}). In this way, investors are able to execute much of the large order, which is called \textit{metaorder}, minimizing the price impact and the leakage of information on their trading activity.

A possible mechanism for efficiency\footnote{In this paper we use the term “efficient” to indicate a price process that follows a random walk behaviour, in which therefore returns are not forecastable. We are not considering here the definition in which prices reflect fundamental values. See~\cite{bouchaud2009markets} for a discussion of this point.} is the \textit{asymmetric liquidity}~\cite{lillo2004long}. Defining price impact of a trade as the difference between the log-price before the next trade and the price before the current trade, asymmetric liquidity states that the price impact of a type of order (buy or sell) is inversely related to the probability of its occurrence. This means that if at a certain point in time it is more likely that the next trade is a buy rather than a sell, a buyer initiated trade will have a smaller impact than a seller initiated trade. There is therefore a compensation between probability of an event and its effect on the price.

\begin{figure}[t]
\centering
\includegraphics[width=7.5cm]{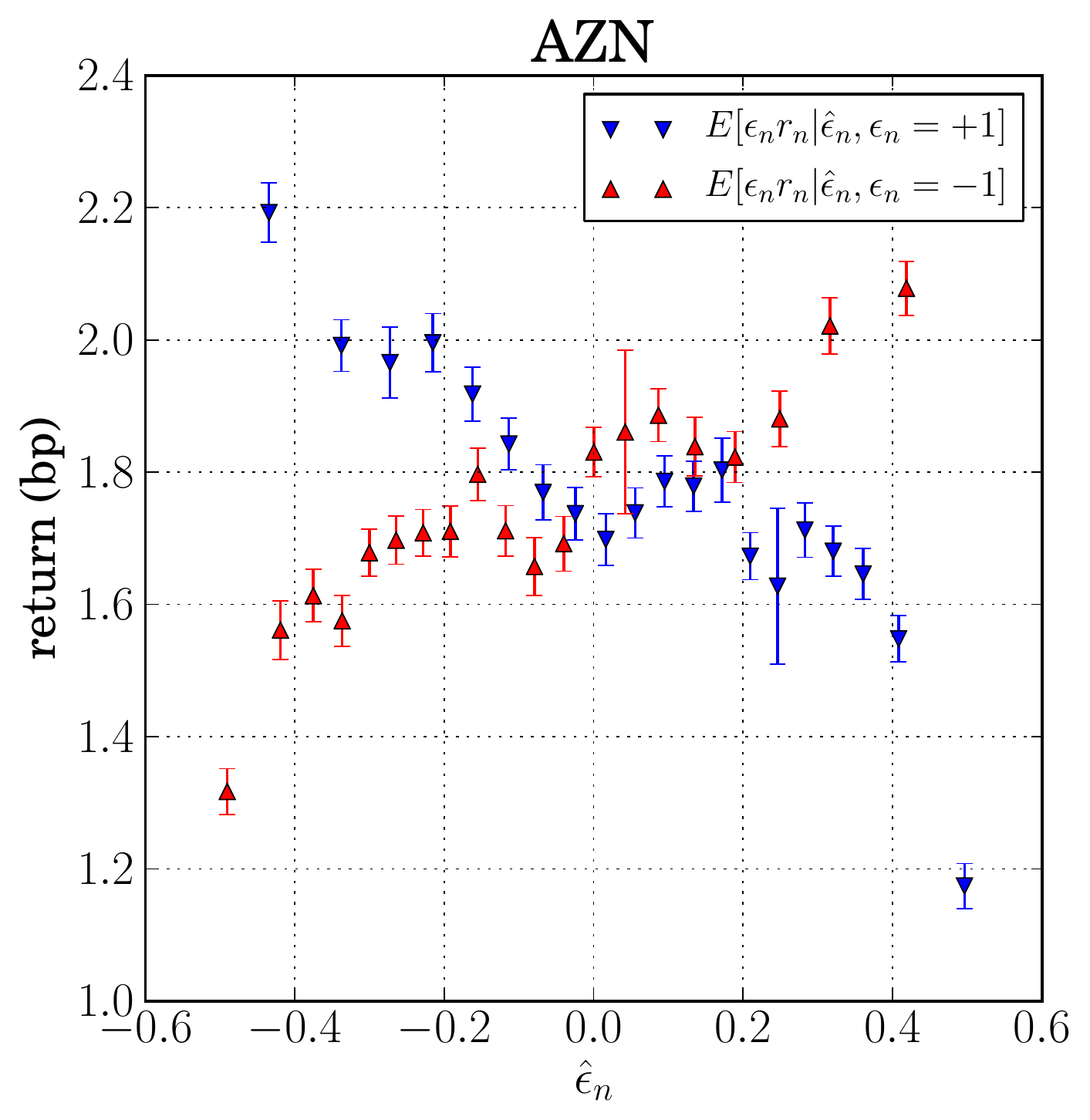}
\caption{Expected value of the product of the tick by tick return times the sign of the triggering order as a function of the order sign predictor $\hat{\epsilon}_n$ for the asset Astrazeneca in 2004.}
\label{fig:ret_vs_correctness}
\end{figure}

An empirical demonstration of the asymmetric liquidity is given in Figure~\ref{fig:ret_vs_correctness}. We indicate with $\epsilon_n$ the sign of the $n$-th trade, where $\epsilon_n=+1$ ($-1$) for a buyer (seller) initiated trade. Moreover $r_n$ is the observed price impact due to the $n-$th trade (according to the definition above). We construct an autoregressive predictor $\hat{\epsilon}_n=\mathbb{E}[\epsilon_n|\Omega_{n-1},\mathcal{M}]$ of the order flow, where $\Omega_{n-1}$ and $\mathcal{M}$ are, respectively, the information set used and the particular model used to describe the order flow (see below for more details on the predictor). We compute the average signed stock return $\epsilon_n r_n$ conditional on the sign predictor $\hat{\epsilon}_n$ and on being triggered by a buy $(\epsilon_n=+1)$ or sell $(\epsilon_n=-1)$ initiated trades. The investigated stock is Astrazeneca traded on the London Stock Exchange in the whole year 2004. From the buyer initiated trades curve (blue triangles), we observe that when the next order is more likely to be a buy (essentially due to an excess of buys in the recent past), \textit{i.e.} $\hat{\epsilon}_n>0$, a buy trade moves on average the price less than a sell trade. The opposite occurs when the next order is more likely to be a sell ($\hat{\epsilon}_n<0$). This is exactly what asymmetric liquidity prescribes and in this case the mechanism is at work even at the level of individual transactions. 

The asymmetric liquidity mechanism is conceptually clear, but it does not give any indication about the microstructural mechanisms which are responsible of it. In other words, why does a highly predictable trade impact very marginally prices? There are several possible explanations, which can be for the sake of convenience classified into two categories: The first one includes those mechanisms due to the action of the initiators of the trade and the second one where the liquidity providers are responsible. In fact, in an electronic double auction market, the initiator of the trade (the \textit{liquidity taker}) can control the volume of the market order initiating the trade. In this way she can decrease the probability that her order triggers a price change by using small volumes when her order sign is more predictable.  On the other hand, other agents submitting and cancelling limit orders (the \textit{liquidity providers}) can control the price adjustment between two trades\footnote{In electronic markets the distinction between liquidity takers and providers is a bit artificial since most of the agents use a combination of limit and market orders. However, for exposition convenience we will stick to this terminology to indicate the two types of agents.}. This can be done, for example by reverting, at least partly, the price when a predictable order arrives and moving the price in the same direction of the trade when its sign is unpredictable. In reality both types of agents are partly responsible of asymmetric liquidity. In this paper we want to investigate empirically which microstructural mechanisms enforce efficiency of prices. We will present an extensive empirical analysis aimed at identifying the main contributions to asymmetric liquidity and therefore to price efficiency.

The persistence of the order flow leads to significant challenges also in the modeling of the order book dynamics. Order book modeling is a complex task, especially if one wants to take into account the strategic behaviour of economic agents. For this reason, in recent years there has been a growing interest toward the statistical modeling of order book. This type of modeling, pioneered by~\cite{daniels2003quantitative,smith2003statistical}, drops agent rationality almost completely and describes the different types of orders as  random variables. Although no one would dispute that agents in financial markets behave strategically, and that for some purposes taking this into account is essential, there are some problems where other factors might be more important. For example, this approach has the merit that can be calibrated and tested against real data~\cite{farmer2005predictive,cont2013price}, because it presents simple quantitative laws that relate one set of market properties to another, placing restrictions on the allowed values of variables. The simplest class of these models are the so called “zero-intelligence” models, where one assumes that limit and market orders arrive randomly according to Poisson processes. Moreover queued limit orders are cancelled according to a Poisson process. To keep the model as simple as possible, there are equal rates for buying and selling, and all these processes are independent. The model just described is a prototypical queueing model of limit order book dynamics which consists in specifying the arrival rates of different types of order book events and the rules of execution of these orders. To the same class of models belong the Markovian queueing models discussed in~\cite{cont2010stochastic,cont2013price}. 

However, all these modeling approaches neglect the persistence of the order flow, which destroys the Markovian feature of the modeling and leads to unrealistic behaviour and wrong predictions on the dynamics of prices. To be specific, we have calibrated a zero intelligence model~\cite{daniels2003quantitative} on the stock Astrazeneca in the whole year 2004. We have then replaced in the model the Poisson market order flow with the one extracted from the real data. We have then studied the diffusivity properties of the generated prices. To this end we computed the “signature plot” of the model, \textit{i.e.} the plot of the quantity
\begin{equation}
\sigma(\ell)=\sqrt{\frac{\mathbb{E}[(p_{n+\ell}-p_n)^2]}{\ell}}
\label{eqn:signplot}
\end{equation}
where the average $\mathbb{E}[(p_{n+\ell}-p_n)^2]$ is done over different instants of time $t_n$, which is the time that immediately precedes the $n$-th transaction. This quantity is a measure of the volatility of the price process on time scale $\ell$. For a purely diffusive process, $\sigma(\ell)=D$ is constant and independent of $\ell$. If $\sigma(\ell)$ decays when $\ell$ increases, the price motion is “sub-diffusive” and it has a mean-reverting behaviour (there exist negative correlations between lagged returns). On the other hand, if $\sigma(\ell)$ increases when $\ell$ increases, the price process is super-diffusive and it shows a trending behaviour (positive correlations between lagged returns). Thus, a necessary condition for price efficiency is that $\sigma(\ell)=D$ is constant. It is well known that real price time series show a sub-diffusive behaviour for very short lags, and then the price is diffusive at the other lags.

\begin{figure}[t]
\centering
\includegraphics[width=0.75\columnwidth]{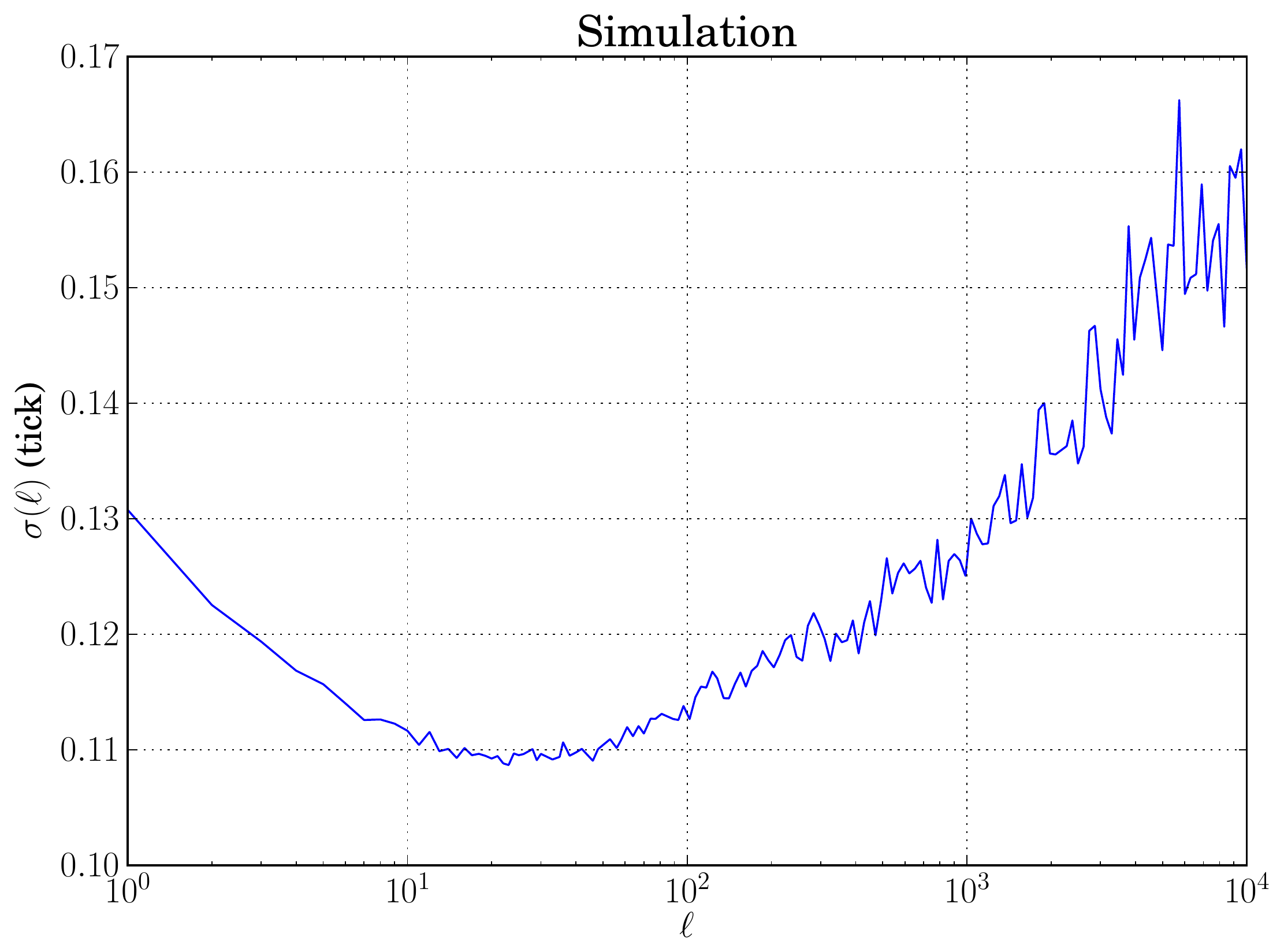}
\caption{Signature plot (see Equation~\ref{eqn:signplot}) of a simulation of the model discussed in~\cite{smith2003statistical}. The parameters are calibrated on Astrazeneca in the whole year 2004 and we used the real market order flow as input of the model.}
\label{fig:smith_real_flow}
\end{figure}

In Figure~\ref{fig:smith_real_flow} we present the result of the above described Monte Carlo simulation. We observe that price is initially sub-diffusive and for lags larger than $\sim 30$ trades it becomes super-diffusive. It is evident that embedding a persistent order flow in the framework developed in~\cite{smith2003statistical} induces a super-diffusive behaviour of prices. An analogous pattern should be expected from any Markovian model. Recent attempts of modeling the limit order book with strongly persistent order flow include~\cite{mike2008empirical,toth2011anomalous,mastromatteo2013agent}. In these papers, however, either the diffusivity is not guaranteed~\cite{mike2008empirical}, or it is attained by fine tuning the value of a parameter describing the counterbalancing reaction to the order flow persistence~\cite{toth2011anomalous}. More importantly, in this last case (detailed below) diffusivity is recovered up to the time scale of the lifetime of limit orders, while for longer time scales the price becomes super-diffusive. 

In this paper, we propose a new statistical model of the limit order book which is able to give diffusive prices at all time scales and to reproduce the empirical statistical properties observed to explain the underlying mechanisms of the asymmetric liquidity. The key ingredient of the modeling is a liquidity dynamics that adapts itself to the degree of predictability of the order flow. In other words, instead of having a fine tuning of a parameter that guarantees (approximate) diffusivity, we model liquidity as an adaptive process that responds to the local predictability of the order flow and gives exact diffusivity.

The paper is organized as follows. Section~\ref{sec:Datasets} describes the investigated data and in Section~\ref{sec:Empirical} we present the empirical findings on the microstructural mechanisms responsible of the restoration of price efficiency and diffusivity when the order flow is correlated. In Section~\ref{sec:Statistical} we discuss the incompatibility between existing limit order book models, correlated order flow, and price diffusion. Section~\ref{sec:Adaptive} presents our model and the main theoretical findings and in Section~\ref{sec:Numerical} we discuss some numerical simulations of the model. Finally, in Section~\ref{sec:Conclusions} we draw some conclusions.

\section{Dataset description}
\label{sec:Datasets}
The data used in our empirical analysis belong to two distinct datasets spanning different time periods and recorded on different markets. The first dataset corresponds to the trading activity of two stocks traded on the London Stock Exchange (LSE) during the whole year 2004. The second one is more recent and records the activity of two stocks traded on the NASDAQ stock exchange in New York. This dataset covers only a short period of time, namely July and August 2009, but the higher trading frequency partially compensates for the shorter horizon.

The LSE dataset includes the limit order book information about the Astrazeneca (AZN) and Vodafone (VOD) stocks. The data come from the Stock Exchange electronic Trading Service (SETS), the LSE's flagship electronic order book, and contain the detailed description of all order book events (submissions of limit and market orders and cancellations of outstanding orders) which occurred in the whole year of 2004 (254 trading days). In particular, the information concerning the market order events report the execution time of the event, the sign of the order (\emph{i.e.} if it is buyer or seller initiated), the traded volume and price. We select AZN and VOD because of the sensible difference in the discretization of the prices. AZN has a tick size-price ratio of few basis points, whereas VOD is characterized by a very large tick size-price ratio (see Table~\ref{tab:info_datasets}). For this reason, we refer to the former as a small-tick stock, while to the latter as a large-tick stock.

The second dataset includes all the executed trades and order book updates of stocks traded at the NASDAQ market in New York. In particular, we analyse two liquid stocks, namely a small-tick stock, Apple (AAPL), and, a large-tick stock (relatively to AAPL), Microsoft (MSFT). The data cover 42 days of trading activity during July and August of 2009. For the two datasets, we have taken care of the possibility that the execution of a single market order hitting several existing limit orders produces many records with the same timestamp. We have aggregated them in a single market order, whose volume is the cumulative volume of the components. A summary of the properties of the four stocks is collected in Table~\ref{tab:info_datasets}.

The empirical analysis has been performed using a code written in the \textit{Python} programming language. Specifically, we have used the scientific \textit{SciPy} libraries, the statistical library \textit{StatsModels}, while all the graphs have been generated by the plotting library \textit{Matplotlib}.

\begin{table}[t]
\caption{\label{tab:info_datasets}Summary of the investigated stocks. The average stock price is expressed in U.S. Dollars for AAPL and MFST, whereas it is expressed in Great Britain Pounds for AZN and VOD. The average intertrade time and tick size-price ratio are given in seconds and in basis points, respectively.}
\begin{indented}
\item[]\begin{tabular}{@{}lrrrrr}
\br
\multirow{2}{*}{Symbol} & \multirow{2}{*}{Year} &Number of & Average & Average & Average \\
& & trades & intertrade time & stock price & tick size-price ratio \\
\mr
AAPL & \multirow{2}{*}{2009} & 857,925 & 1.1  s & 157.17 USD &  0.6 bp\\
MSFT & & 575,040 & 1.7  s & 23.74 USD &  4.2 bp\\
\mr
AZN  & \multirow{2}{*}{2004} &405,481 & 23.1 s & 24.38  GBP &  4.1 bp\\
VOD  & & 411,736 & 22.9 s & 1.34 GBP & 18.7 bp\\
\br
\end{tabular}
\end{indented}
\end{table}

\section{Empirical evidences of the origin of asymmetric liquidity}
\label{sec:Empirical}
In this section we investigate empirically the mechanisms responsible for restoring efficiency and, as a consequence, diffusivity of prices. More specifically, we perform an empirical analysis in order to investigate the origin of the asymmetric liquidity mechanism. We consider the variables of the order book at the instant of time $t_n$ which immediately precedes the $n$-th transaction. The best ask price $A_n$ is the lowest price among the sell limit orders in the book at time $t_n$. Symmetrically, the best bid price $B_n$ is the highest price among the buy limit orders quoted in the book. Then, we can define the midpoint price $P_n=(A_n+B_n)/2$, and introduce the logarithmic quantities, namely, the best ask log-price $a_n=\log A_n$, the best bid log-price $b_n=\log B_n$, and the log-midprice $p_n=\log P_n$. Then, we characterize the limit order book sparsity by means of the bid (ask) gap and the volume at the best bid (ask) quote. The former is the logarithmic price difference between the best bid quote and the second best bid quote, \emph{i.e.} $g_n^B=b_n-b_n^{2nd}$ (between the second best ask quote and the best ask quote, \emph{i.e.} $g_n^A=a_n^{2nd}-a_n$), where the variable $b_n^{2nd}$ ($a_n^{2nd}$) is the second best bid (second best ask) price. The last investigated quantity is the amount of shares available at the best bid $v_n^B$ (at the best ask $v_n^A$). 

\subsection{Predictability of market order flow}
Given the crucial role played by the order flow and following~\cite{lillo2004long} we introduce the \textit{sign predictor} and study the variables characterizing the state of the order book conditioned on its value. Assuming a model for the order flow process, we can compute at each time $t_{n-1}$ the expected value of the future market order sign, $\hat{\epsilon}_n=\mathbb{E}[\epsilon_n|\Omega_{n-1},\mathcal{M}]$, conditional to the information set $\Omega_{n-1}$ (typically the past order flow) and the particular model $\mathcal{M}$ used to describe the order flow. In early works, the order flow was modelled in terms of a real valued autoregressive process, but clearly the order flow takes only discrete values. Thus, we model it by means of a Discrete Autoregressive process of order $p$ (\textit{DAR(p)}), which is an integer valued process easy to calibrate on real data. The \textit{DAR(p)} was introduced in a series of papers~\cite{jacobs1978discrete,jacobs1983stationary} and describes a sequence of stationary discrete random variables with the properties of a Markov process of order $p$. In~\ref{app:DAR} we review the main properties of the \textit{DAR(p)} process, its auto-covariance structure and how to forecast it. 

The first step in our analysis is the estimation of the \textit{DAR(p)} model by using the observed order flow $\epsilon_n$. When performing estimation we discard the first $p$ trades of each trading day to avoid spurious overnight effect. The idea is that a sign predictor obtained with the order flow partially observed the previous trading day is less significant than a sign predictor computed with the order flow which belongs entirely to the same trading day.

In order to evaluate the predictability of order flow, we rely on the sign predictor defined by Equation~\ref{eqn:epshat} in~\ref{app:DAR}, $\hat{\epsilon}_n=\mathbb{E}[\epsilon_n|\Omega_{n-1},\mathrm{DAR(p)}]$, and specify a loss function indicating how much our prediction is correct. We employ the Mean Square Error (MSE) defined as
\begin{equation}
  \mbox{MSE}(\hat{\epsilon}_n)=\mathbb{E}\left[(\epsilon_n-\hat{\epsilon}_n)^2\right]\,.
\end{equation}

This function has an upper bound equal to $\mbox{MSE}(\hat{\epsilon}_n)=4$, when the prediction is always wrong, and a lower bound $\mbox{MSE}(\hat{\epsilon}_n)=0$ when the prediction is systematically correct. When $\mathbb{E}[\epsilon_n\cdot\hat{\epsilon}_n]=0$, the MSE is equal to $1+\mathbb{E}[\hat{\epsilon}_n^2]$ and this value is the upper bound in case of no predictability of the model.

\begin{figure}[t]
  \begin{minipage}[c]{0.5\columnwidth}
    \includegraphics[width=\columnwidth]{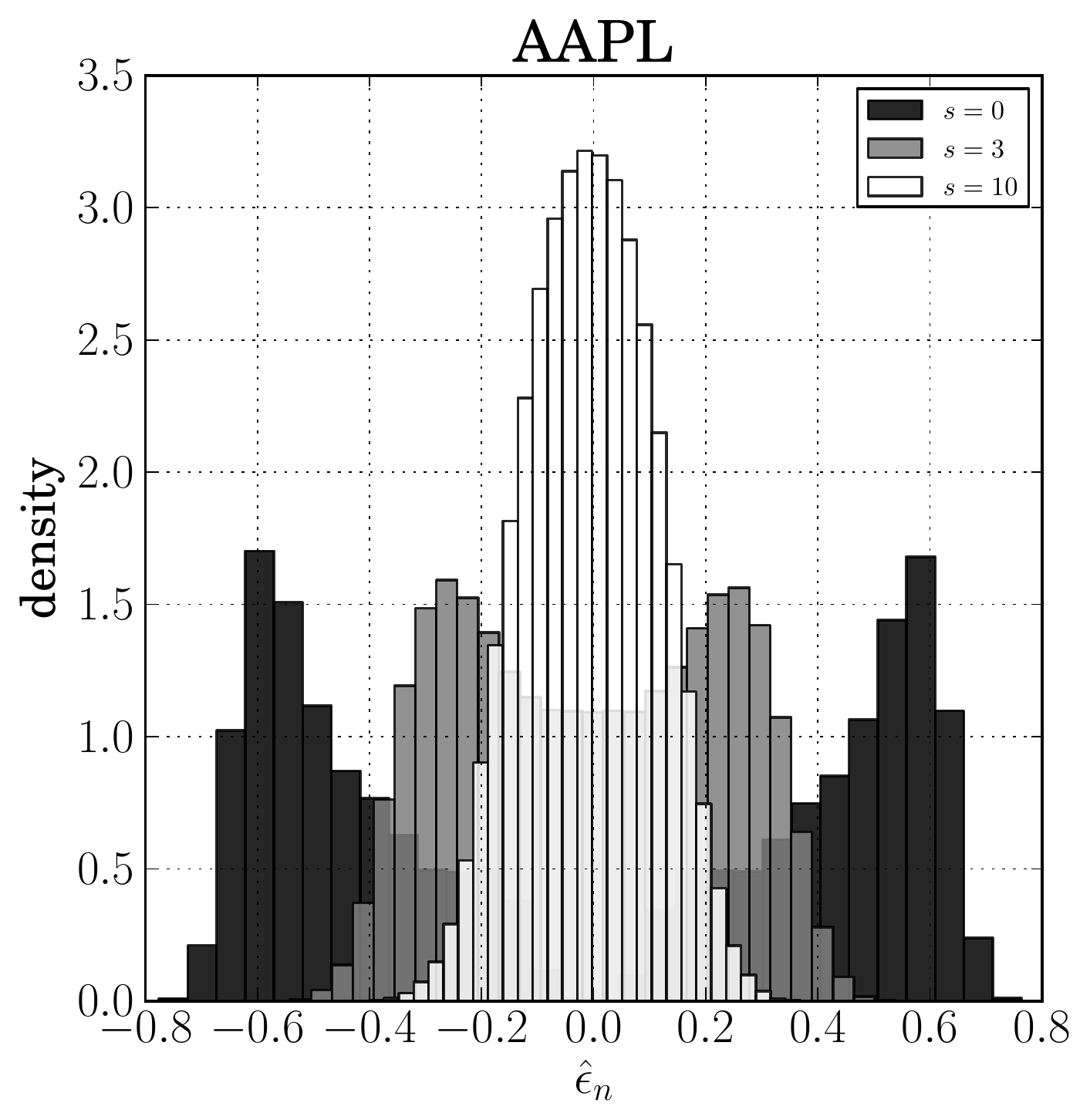}
  \end{minipage}%
  \begin{minipage}[c]{0.5\columnwidth}
    \includegraphics[width=\columnwidth]{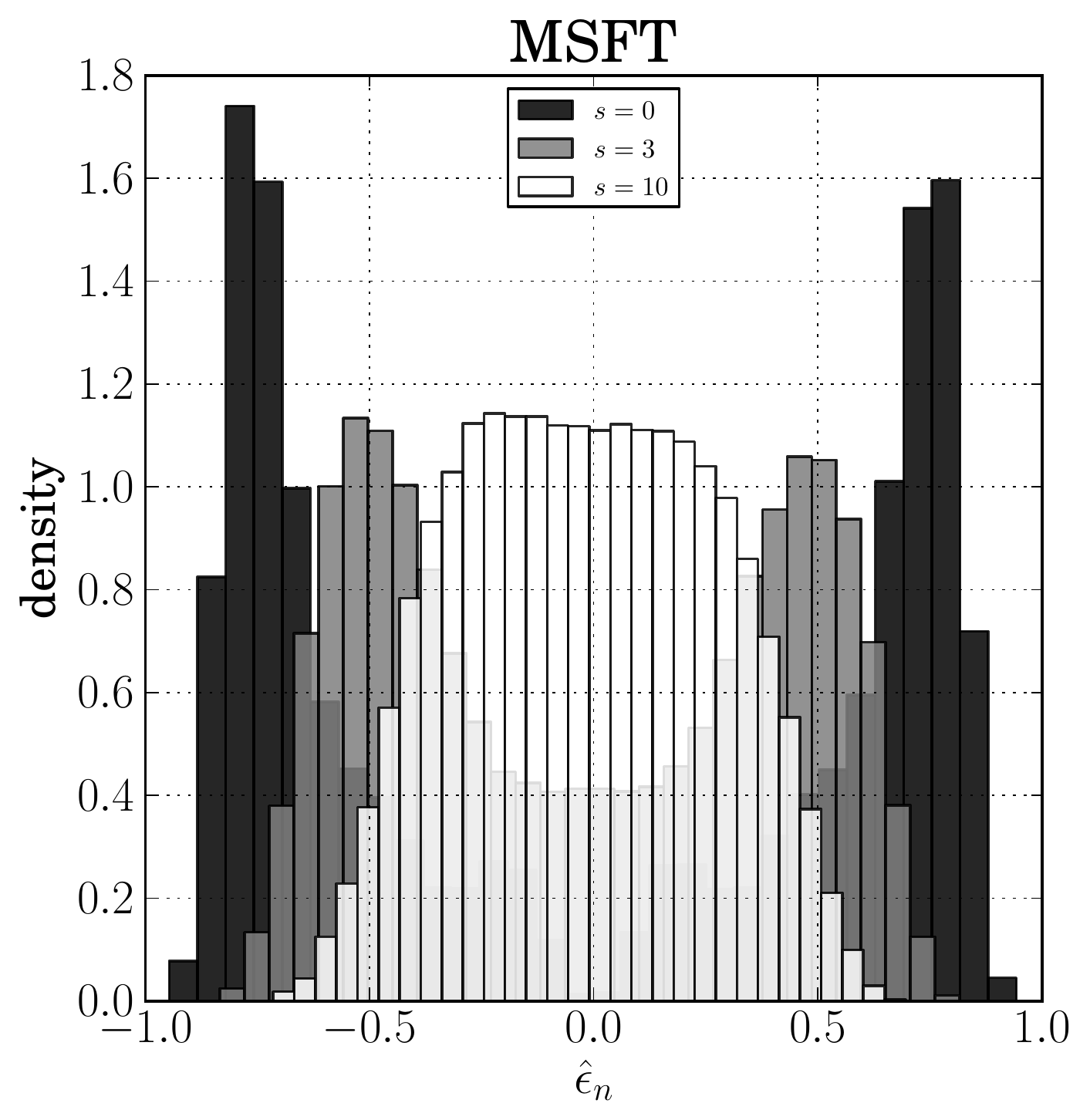}
  \end{minipage} \\
  \begin{minipage}[c]{0.5\columnwidth}
    \includegraphics[width=\columnwidth]{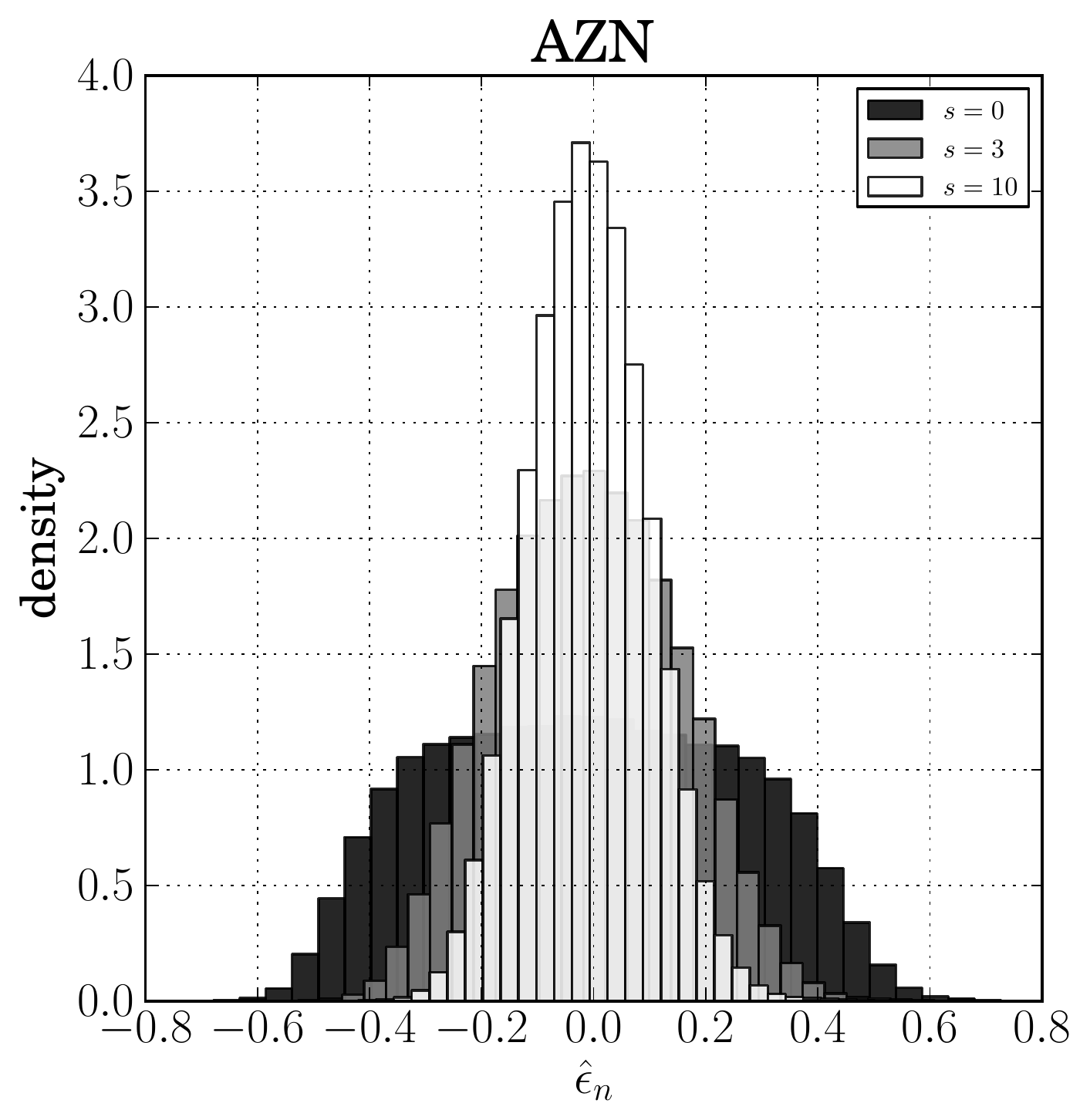}
  \end{minipage}%
  \begin{minipage}[c]{0.5\columnwidth}
    \includegraphics[width=\columnwidth]{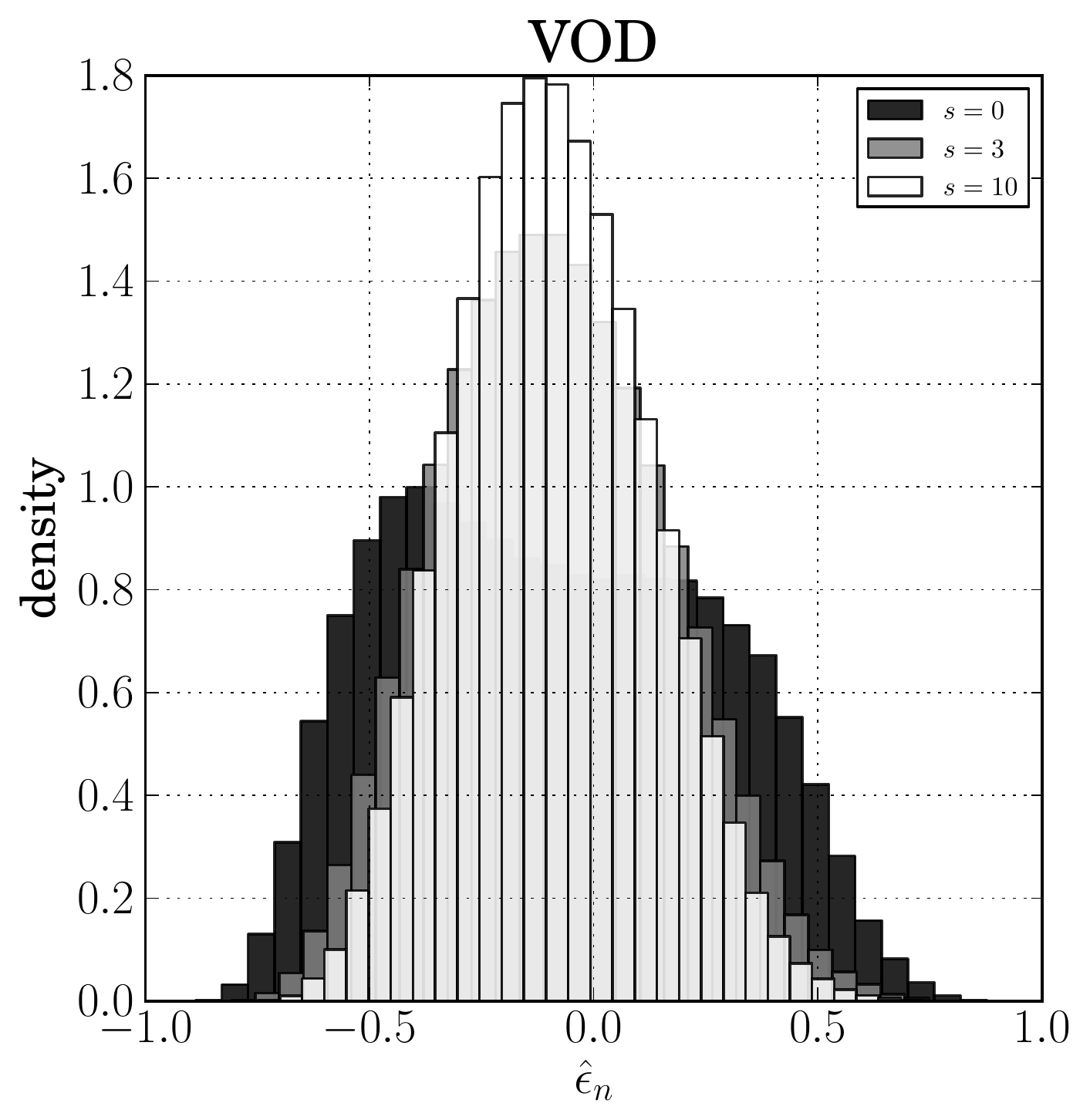}
  \end{minipage}
  \caption{Distributions of the sign predictor for the stocks AAPL, MSFT, AZN, VOD and $s=0,3,10$ trades.}
  \label{fig:dist_pred}
\end{figure}

In Table~\ref{tab:mse} we list the MSE values computed for AAPL, MSFT, AZN, VOD, and three different values $p=100,500,700$. As anticipated, we see that the MSE values obtained for each stock are almost independent from the order $p$ of the auto-regression\footnote{This might depend on the choice of MSE as a loss function.}. Thus, for the rest of the analysis we fix it equal to 500. 

\begin{table}[t]
\caption{\label{tab:mse}MSE values and standard errors for AAPL, MSFT, AZN, VOD, and for three different values $p=100,500,700$. Last three columns: upper bound for MSE in case of absence of predictability.}
\small
\begin{tabular}{@{}lcccccc}
\br
\multirow{2}{*}{Symbol} & \multicolumn{3}{c}{MSE DAR(p)} & \multicolumn{3}{c}{$1+\mathbb{E}[\hat{\epsilon}_n^2]$ DAR(p)}\\ 
& $p=100$ & $p=500$ & $p=700$ & $p=100$ & $p=500$ & $p=700$\\
\mr
AAPL & $0.7692 \pm 0.0009$ & $0.7686 \pm 0.0009$ & $0.7684 \pm 0.0009$ & $1.2308$ & $1.2313$ & $1.2315$ \\
MSFT & $0.5660 \pm 0.0012$ & $0.5651 \pm 0.0012$ & $0.5649 \pm 0.0012$ & $1.4336$ & $1.4344$ & $1.4346$ \\
\mr
AZN  & $0.9332 \pm 0.0008$ & $0.9321 \pm 0.0008$ & $0.9317 \pm 0.0008$ & $1.0667$ & $1.0678$ & $1.0683$ \\
VOD  & $0.8722 \pm 0.0010$ & $0.8709 \pm 0.0010$ & $0.8705 \pm 0.0010$ & $1.1198$ & $1.1211$ & $1.1215$ \\
\br
\end{tabular}
\end{table}

We notice that the MSE values of small-tick stocks are always higher than the values of large-tick stocks within the same dataset. More interestingly, there are substantial differences between the MSE of the stocks belonging to the LSE and NASDAQ datasets. The MSE for AAPL and MSFT are smaller than those of AZN and VOD, which are closer to the value $1+\mathbb{E}[\hat{\epsilon}_n^2]$. We also compute the lagged sign predictor $\hat{\epsilon}_{n+s}=\mathbb{E}[\epsilon_{n+s}|\Omega_{n-1},\mathrm{DAR(500)}]$ for $s\geq 0$ and in Figure~\ref{fig:dist_pred} we show the distributions of the sign predictor values corresponding to $s=0,3,10$ trades. Two distinct regions characterize the predictor distributions: The region where the values of the sign predictor are close to the extrema of the support, and the region where the predictor is close to zero. The former is the high predictability region and indicates that the corresponding market operates in a high predictable regime. This is the case for the NASDAQ dataset for $s=0$ and the effect is more intense for the large-tick stock MSFT than for AAPL. The latter is the low predictability region where typically $\hat{\epsilon}_{n+s}\approx 0$. The LSE market for both AZN and VOD operates under this regime confirming the previous findings about the MSE values. It is worth noticing that when the value of $s$ increases from 0 to 10, the predictor distribution converges to the low predictability region also for the assets belonging to the NASDAQ dataset. This convergence is faster for the small-tick asset AAPL than for MSFT, thus we expect that possible divergences from an efficient behaviour should be more evident for large-tick stocks.

\subsection{Best bid and ask volume conditional expectation}
Equipped with the order sign predictor, we measure the order book state variable at the instant of time which immediately precedes the $n$-th transaction ($t_n^-$) conditional on the predictor value $\hat{\epsilon}_n$. We use the \textit{DAR(500)} model to construct a sign predictor and we split the range of $\hat{\epsilon}_n$ into a finite number of bins. We do not evenly sample the bins, but we fix the bins according to the empirical quantiles requiring that the number of empirical sign predictors falling within each bin is the same.

\begin{figure}[t]
  \begin{minipage}[c]{0.5\columnwidth}
    \includegraphics[width=\columnwidth]{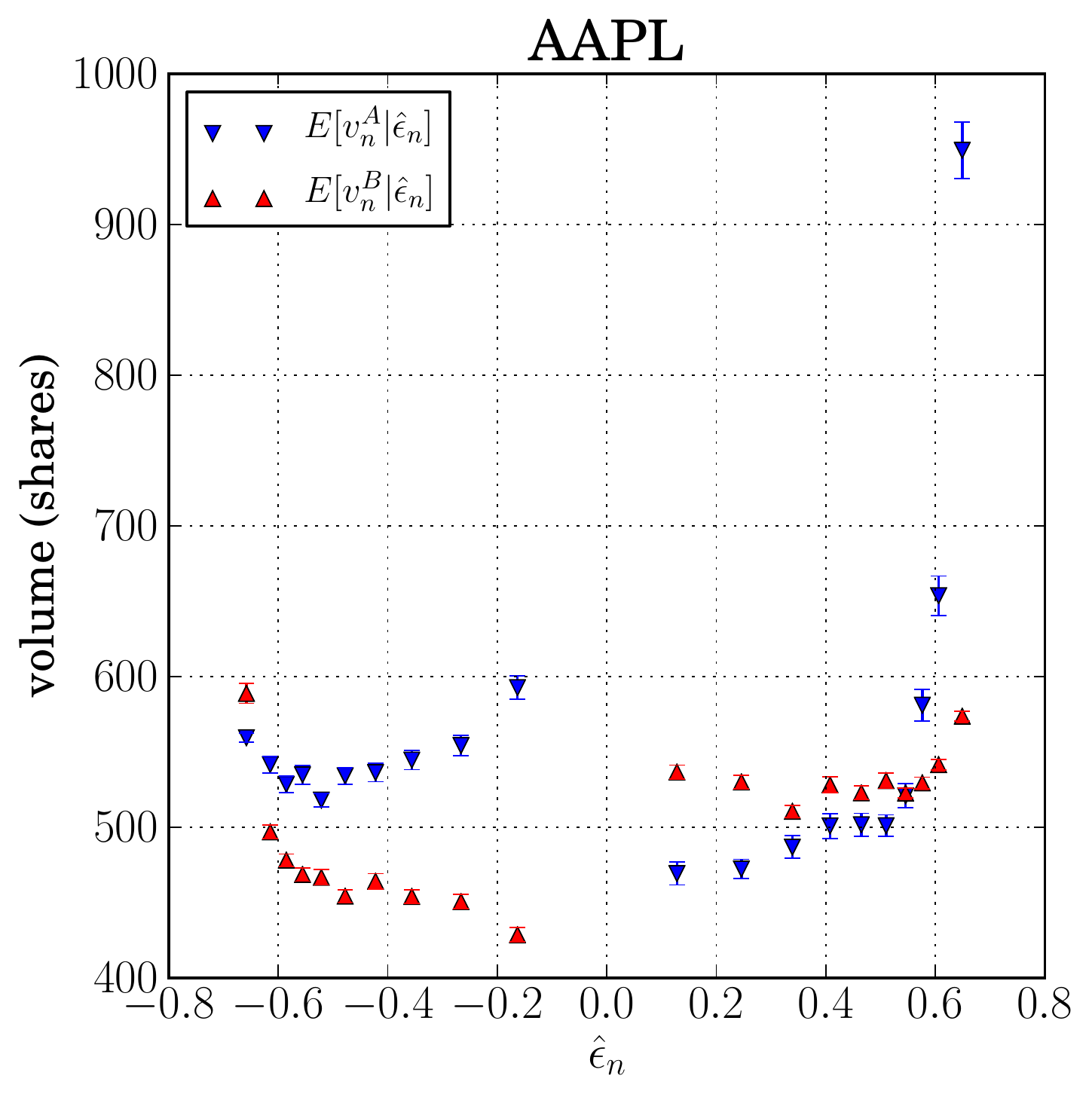}
  \end{minipage}%
  \begin{minipage}[c]{0.5\columnwidth}
    \includegraphics[width=\columnwidth]{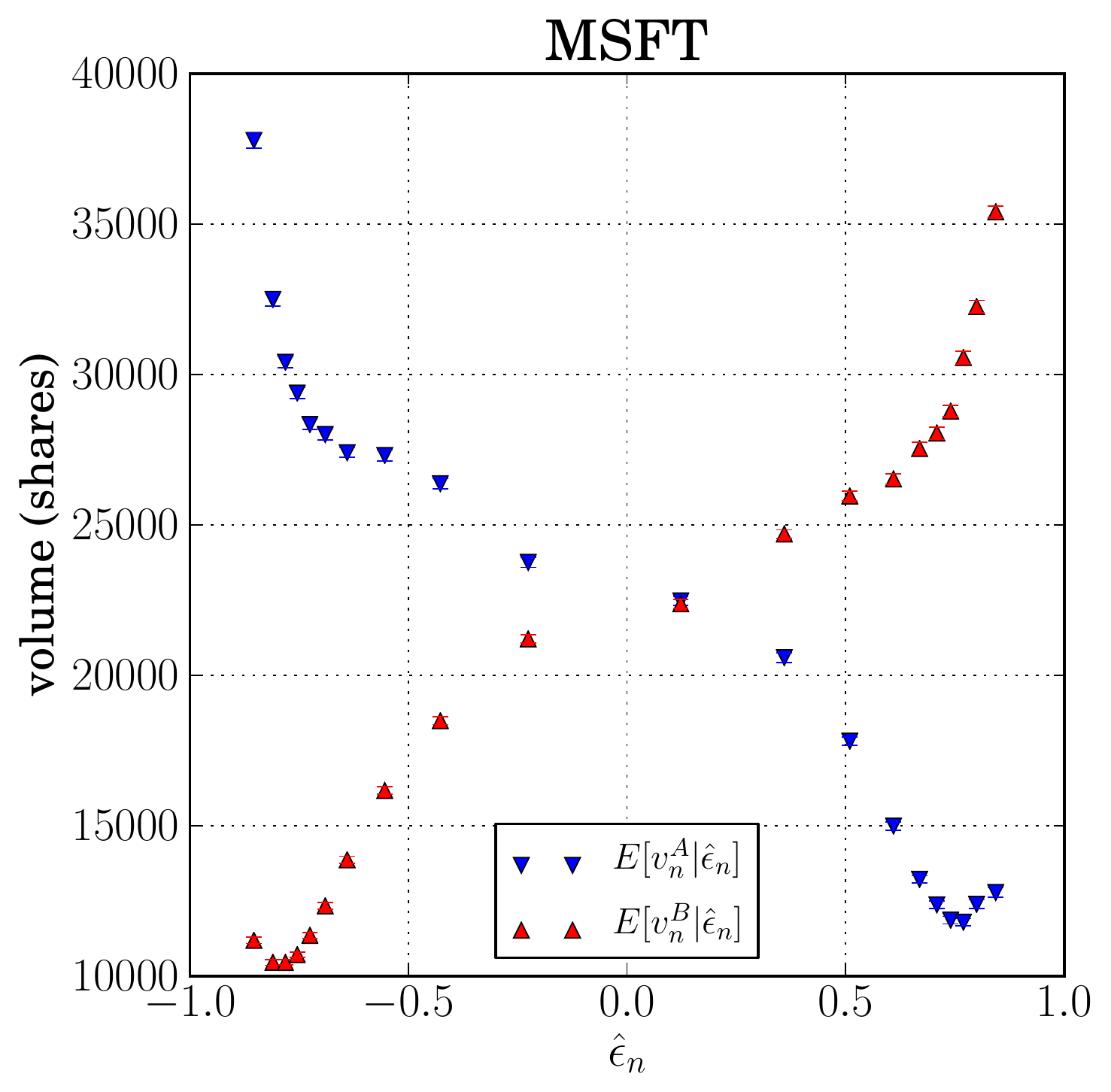}
  \end{minipage} \\
  \begin{minipage}[c]{0.5\columnwidth}
    \includegraphics[width=\columnwidth]{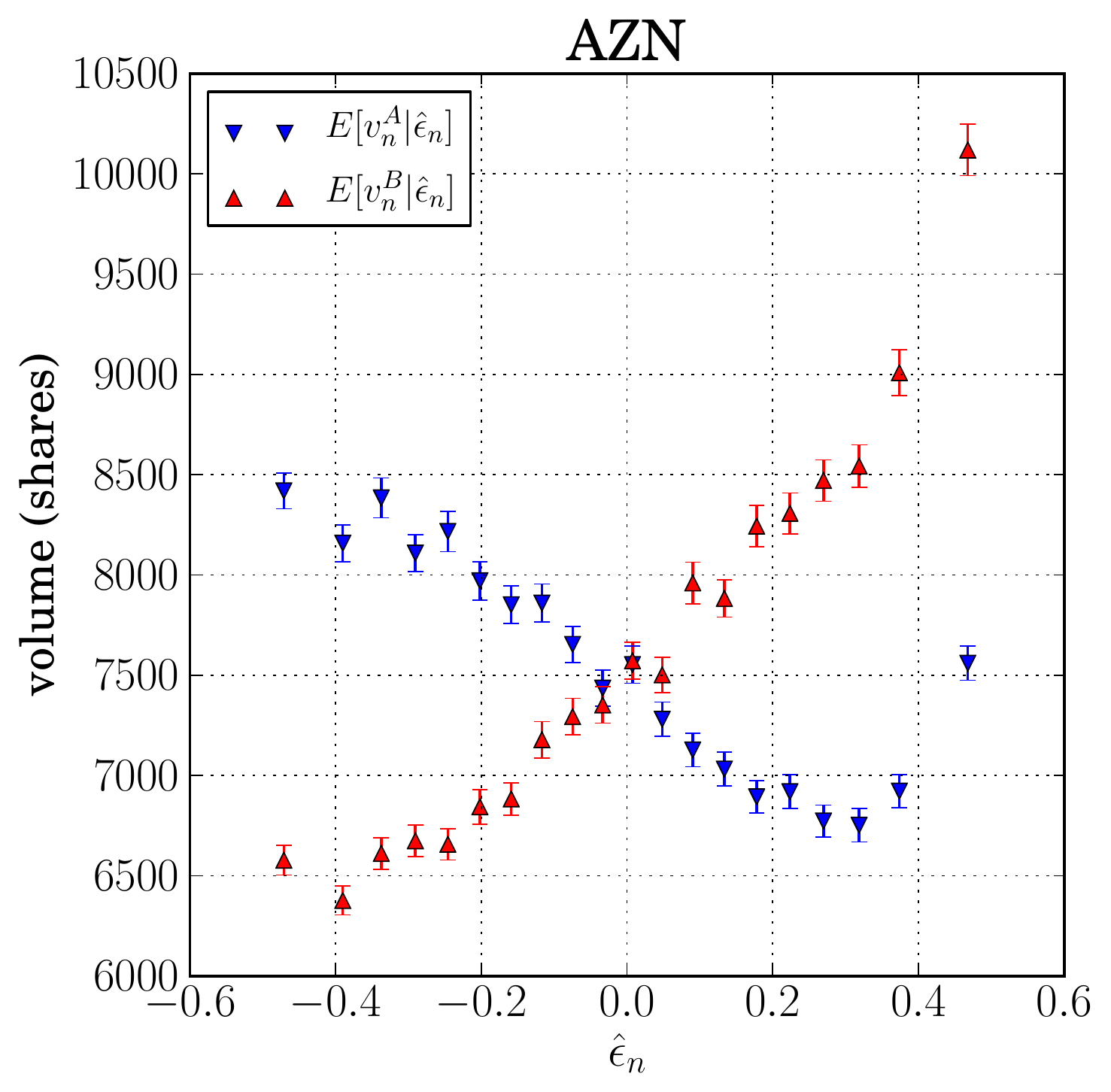}
  \end{minipage}%
  \begin{minipage}[c]{0.5\columnwidth}
    \includegraphics[width=\columnwidth]{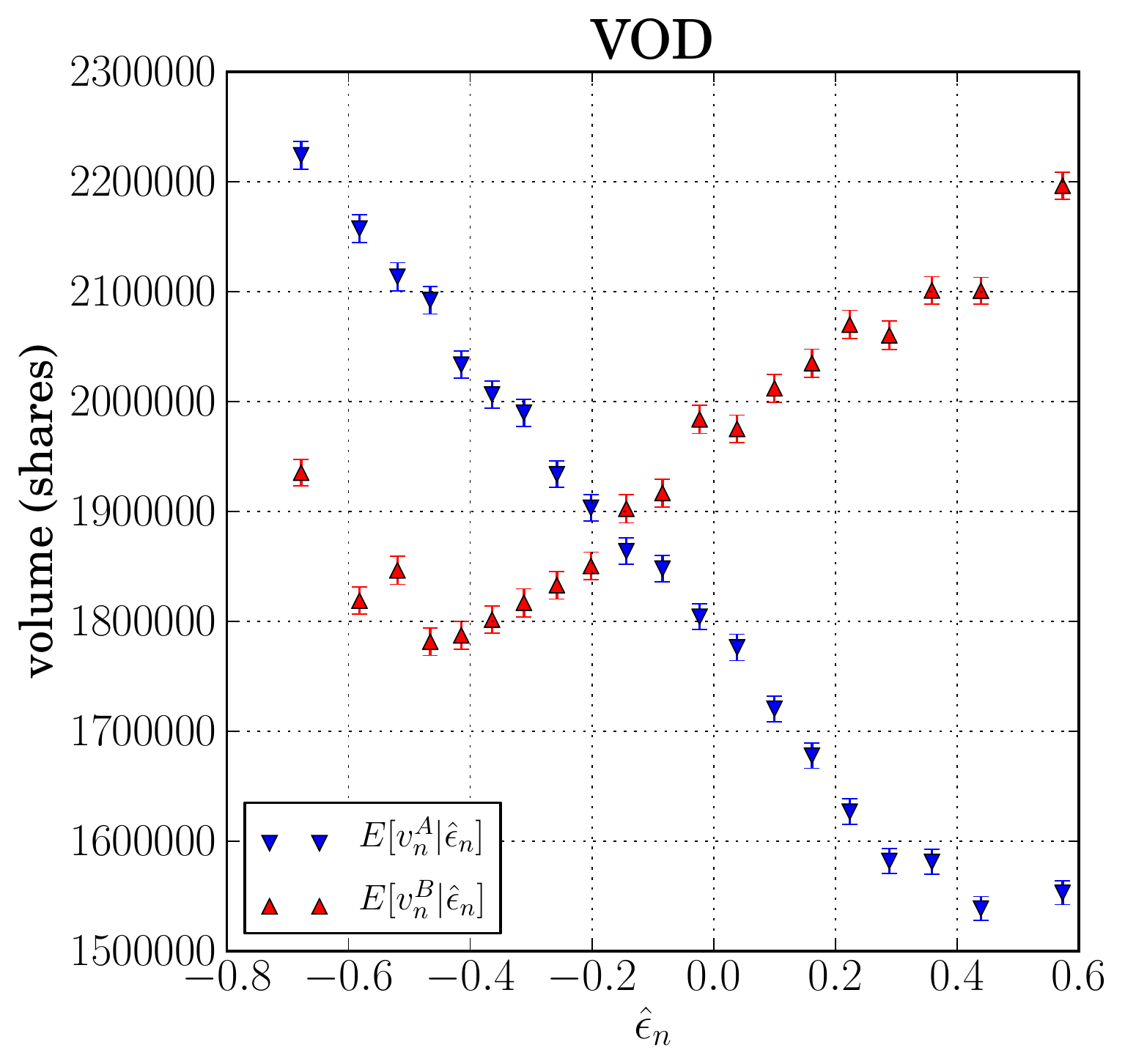}
  \end{minipage}
  \caption{Conditional best ask volumes $\mathbb{E}[v_n^A|\hat{\epsilon}_n]$ and conditional best bid volumes $\mathbb{E}[v_n^B|\hat{\epsilon}_n]$ on different sign predictor values, for four stocks (AAPL, MSFT, AZN, VOD). The error bars are standard errors.}
  \label{fig:volbest}
\end{figure}

The first quantity that we consider is the volume outstanding at the best quotes. The best bid and ask volumes are natural indicators of the liquidity available on each side of the order book. We condition the volume at the best ask $v_n^A$ and the volume at the best bid $v_n^B$ on the level of the sign predictor, and we take their conditional expectation, $\mathbb{E}[v_n^A|\hat{\epsilon}_n]$ and $\mathbb{E}[v_n^B|\hat{\epsilon}_n]$. In Figure~\ref{fig:volbest} we show the conditional average of the best volumes as a function of $\hat{\epsilon}_n$ for the assets AAPL, MSFT, AZN, and VOD. 

We start commenting on the stocks which belong to the LSE dataset. We recall from Figure~\ref{fig:dist_pred} that for AZN and VOD the sign predictor is mainly distributed in the low predictability region. We focus on the behaviour of the volumes at the best ask, for those at the bid side similar comments apply. When buy orders are more likely than sell orders ($\hat{\epsilon}_n>0$) the average volume outstanding at the ask side is smaller than the volume outstanding at the bid side. Moreover, when the sign predictor increases, the best ask volumes decrease and the best bid volumes increase. This behaviour is compatible with a model where liquidity takers mechanically erode the liquidity available at the opposite side of the book. Indeed, a positive sign predictor means that the recent order flow has been dominated by a sequence of buy orders and the volume outstanding at ask side of the book has been eroded by market orders. However, when the predictor is approaching the upper bound ($\hat{\epsilon}_n=1$), the volume at the ask side starts to increase. Indeed, high predictability of the order flow means that significant information about the intentions of the liquidity taker has been released to the market and a large fraction of her metaorder has been executed. Thus, the probability that the metaorder is close to expiration is high and it becomes pressing for liquidity providers to refill the ask side of the order book at the best price. For the LSE dataset both the large-tick and the small-tick assets manifest the same behaviour. When we switch to the NASDAQ dataset the picture is less clear. While the large-tick asset MSFT follows the same pattern of the LSE assets, for AAPL the situation is different. When buy orders are very likely ($\hat{\epsilon}_n \approx 1$) the volume refill of the liquidity providers dominate and the average outstanding volume increases with $\hat{\epsilon}_n$. However, since we know that the sign predictor is concentrated in the high predictable region conclusions about the behaviour of the average volume in the central region are less definitive.

Finally, we notice that the volumes at the best quotes are higher in large-tick stocks than in small-tick stocks, not only in the number of shares but also in dollar (pound) value. In fact, if we multiply the average volume by the average price, we find that there is a difference of one (two) order of magnitude between large-tick and small-tick stocks of the NASDAQ (LSE) dataset. This difference is likely caused by the discretization effect of limit prices since liquidity providers suffer less availability of quotes in large-tick than in small tick stocks and they pile up volumes at the same price.

\subsection{Bid and ask gap conditional expectation}
Following the same procedure described in the previous section, we consider the conditional distribution of the bid (ask) logarithmic gap between the best bid price and the second best quote (the best ask price and the second best ask quote) immediately before the transaction time $t_{n}$. As before we compute the expectation of these quantities conditioning on the level of the sign predictor, \emph{i.e.} we compute $\mathbb{E}[g_n^A|\hat{\epsilon}_n]$ and $\mathbb{E}[g_n^B|\hat{\epsilon}_n]$.

\begin{figure}[t]
  \begin{minipage}[c]{0.5\columnwidth}
    \includegraphics[width=\columnwidth]{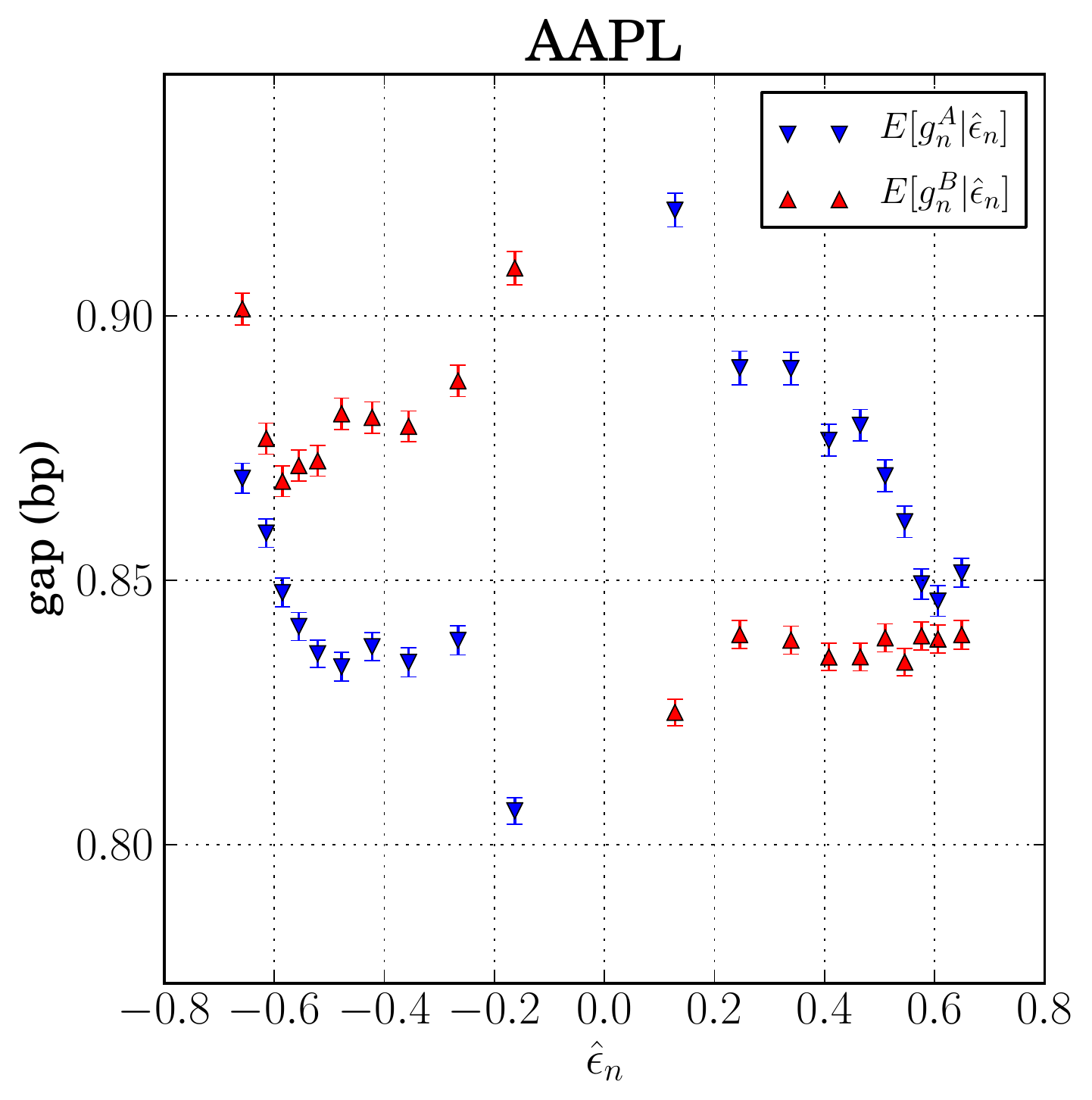}
  \end{minipage}%
  \begin{minipage}[c]{0.5\columnwidth}
    \includegraphics[width=\columnwidth]{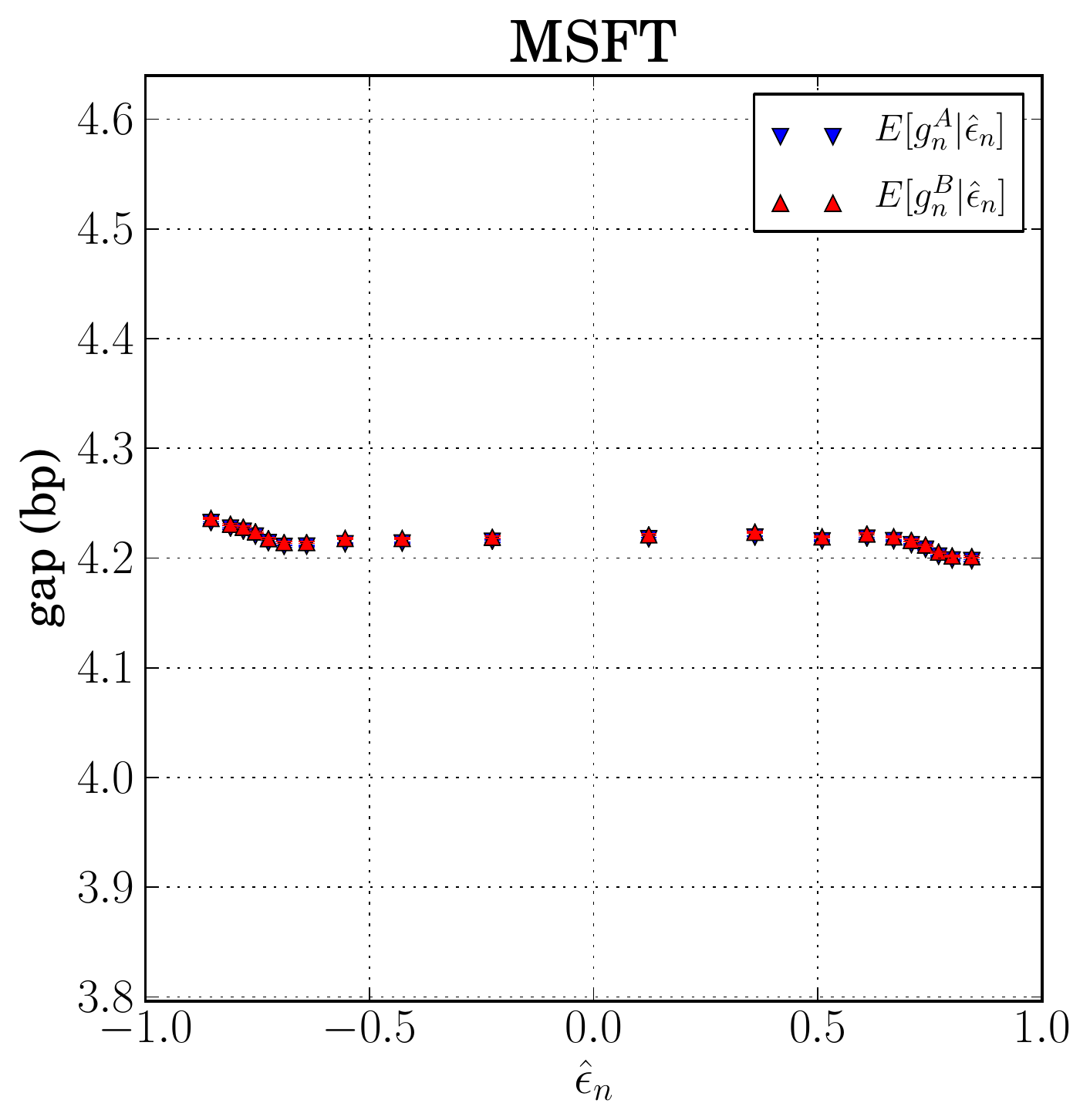}
  \end{minipage} \\
  \begin{minipage}[c]{0.5\columnwidth}
    \includegraphics[width=\columnwidth]{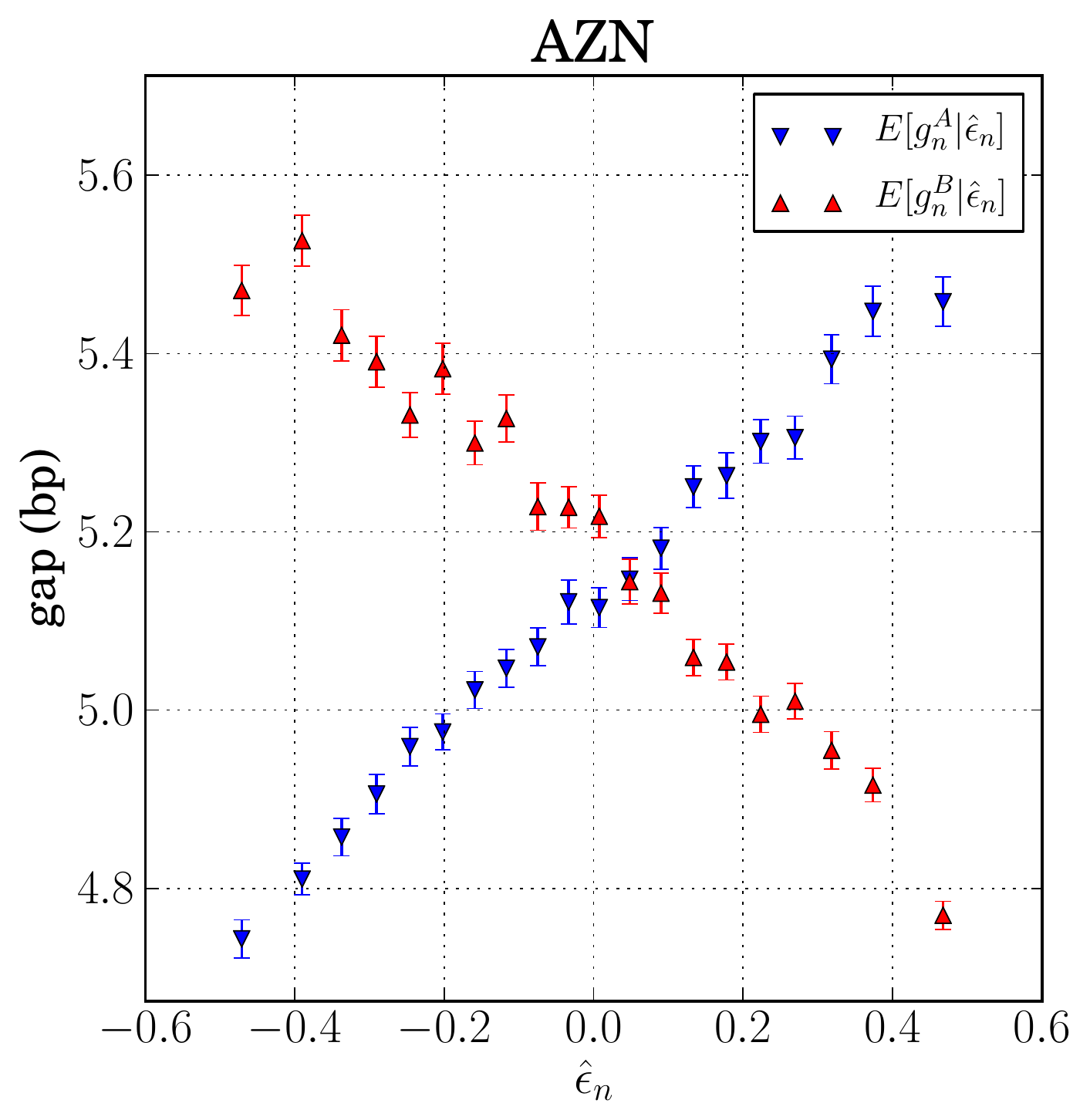}
  \end{minipage}%
  \begin{minipage}[c]{0.5\columnwidth}
    \includegraphics[width=\columnwidth]{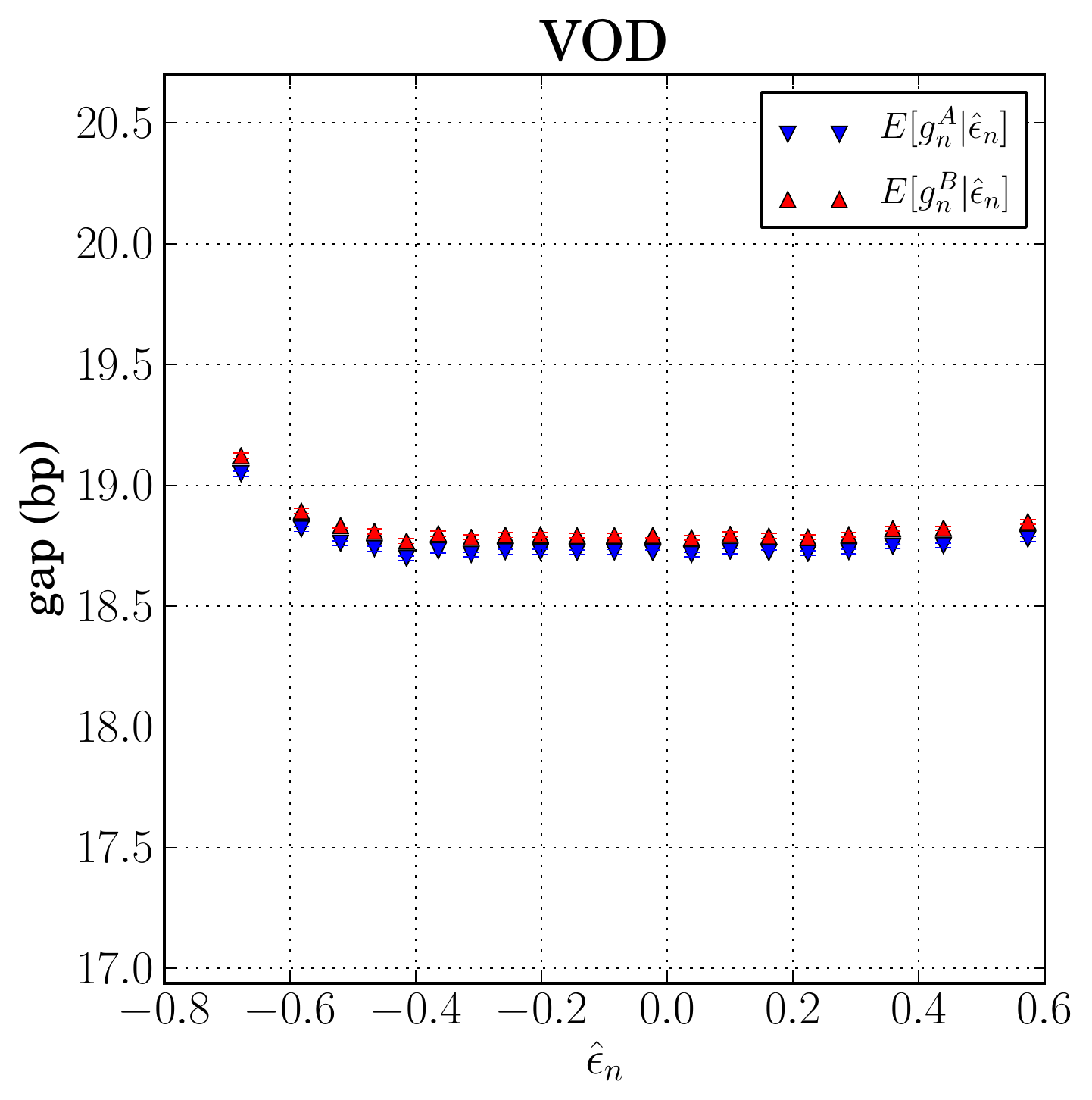}
  \end{minipage}
  \caption{Conditional ask gap $\mathbb{E}[g_n^A|\hat{\epsilon}_n]$ and conditional bid gap $\mathbb{E}[g_n^B|\hat{\epsilon}_n]$ as a function of the sign predictor for the four stocks (AAPL, MSFT, AZN, VOD). The error bars are standard errors.}
  \label{fig:gap}
\end{figure}

In Figure~\ref{fig:gap} we plot the conditional mean of the bid and ask gap as a function of the sign predictor for the stocks AAPL, MSFT, AZN, and VOD. We observe that for large-tick stocks, independently of the dataset, the bid and ask gap are approximately constant and equal to one tick for all sign predictor values. This is largely expected and is due to the high level of discretization of limit prices. For small-tick stocks the bid gap is larger than the ask gap when sell orders are more likely ($\hat{\epsilon}_n<0$), whereas the ask gap is larger than the bid gap when buy orders are more likely ($\hat{\epsilon}_n>0$). The slope of the curves strongly change if we move from the LSE asset to the NASDAQ asset. For AZN the bid gap monotonically decreases and the ask gap monotonically increases when the sign predictor value increases. For AAPL when the sign predictor increases and a buy order is more probable the ask gap decreases, whereas when the sign predictor goes from zero to the minimum value the ask gap increases. The opposite behaviour holds for the bid gaps.

In conclusion, for large-tick assets the conditional distribution of the gaps is not informative. Conversely, for small-tick assets figures show an interesting behaviour. If buy orders are very likely at a given time it means that many buy orders have taken place in the recent past and they have eroded liquidity and increased the sparsity of the ask side of the book. Therefore, the slope of the gap distribution for AZN could be consistent with a purely mechanical effect due to the erosion of the market orders. However, this explanation neglects the possible presence of liquidity providers refilling the order book. Moreover, as it will be clarified in the next section, liquidity takers adjust their trades in order to minimize the price impact, they do not penetrate the opposite side of the order book, and thus the impact of the erosion can not be the only mechanism which determines sparsity of the order book. Finally, the slope of the curves for AAPL cannot be explained without considering the interplay with market makers. Indeed, the negative slope of the curve for the ask gap when predictability increases suggests that the extreme probability of a buy order stimulates the liquidity providers to refill the ask side of the book. For AAPL the figure is consistent with a refill taking place not only at the opposite best, as already confirmed by the volume curve, but also at quotes inside the order book and close to the best price.    

\subsection{Mechanical and quote revision impact}
We now ask how market orders, limits and cancellations determine the price impact. We define the returns as the difference of the logarithmic mid-prices measured immediately before the $n$-th and the $n+1$-th trades, $r_n=p_{n+1}-p_n$, and we decompose them in two components. The first component is due to the mechanical impact of market orders, $r_n^M$, and is given by the difference between the log-price observed immediately after and the one observed before the trade. The second component is the aggregate effect  of the quote revision $r_n^Q$ and cumulates the effect of all the limit orders and cancellations placed in the order book immediately after the $n$-th trade and before the next trade. Thus, we have
\begin{equation*}
  r_n=r_n^M+r_n^Q. 
\end{equation*}
Then, we introduce the quantity $\epsilon_n\cdot\hat{\epsilon}_n$ which measures the correctness of a prediction at a given trade time $t_n$ and quantifies the surprise of the transaction sign given the level of the predictor. The former information is delivered by the sign of $\epsilon_n\cdot\hat{\epsilon}_n$ since when $\epsilon_n\cdot\hat{\epsilon}_n$ is positive we can conclude that the prediction was correct, whereas when $\epsilon_n\cdot\hat{\epsilon}_n$ is negative the prediction was wrong. The amount of surprise associated to the realized order sign is instead related to the absolute value of $\epsilon_n\cdot\hat{\epsilon}_n$. For instance, a large negative value of $\epsilon_n\cdot\hat{\epsilon}_n$ is more informative than a negative value close to zero since it implies that the order sign was largely unexpected by the market. We are interested in the conditional expectation of the return components
\begin{equation*}
  \mathbb{E}[\epsilon_n r_n^M|\epsilon_n\cdot\hat{\epsilon}_n], \qquad \mathbb{E}[\epsilon_n r_n^Q|\epsilon_n\cdot\hat{\epsilon}_n],
\end{equation*}
which show how the correctness of the sign prediction determines the mechanical and quote revision impact, respectively.

The first term is the conditional expectation of the mechanical impact and depends on the probability of an order to penetrate the opposite best price and on the distribution of the gaps on the opposite side of the book, $g_n^{OB}$. This expectation satisfies the approximate relation 
\begin{eqnarray}
\fl  \mathbb{E}[\epsilon_n r_n^M|\epsilon_n\cdot\hat{\epsilon}_n]&\simeq&\sum_{r_n^M \neq 0}\epsilon_n r_n^M P(v_n \geqslant v_n^{OB},\epsilon_n g_n^{OB} \simeq 2r_n^M|\epsilon_n\cdot\hat{\epsilon}_n,\mathcal{C}_n^v) \nonumber \\
  &=&\sum_{r_n^M \neq 0}\epsilon_n r_n^M P(v_n \geqslant v_n^{OB}|\epsilon_n\cdot\hat{\epsilon}_n,\mathcal{C}_n^v)P(\epsilon_n g_n^{OB} \simeq 2r_n^M|v_n \geqslant v_n^{OB},\epsilon_n\cdot\hat{\epsilon}_n,\mathcal{C}_n^v) \nonumber \\
  &\simeq&P(v_n \geqslant v_n^A|\epsilon_n\cdot\hat{\epsilon}_n,\mathcal{C}_n^v,\epsilon_n=1)\mathbb{E}\left[\frac{g_n^A}{4}\bigg|v_n \geqslant v_n^A,\epsilon_n\cdot\hat{\epsilon}_n,\mathcal{C}_n^v,\epsilon_n=1\right] \nonumber \\
  &&+\,P(v_n \geqslant v_n^B|\epsilon_n\cdot\hat{\epsilon}_n,\mathcal{C}_n^v,\epsilon_n=-1)\mathbb{E}\left[\frac{g_n^B}{4}\bigg|v_n \geqslant v_n^B,\epsilon_n\cdot\hat{\epsilon}_n,\mathcal{C}_n^v,\epsilon_n=-1\right] \nonumber \\
  &\equiv& r_n^M(\hat{\epsilon}_n)\,,\label{eqn:mec_impact}
\end{eqnarray}
where $v_n^{OB}$ and $v_n^{OB-2nd}$ are the opposite best and second opposite best volumes, respectively,  $v_n$ is the volume of the market order, and  $\mathcal{C}_n^v$ corresponds to the condition $v_n<v_n^{OB}+v_n^{OB-2nd}$.

The approximate equality in the first line is due to two distinct effects. First, we neglect the possibility that the market order volume is greater than the sum of the best and second opposite best volumes. In real datasets this condition is verified for the large majority of the transactions and only in a very small fraction of trades ($<1\%$) a market order penetrates the opposite side of the book deeper than the first price level (see also~\cite{farmer2004really}). This is in particular true for large-tick assets. Then, we assume that $\epsilon_n g_n^{OB} \simeq 2r_n^M$ when $v_n \geqslant v_n^{OB}$, which is exactly true for linear gaps and linear returns and holds only approximately for logarithmic quantities. For instance, for buy orders the relation between log-gaps and mechanical log-returns is given by 
\begin{eqnarray*}
\fl r_n^M &=& \log{(A_n+B_n+A_n^{2nd}-A_n)}-\log{(A_n+B_n)}=\log{\left(1+\frac{A_n^{2nd}-A_n}{A_n+B_n}\right)} \nonumber \\
\fl & \approx & \frac{A_n^{2nd}-A_n}{A_n+B_n} \approx \frac{g_n^A}{1+B_n/A_n}\approx\frac{g_n^A}{2}\,,
\end{eqnarray*}
where $B_n/A_n \approx 1$. Finally, the approximation in the third line of Equation~\ref{eqn:mec_impact} follows from the realistic assumption that $P(\epsilon_n=1|\epsilon_n\cdot\hat{\epsilon}_n,\mathcal{C}_n^v)=P(\epsilon_n=-1|\epsilon_n\cdot\hat{\epsilon}_n,\mathcal{C}_n^v)\simeq 1/2$. The quantity $P(v_n \geqslant v_n^{OB}|\epsilon_n\cdot\hat{\epsilon}_n,\mathcal{C}_n^v)$ corresponds to the conditional probability that the volume of a market order is larger than the liquidity available at the opposite best. Thus, it represents the probability that a market order immediately triggers a mid-price change. We estimate the penetration probability on the real datasets and condition it on the correctness of the order sign predictor $\epsilon_n\cdot\hat{\epsilon}_n$, but we remove the mild conditioning $\mathcal{C}_n^v$. Starting from the quantities 
\begin{equation*}
  P(v_n \geqslant v_n^A|\epsilon_n\cdot\hat{\epsilon}_n,\epsilon_n=+1), \qquad P(v_n \geqslant v_n^B|\epsilon_n\cdot\hat{\epsilon}_n,\epsilon_n=-1)\,,
\end{equation*}
we express the total penetration probability of market orders as
\begin{equation*}
\fl  P(v_n \geqslant v_n^{OB}|\epsilon_n\cdot\hat{\epsilon}_n)\simeq \frac{1}{2}\left[P(v_n \geqslant v_n^A|\epsilon_n\cdot\hat{\epsilon}_n,\epsilon_n=1)+P(v_n \geqslant v_n^B|\epsilon_n\cdot\hat{\epsilon}_n,\epsilon_n=-1)\right] \,,
\end{equation*}
where we have assumed that $P(\epsilon_n=+1|\epsilon_n\cdot\hat{\epsilon}_n) = P(\epsilon_n=-1|\epsilon_n\cdot\hat{\epsilon}_n) \simeq 1/2$. We also compute the conditional average fraction of liquidity eroded by a market order $\mathbb{E}[f|\epsilon_n\cdot\hat{\epsilon}_n]$ with $f=v_n/v_n^{OB}$.

\begin{figure}[t]
\begin{minipage}[c]{0.5\columnwidth}
\includegraphics[width=\columnwidth]{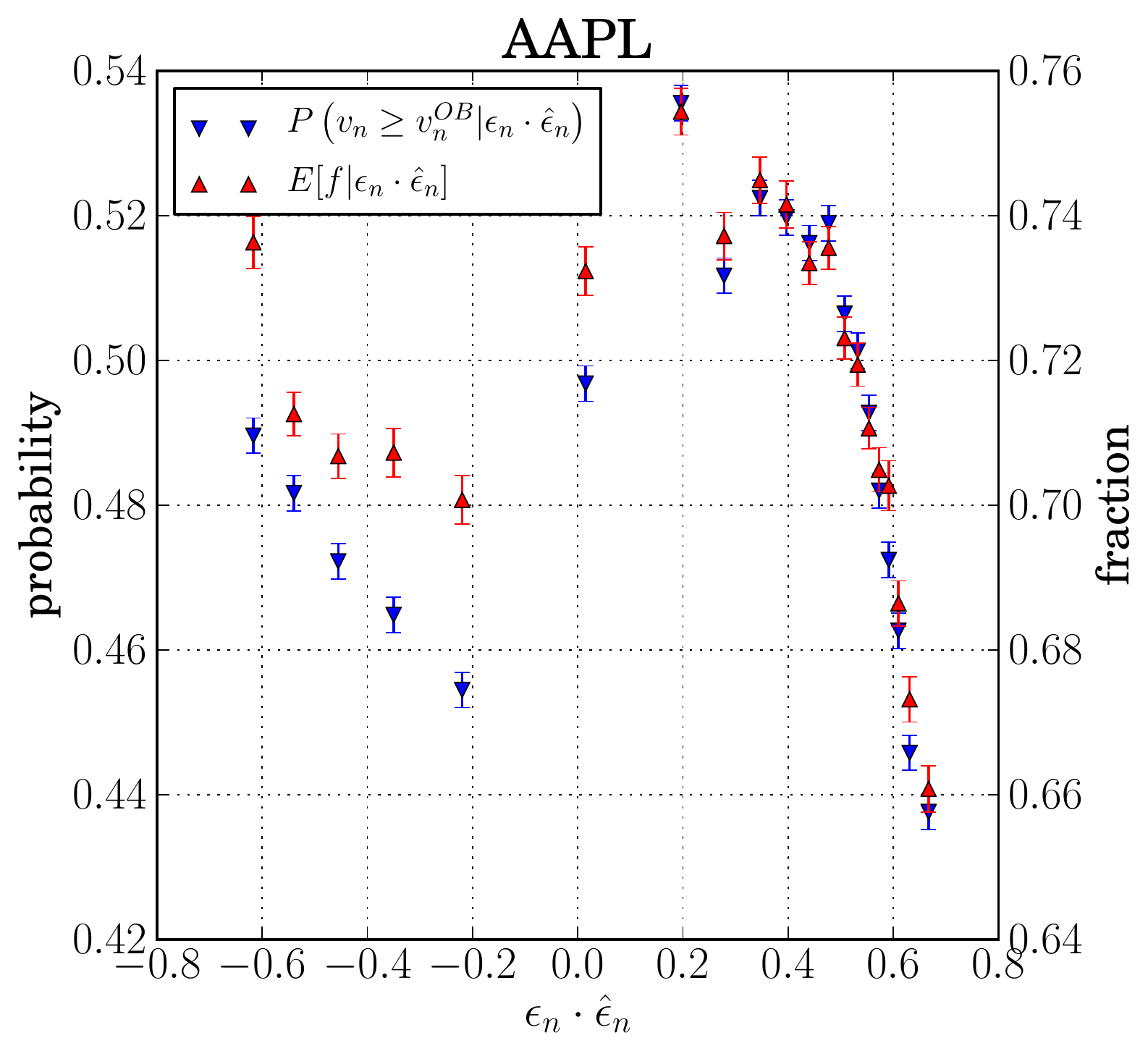}
\end{minipage}%
\begin{minipage}[c]{0.5\columnwidth}
\includegraphics[width=\columnwidth]{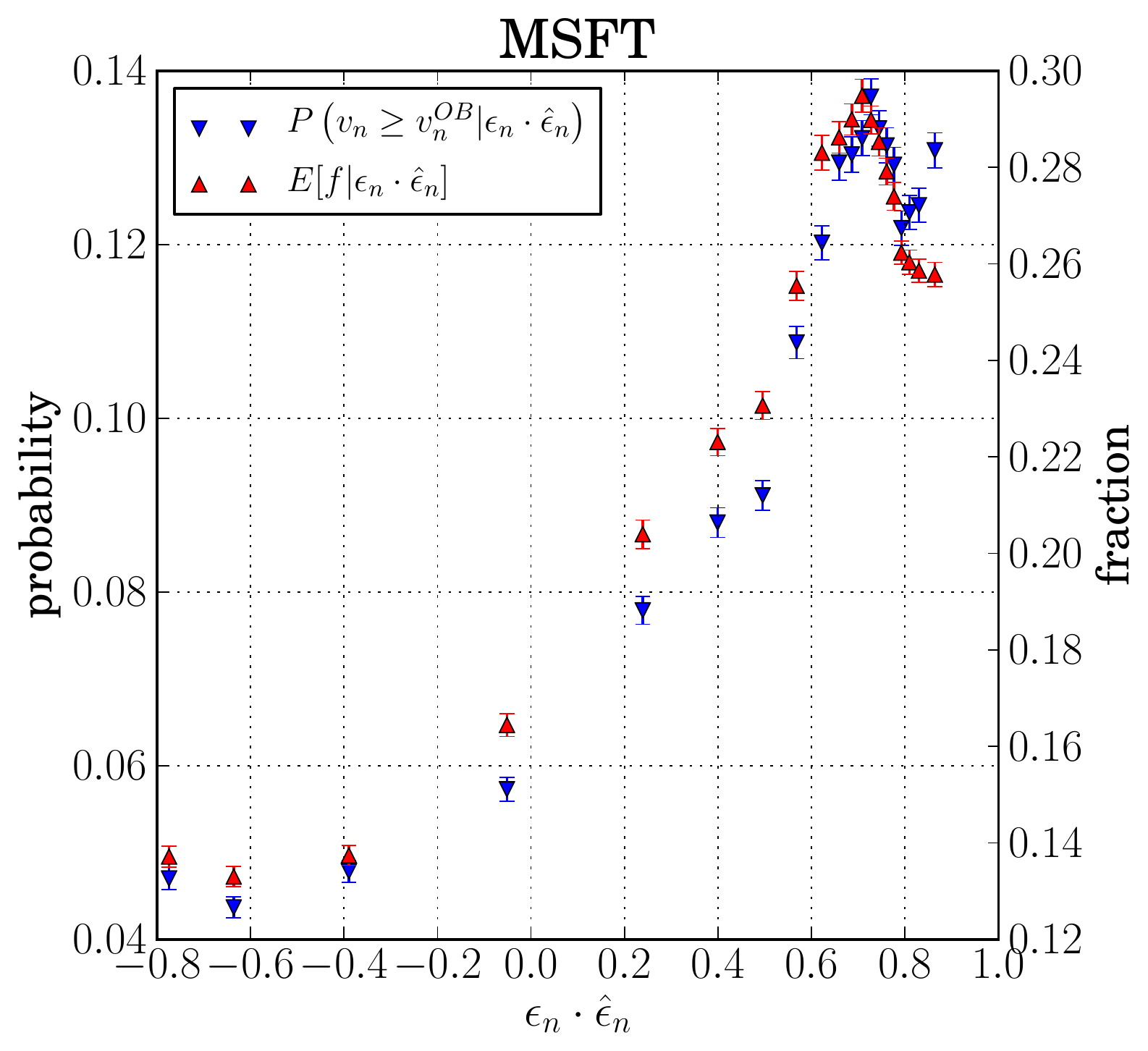}
\end{minipage} \\
\begin{minipage}[c]{0.5\columnwidth}
\includegraphics[width=\columnwidth]{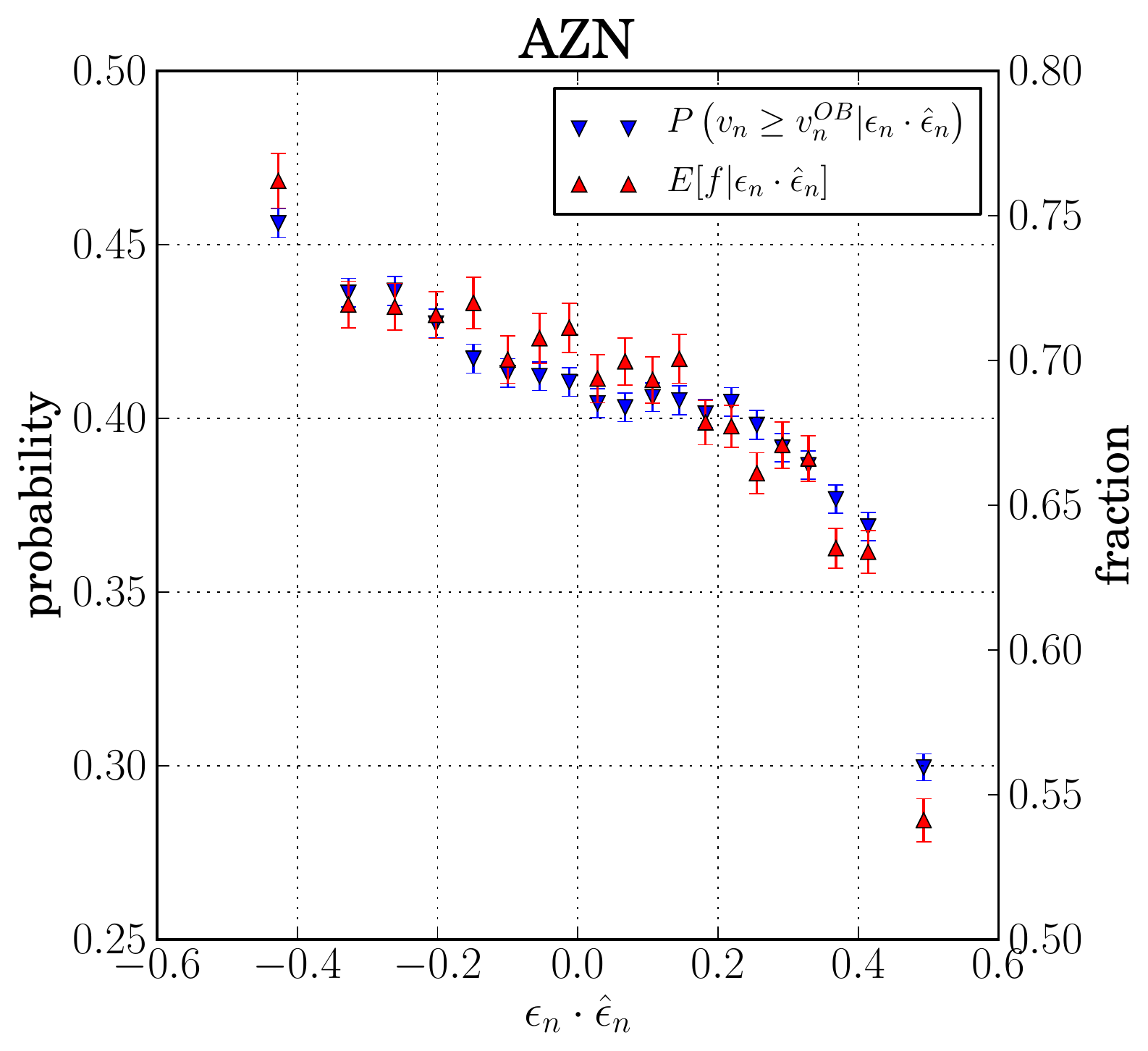}
\end{minipage}%
\begin{minipage}[c]{0.5\columnwidth}
\includegraphics[width=\columnwidth]{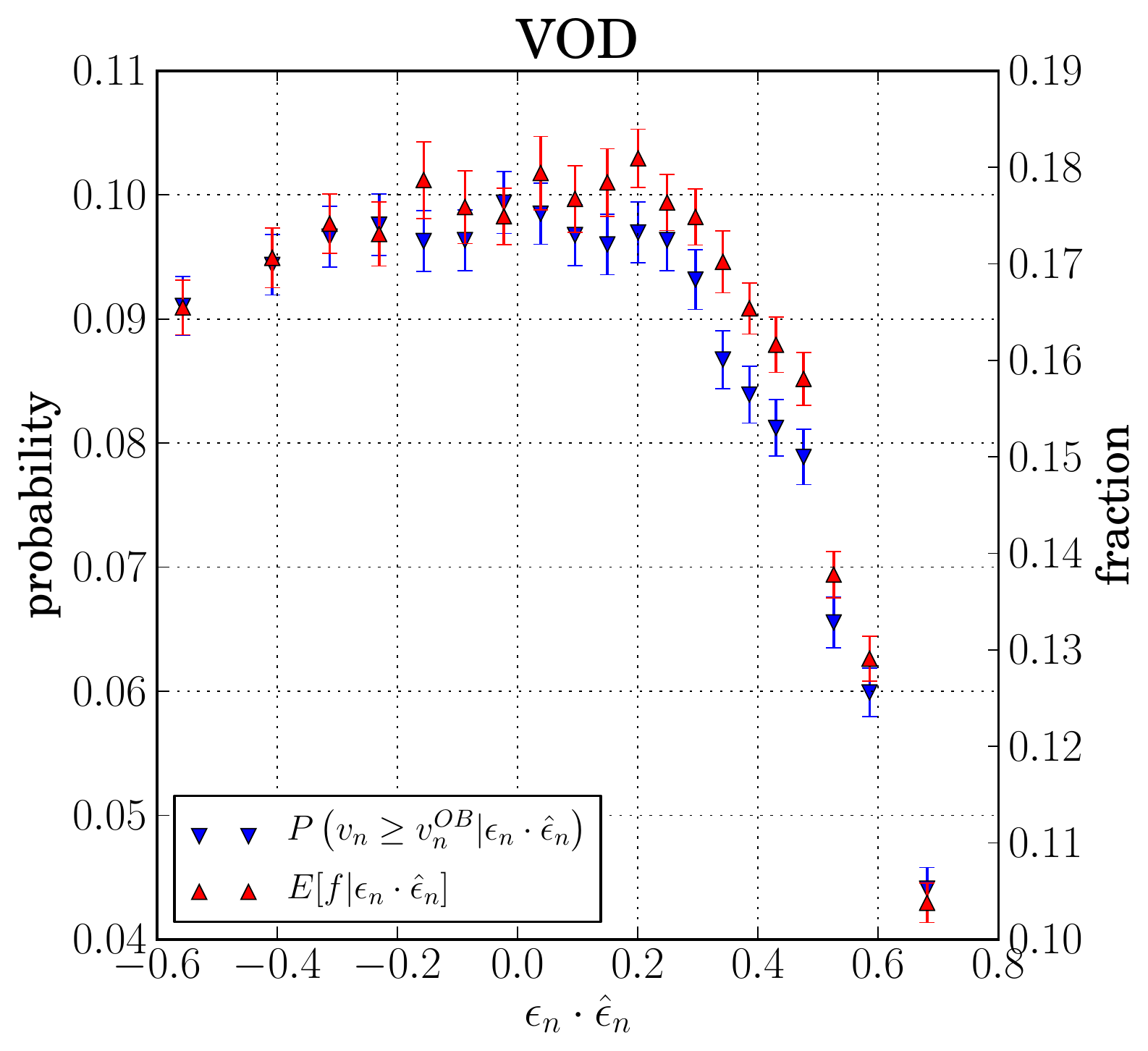}
\end{minipage}
\caption{Conditional penetration probabilities of the market orders and conditional average ratio between market order volumes and best opposite volumes for AAPL, MSFT, AZN, VOD as a function of $\epsilon_n\cdot\hat{\epsilon}_n$. The error bars are standard errors.}
\label{fig:penetration_fraction}
\end{figure}

Figure~\ref{fig:penetration_fraction} shows the conditional penetration and fraction for AAPL, MSFT, AZN and VOD. We see that for all the stocks the eroded fraction tracks quite well the behaviour of the penetration probability, and we observe the largest discrepancies for the NASDAQ assets in a region $\epsilon_n\cdot\hat{\epsilon}_n<0$ which is scarcely populated. For the LSE dataset the penetration probability is consistent with previous findings in the literature~\cite{lillo2004long}, \emph{i.e.} when the order sign predictability increases and the prediction is correct, the probability of penetration drops. The stocks of the NASDAQ dataset (AAPL, MSFT) show deviations from a monotonic behaviour. MSFT shows an increasing penetration probability when the order sign predictability increases and the prediction is correct up to $\epsilon_n\cdot\hat{\epsilon}_n\simeq 0.7$ then drops, whereas for AAPL deviations from a decreasing behaviour are relevant in the region where $\epsilon_n\cdot\hat{\epsilon}_n \lesssim 0.3$. This effect leads to inefficiencies of the market that we will comment about more extensively in the next section. Finally, as expected the penetration probability of large-tick stocks (MSFT,VOD) is lower than the probability of small-tick stocks (AAPL, MSFT). Indeed, for large-tick stocks market orders eroding the opposite liquidity are less frequent since they trigger a large impact on the mid-price.

\begin{figure}[t]
  \begin{minipage}[c]{0.5\columnwidth}
    \includegraphics[width=\columnwidth]{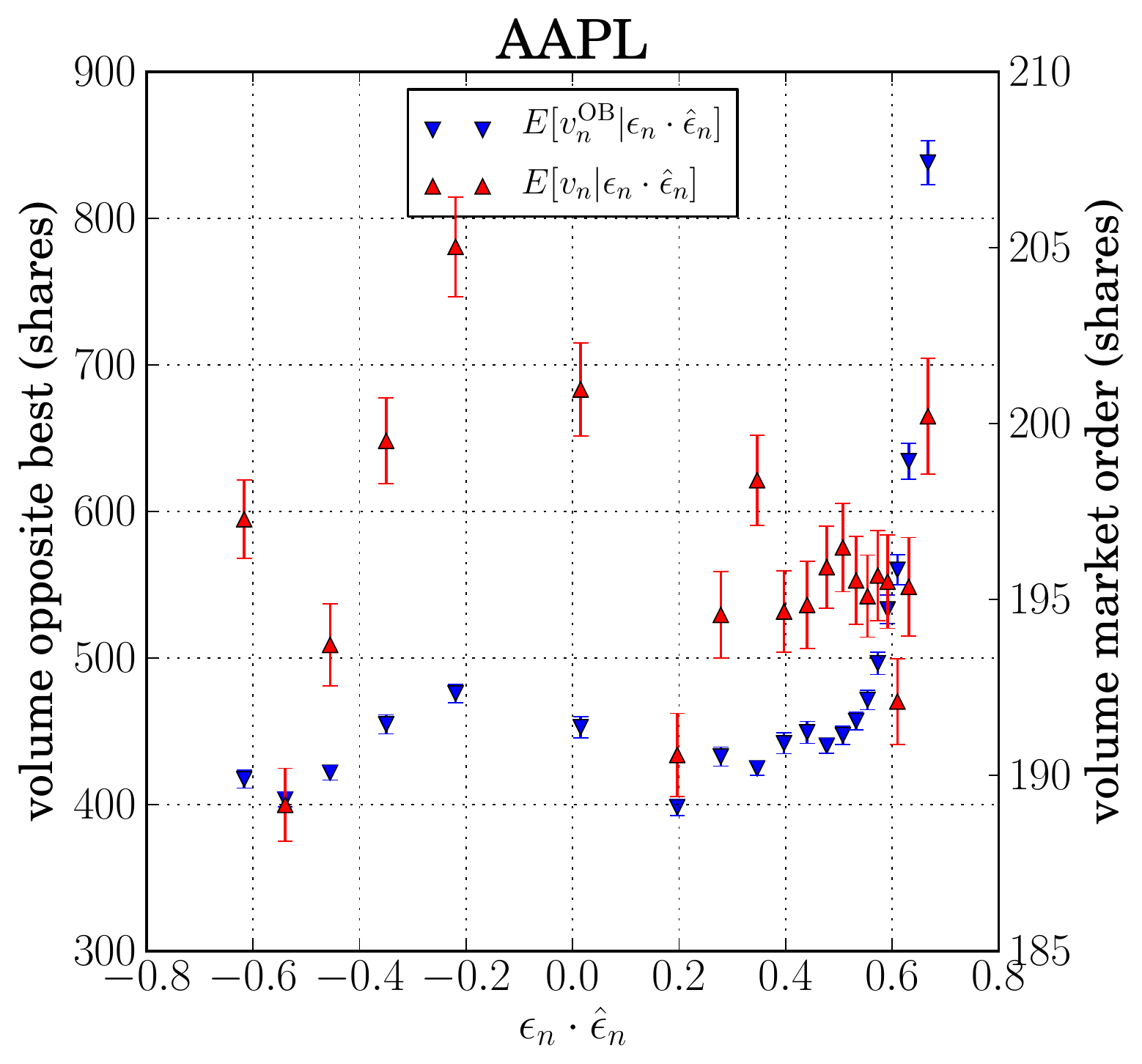}
  \end{minipage}%
  \begin{minipage}[c]{0.5\columnwidth}
    \includegraphics[width=\columnwidth]{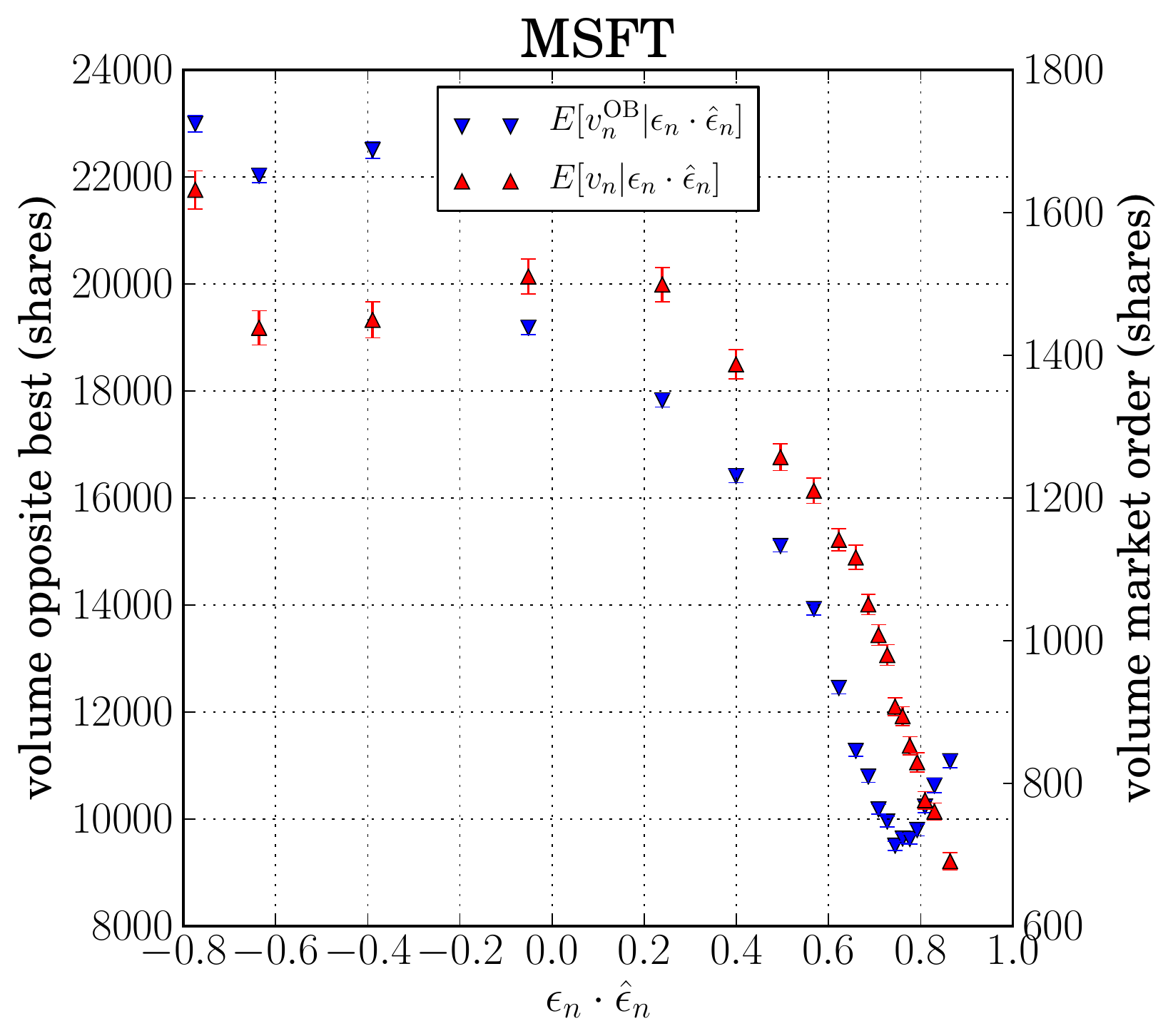}
  \end{minipage} \\
  \begin{minipage}[c]{0.5\columnwidth}
    \includegraphics[width=\columnwidth]{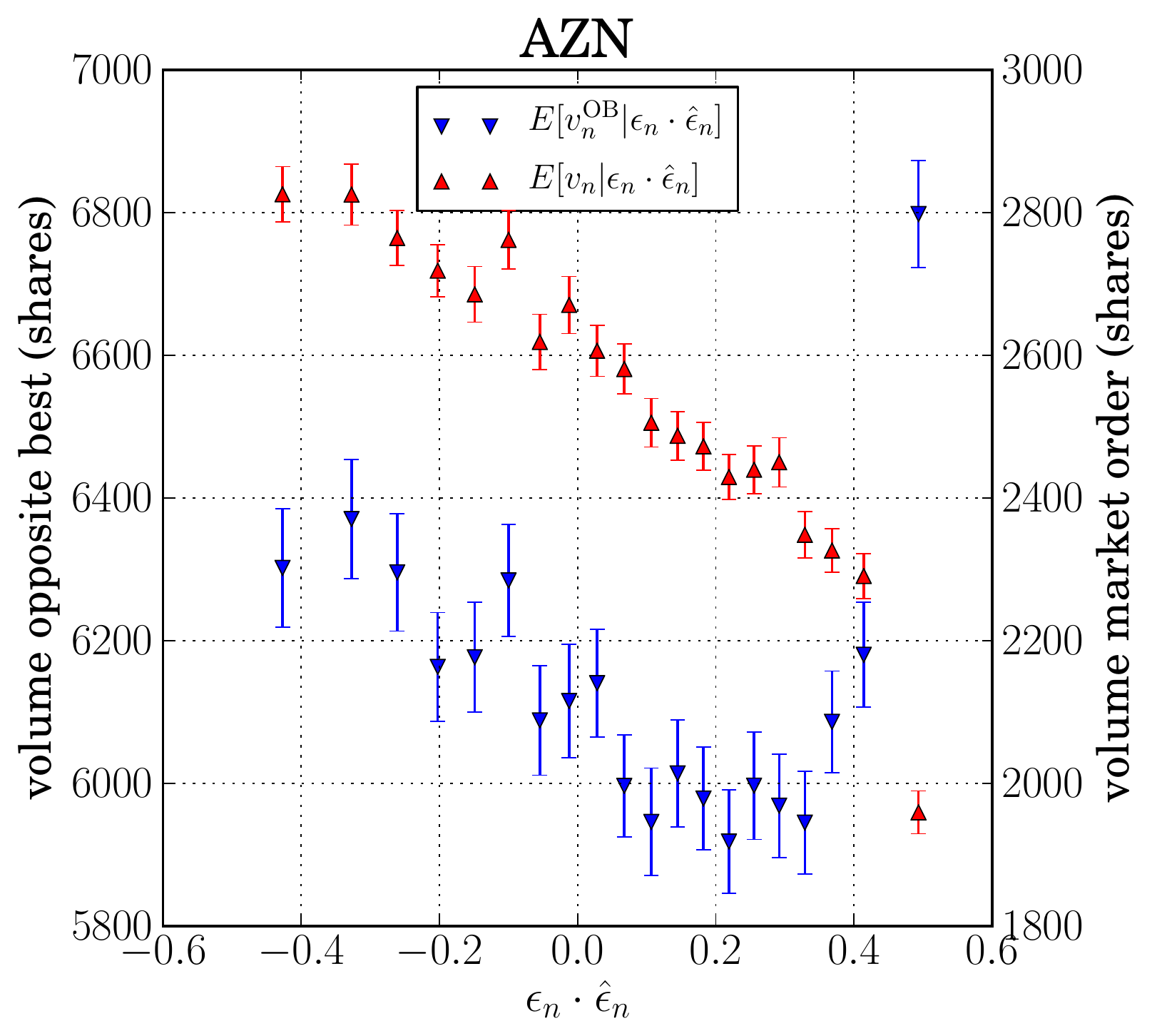}
  \end{minipage}%
  \begin{minipage}[c]{0.5\columnwidth}
    \includegraphics[width=\columnwidth]{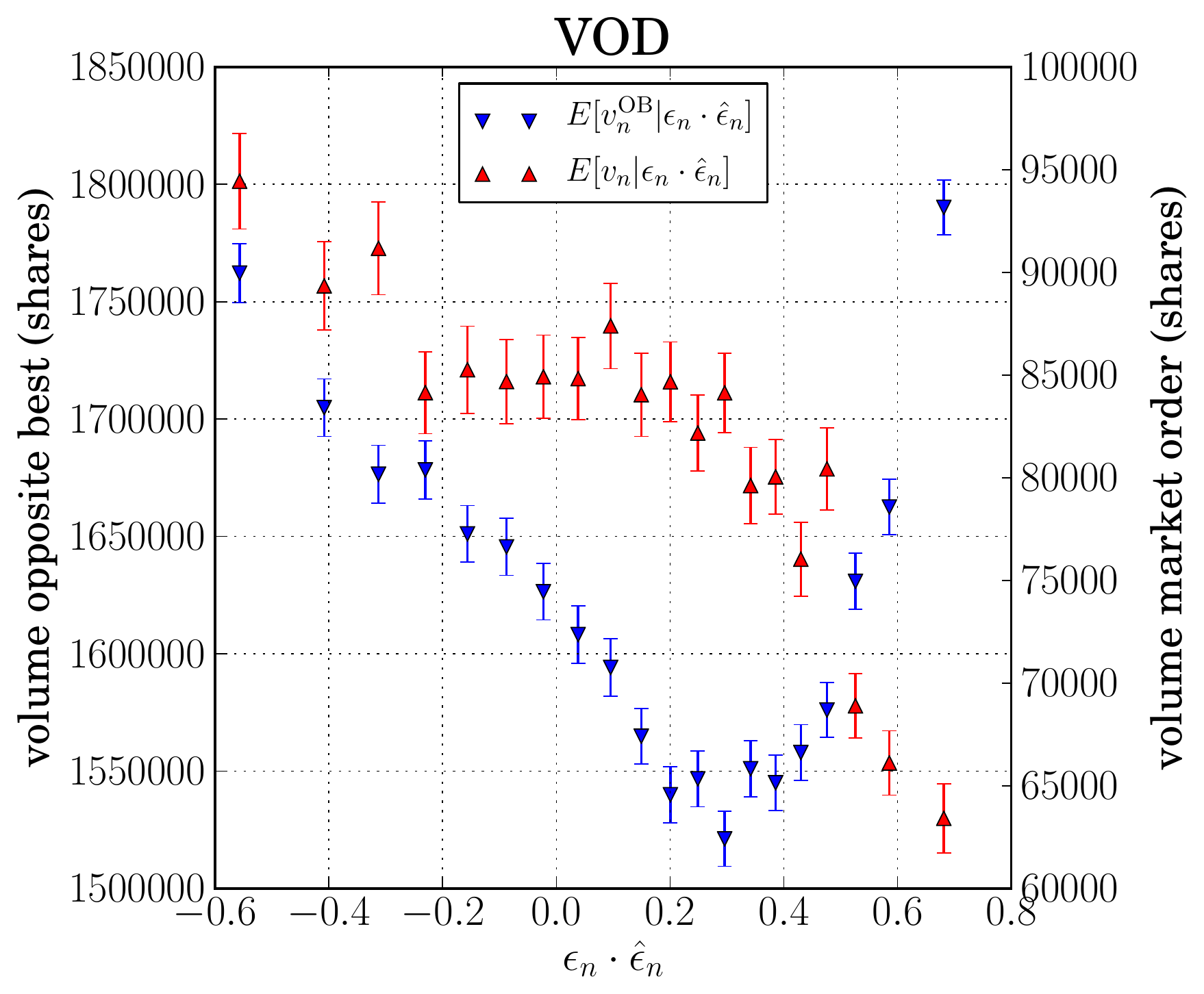}
  \end{minipage}
  \caption{Conditional best opposite volumes $\mathbb{E}[v_n^{OB}|\epsilon_n\cdot\hat{\epsilon}_n]$ and conditional market order volumes $\mathbb{E}[v_n|\epsilon_n\cdot\hat{\epsilon}_n]$ for AAPL, MSFT, AZN, and VOD. The error bars are standard errors.}
  \label{fig:vol_mo_ob}
\end{figure}

In Figure~\ref{fig:vol_mo_ob} we examine in more detail the empirical behaviour of the opposite best volume and market order volume whose ratio leads to the fraction of eroded liquidity. For MSFT, AZN, and VOD two aspects are common: The conditional average market order volume decreases with  $\epsilon_n\cdot\hat{\epsilon}_n$. The second striking feature is the behaviour of the average amount of volume available at the opposite side. It decreases when the correctness increases, then, for $\epsilon_n\cdot\hat{\epsilon}_n \approx 1$ it increases quickly. Thus, up to moderate values of $\epsilon_n\cdot\hat{\epsilon}_n$ liquidity is removed from the opposite best either because of a mechanical erosion or because liquidity providers revise their limit orders. Then, finally, the high predictability stimulates the liquidity refill, a liquidity barrier piles up at the opposite best and the penetration probability drops. For AAPL the conditional average market order volume is independent from the predictability of the order flow. However, there are clear signs of the liquidity refill effect. We can therefore interpret these effects concluding that liquidity takers adapt their orders to the outstanding liquidity only when correctness is not too high, because in the extreme region of predictability liquidity takers shrink the volume of their markets orders, though the available volume at the opposite side is high.

\begin{figure}[t]
  \begin{minipage}[c]{0.5\columnwidth}
    \includegraphics[width=\columnwidth]{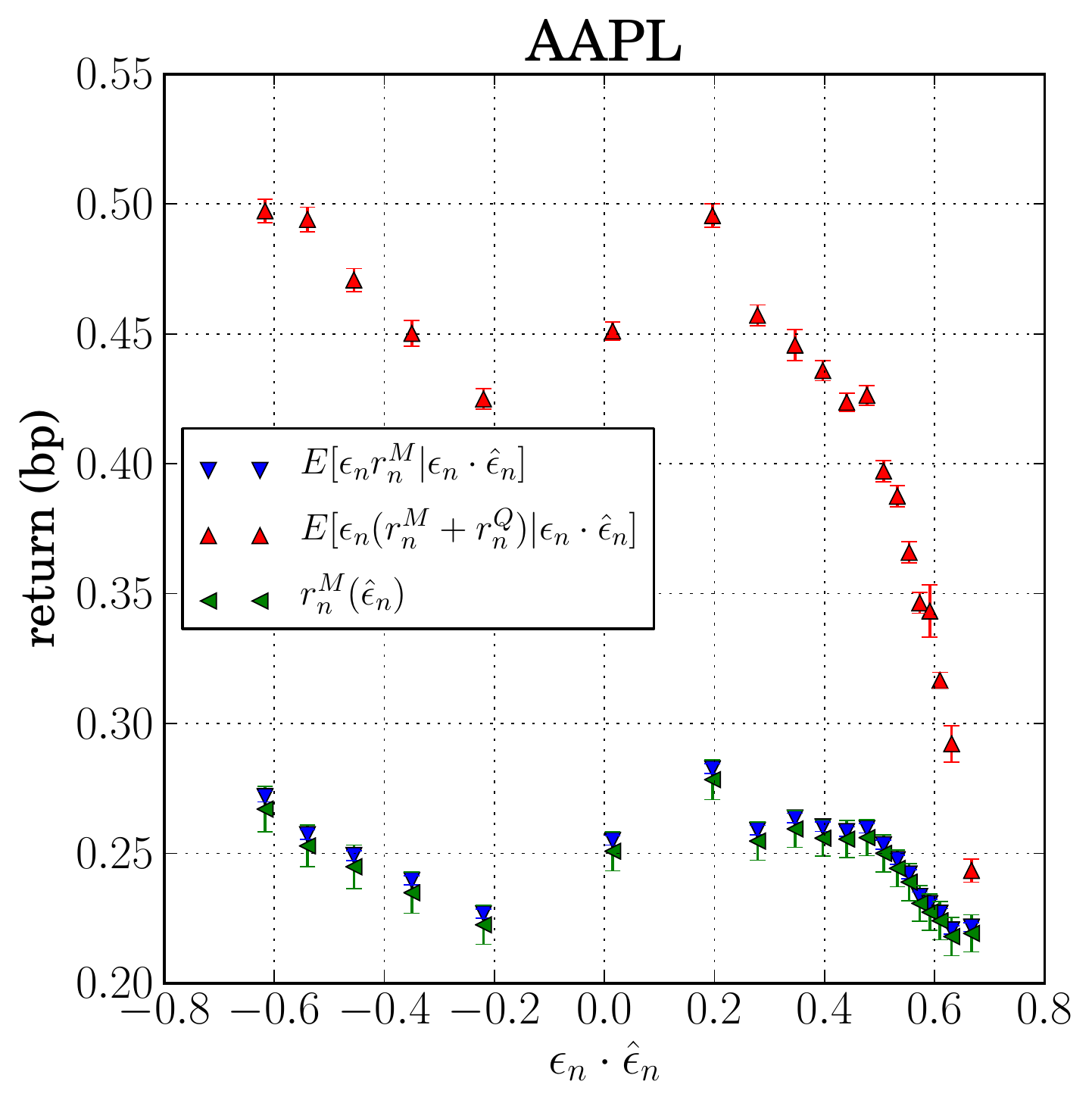}
  \end{minipage}%
  \begin{minipage}[c]{0.5\columnwidth}
    \includegraphics[width=\columnwidth]{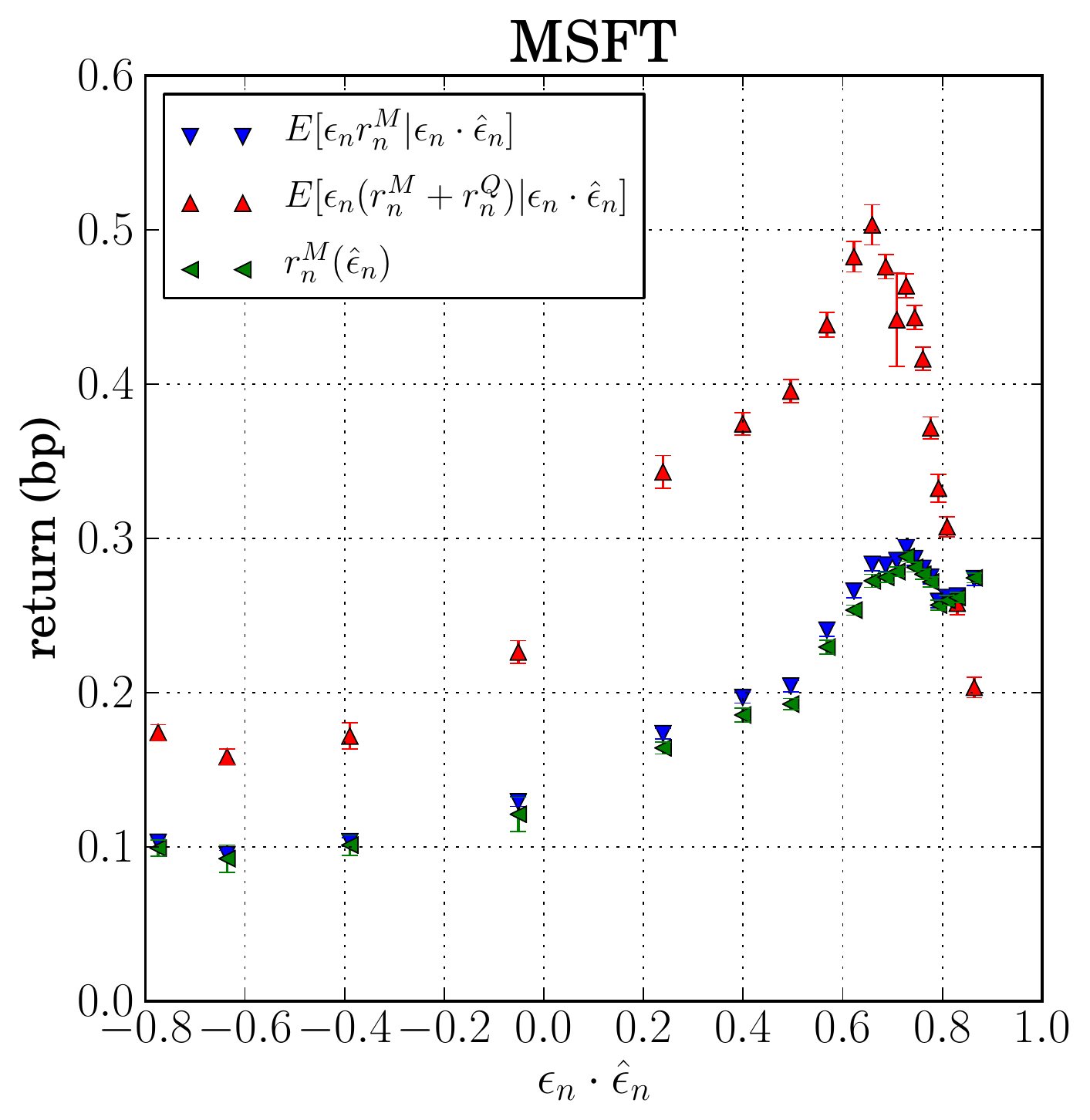}
 \end{minipage} \\
  \begin{minipage}[c]{0.5\columnwidth}
    \includegraphics[width=\columnwidth]{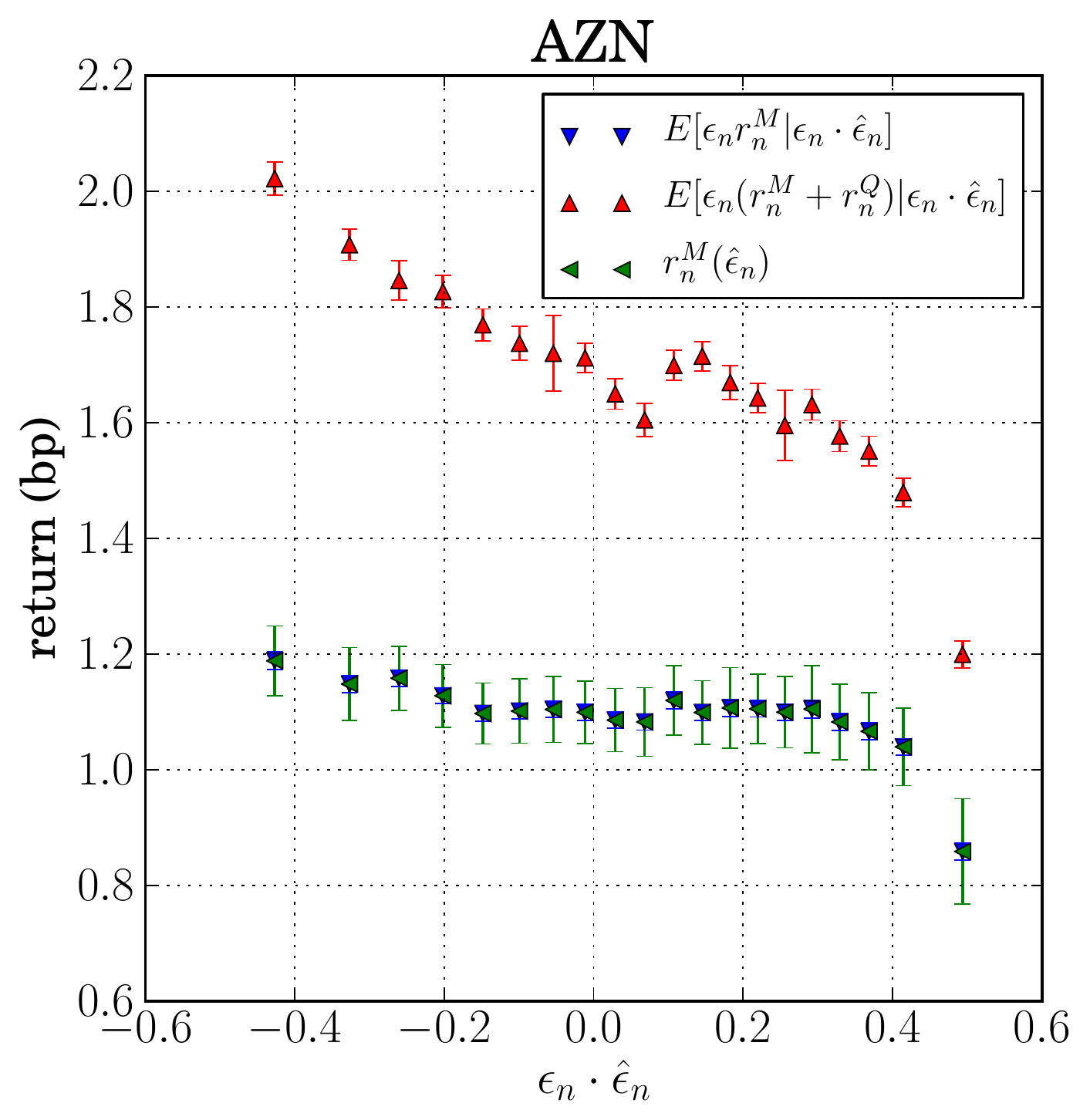}
  \end{minipage}%
  \begin{minipage}[c]{0.5\columnwidth}
    \includegraphics[width=\columnwidth]{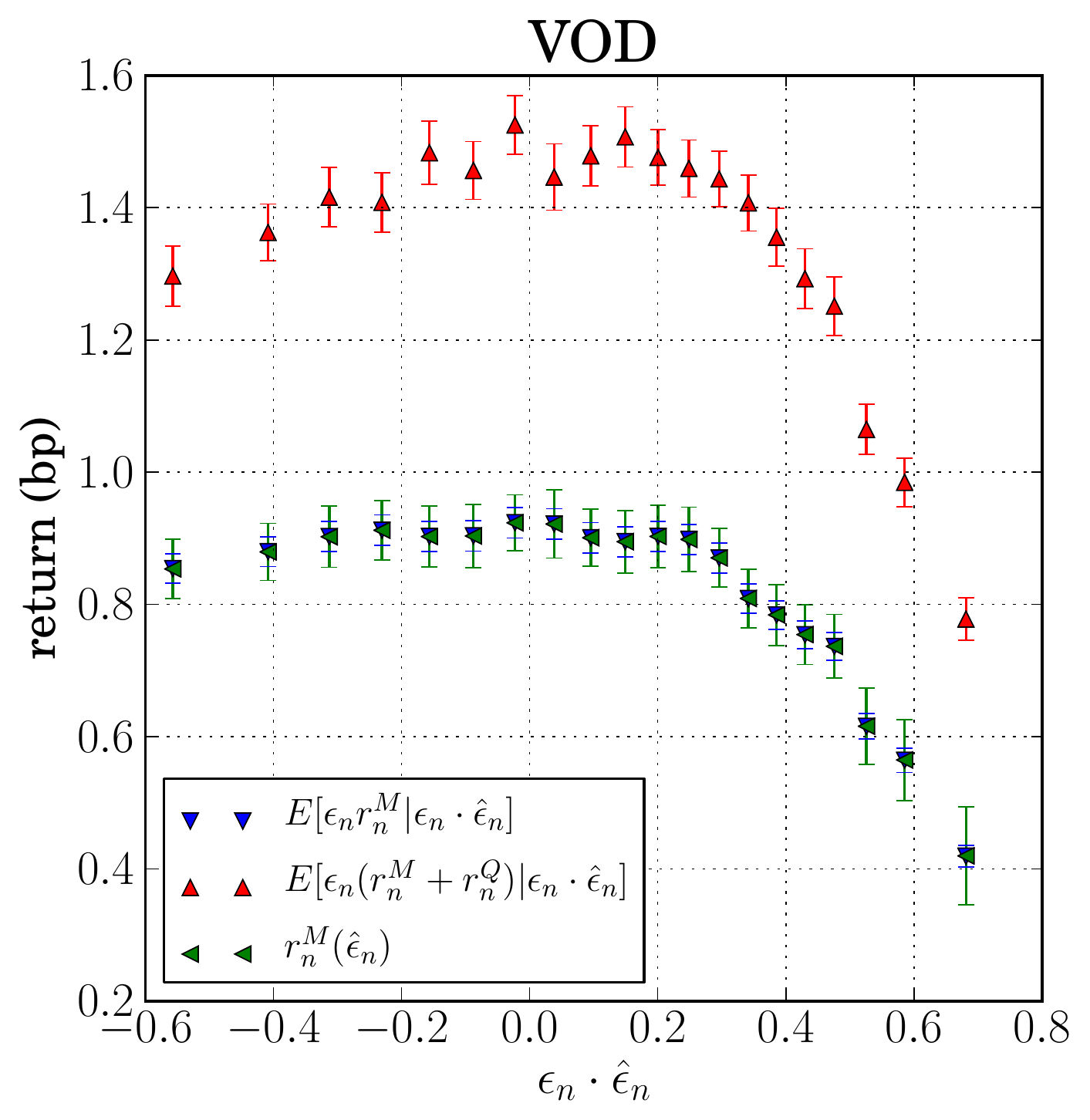}
  \end{minipage}
  \caption{Conditional mechanical impact $\mathbb{E}[\epsilon_n r_n^M|\epsilon_n\cdot\hat{\epsilon}_n]$, the approximate expression $r_n^M(\hat{\epsilon}_n)$, and conditional returns $\mathbb{E}[\epsilon_n(r_n^M+r_n^Q)|\epsilon_n\cdot\hat{\epsilon}_n]$ for the stocks AAPL, MSFT, AZN, VOD. The error bars are standard errors.}
  \label{fig:return}
\end{figure}

In light of above considerations about outstanding and market volumes, and gap distributions, we can now discuss the observed behaviour of the price impact. In Figure~\ref{fig:return} we show the conditional mechanical impact $r_n^M$, the approximate expression $r_n^M(\hat{\epsilon}_n)$ derived in Equation~\ref{eqn:mec_impact}, and the whole impact $r_n$ conditioned on $\epsilon_n\cdot\hat{\epsilon}_n$ for AAPL, MSFT, AZN, and VOD.

From this figure we notice that the approximate quantity $r_n^M(\hat{\epsilon}_n)$ reproduces extremely well the mechanical impact. This evidence supports the idea that the mechanical component of the impact is mainly determined by the gap distribution and by the penetration probability of a market order. The second relevant message is that whenever the correctness of the order sign increases the mechanical impact of the order decreases. This result confirms the results in~\cite{lillo2004long}, but now we can better understand what is happening in the order book. Indeed, for large-tick assets the gap is basically constant, so the main determinant of the impact is the penetration probability. While for VOD the decrease of the mechanical impact is evident for all values of $\epsilon_n\cdot\hat{\epsilon}_n$, for MSFT the drop of the impact becomes clearer when we reach large values of $\epsilon_n\cdot\hat{\epsilon}_n$. The penetration probability is determined by the interplay between the volume outstanding at the opposite side and the volume of the market order. From Figure~\ref{fig:vol_mo_ob} we know that the volume at the opposite best drops when $\epsilon_n\cdot\hat{\epsilon}_n$ increases and increases again only for very large values of the correctness. Thus, the reduction of the impact has to be given by the decrease of the penetration probability which is determined for high values of $\epsilon_n\cdot\hat{\epsilon}_n$ by liquidity takers placing market orders of decreasing volumes and by liquidity providers placing limit orders at best opposite quotes. From the difference between the return impact and the mechanical return we can also infer the impact of quote revisions and draw conclusions about the adaptive behaviour of liquidity providers. 

Figure~\ref{fig:return} shows that the quote revision always acts in the same direction of the mechanical impact (the only exception is represented by the two extreme bins in the MSFT plot, but such evidence should be confirmed with a more systematic analysis of large-tick stocks from the NASDAQ dataset). This suggests that when the correctness increases, liquidity providers tend to cancel their old limit orders and place new orders at quotes beyond the best price. However, this effect becomes less and less severe when the correctness of the sign predictor is very high, since the impact of the quote revision shrinks to zero, and liquidity providers increase the volume of limit orders outstanding at the opposite best. For small-tick assets the empirical analysis gives similar results. The mechanical impact still decreases when $\epsilon_n\cdot\hat{\epsilon}_n$ increases for both AZN and AAPL. From the analysis of the best volume and market order volume profiles we conclude that this effect is due to liquidity takers which adjust their order volume to the outstanding liquidity and thus drop the penetration probability. The quote revision acts as for the large-tick assets in a similar way: For moderate levels of $\epsilon_n\cdot\hat{\epsilon}_n$ the liquidity providers revise their position, whereas for extreme values the revision stops and liquidity piles up at the opposite best. The major difference between AZN and AAPL emerges looking at the gap distribution. Indeed, for AZN it increases monotonically with $\epsilon_n\cdot\hat{\epsilon}_n$, whilst for AAPL it diminishes (see Figure~\ref{fig:gap}). Since liquidity takers act in a similar fashion for both assets, the cause of the different behaviour has to be attributed to the different way liquidity providers revise their position. However, a precise answer to this question is beyond the scope of the current analysis and should deserve further investigation.

\subsection{The route to market efficiency}
The analysis of empirical data discussed in the previous sections largely confirms that asymmetric liquidity is present in financial markets at the transaction by transaction level. However, our analysis clarifies that the drop of the price impact when the order sign predictability increases is the result of the adaptive behaviour of both liquidity takers and liquidity providers acting on the market. In fact, the former adjust their market order volume at the outstanding liquidity, while the latter revise their limit orders and refill liquidity at the best quotes or within the order book as an adaptive answer to the order flow predictability. How are these results related to market efficiency? By observing Figure~\ref{fig:return}, we notice that for AZN and VOD the return is a non increasing function of $\epsilon_n\cdot\hat{\epsilon}_n$, \emph{i.e.} more predictable trades have a smaller impact\footnote{For VOD there is an anomalous behaviour for large negative values of $\epsilon_n\cdot\hat{\epsilon}_n$, but this effect is relatively small.}. AAPL shows more significant deviations around $\epsilon_n\cdot\hat{\epsilon}_n=0$, while MSFT shows a pattern, which is clearly inconsistent with market efficiency. We therefore argue that there is room for some inefficiency in the market. More quantitatively, from the figures for the NASDAQ assets we observe that for some $\epsilon_n\cdot\hat{\epsilon}_n>0$
\begin{equation}
\mathbb{E}[\epsilon_n(r_n^M+r_n^Q)|-\epsilon_n\cdot\hat{\epsilon}_n]<\mathbb{E}[\epsilon_n(r_n^M+r_n^Q)|\epsilon_n\cdot\hat{\epsilon}_n]\,. \label{eqn:cond_inefficiency}
\end{equation}
This inequality means that if we use the information set $\Omega_{n-1}$ at time $t_{n-1}$, a non vanishing predictability of return $r_n$, $\mathbb{E}[r_n|\Omega_{n-1}, \mathcal{M}] \neq 0$, still persists. Obviously this condition is necessary but not sufficient for the inefficiency of markets. In the following we use this condition to test for inefficiency of returns.

\begin{figure}[t]
  \begin{minipage}[c]{0.5\columnwidth}
    \includegraphics[width=\columnwidth]{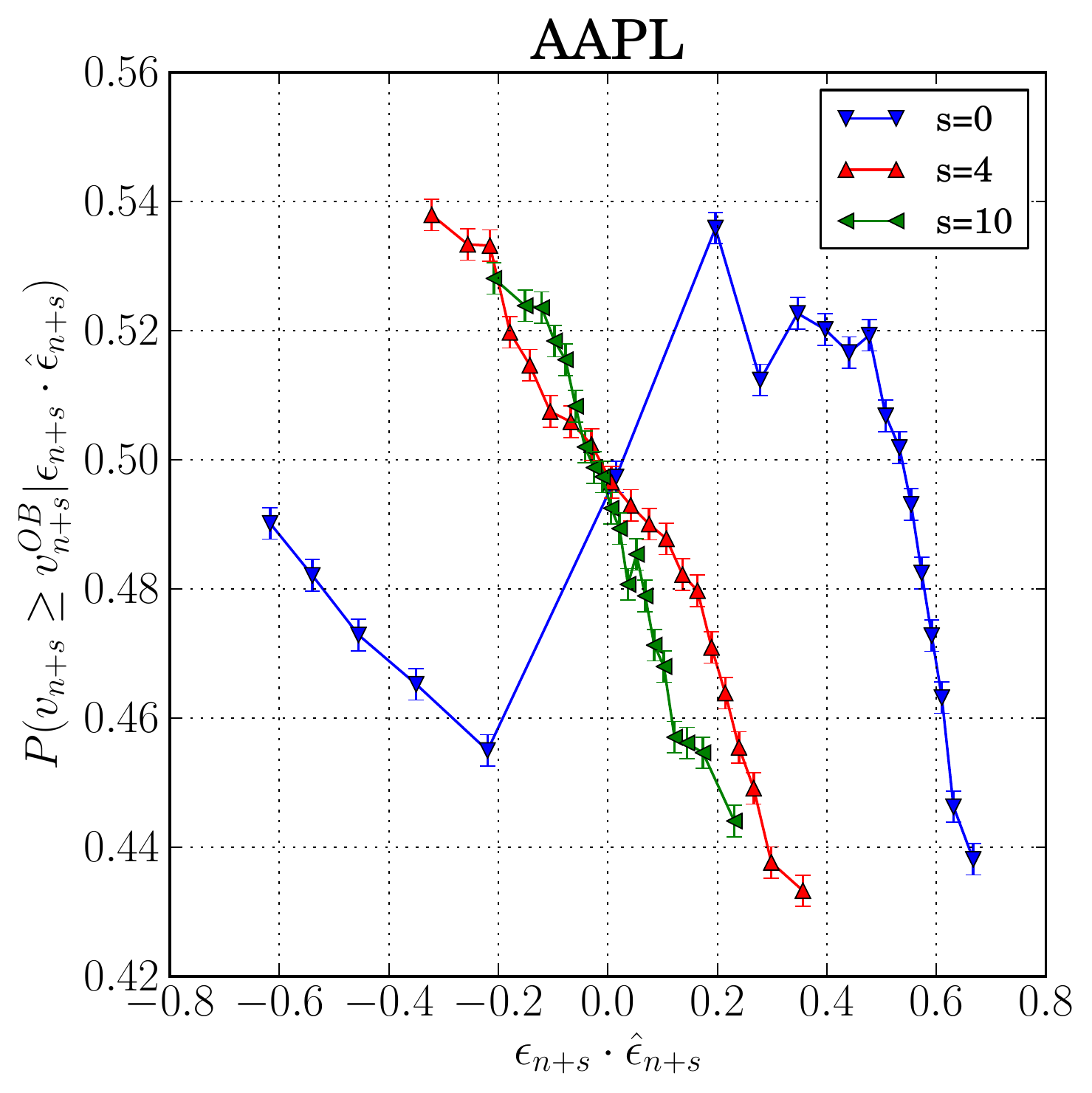}
  \end{minipage}%
  \begin{minipage}[c]{0.5\columnwidth}
    \includegraphics[width=\columnwidth]{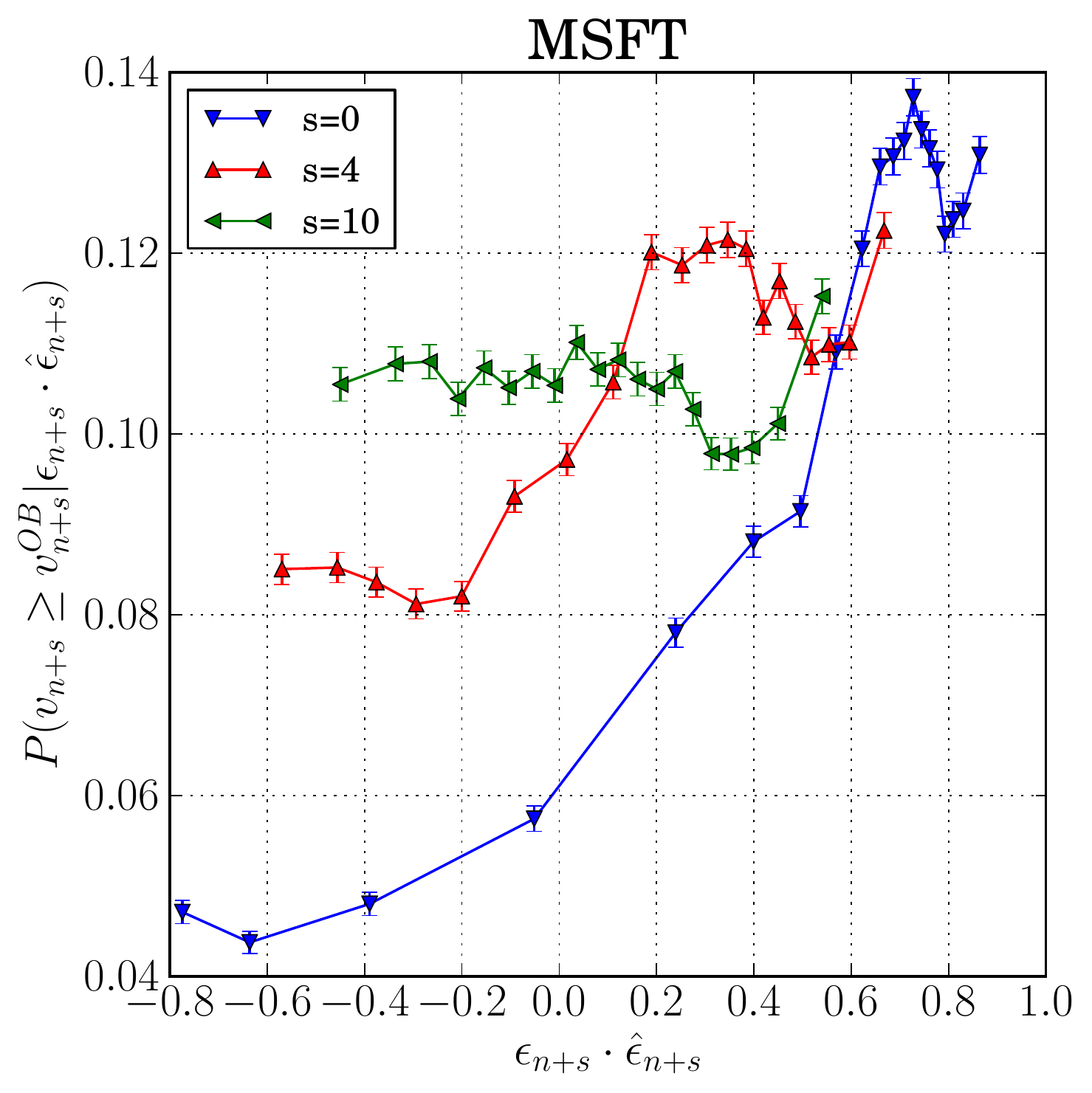}
  \end{minipage} \\
  \begin{minipage}[c]{0.5\columnwidth}
    \includegraphics[width=\columnwidth]{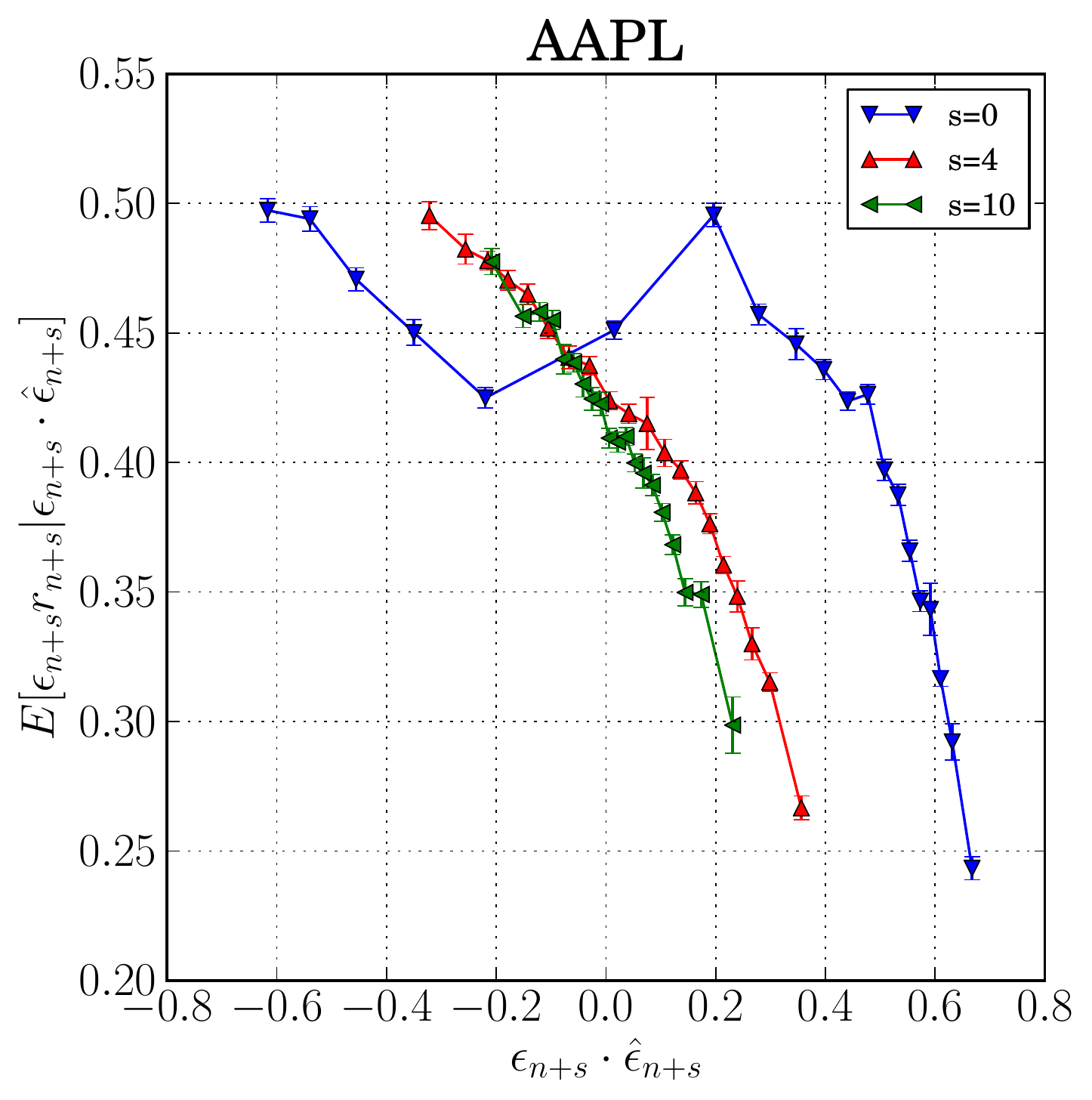}
  \end{minipage}%
  \begin{minipage}[c]{0.5\columnwidth}
    \includegraphics[width=\columnwidth]{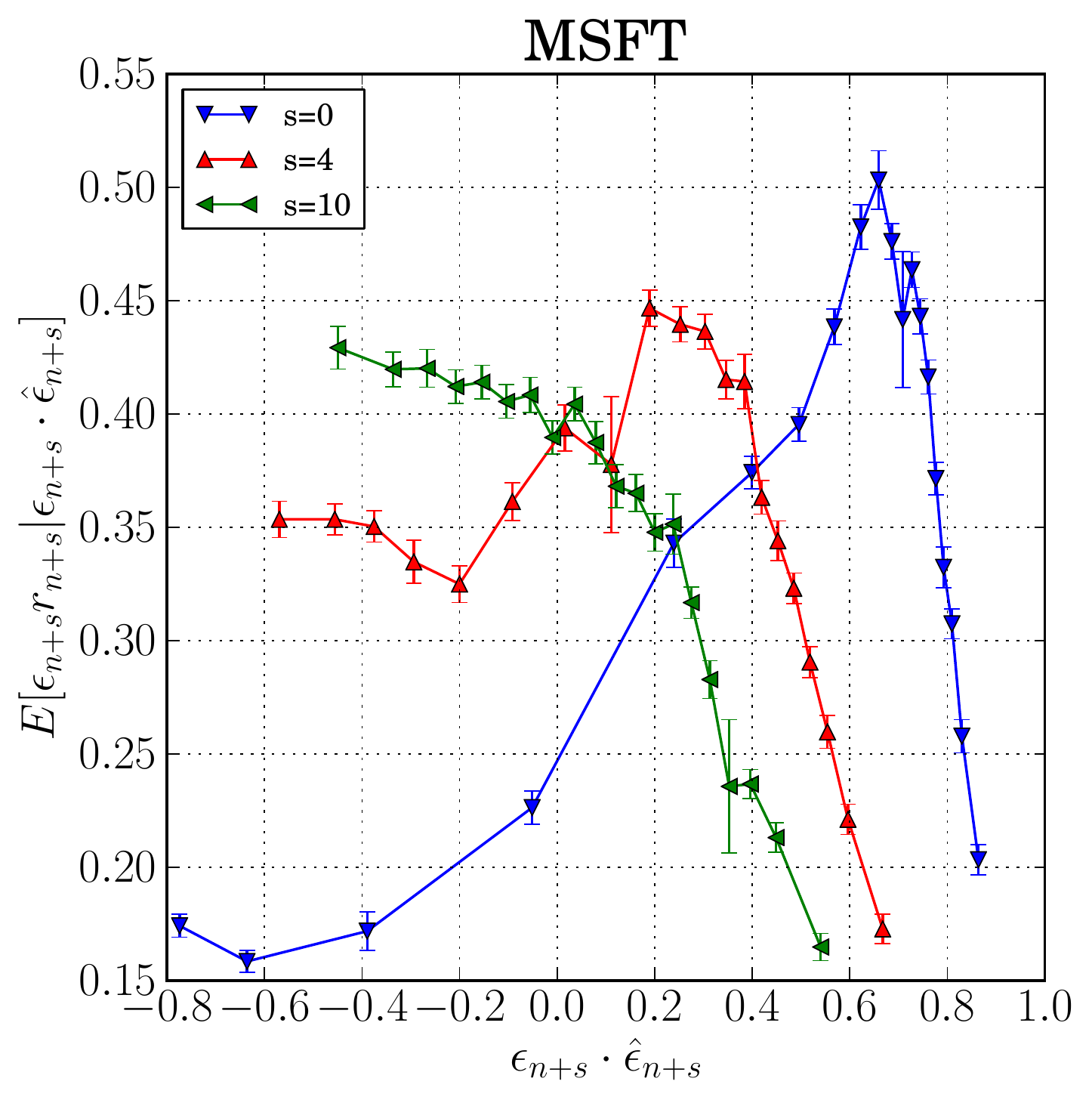}
  \end{minipage}
  \caption{Top row: Lagged probability of penetration $P(v_{n+s} \geqslant v_{n+s}^{OB}|\epsilon_{n+s}\cdot\hat{\epsilon}_{n+s})$. Bottom row: Returns $\mathbb{E}[\epsilon_{n+s}r_{n+s}|\epsilon_{n+s}\cdot\hat{\epsilon}_{n+s}]$ for AAPL and MSFT and different step values $s$. The error bars are standard errors.}
  \label{fig:lagged_penetration_return1}
\end{figure}

\begin{figure}[t]
  \begin{minipage}[c]{0.5\columnwidth}
    \includegraphics[width=\columnwidth]{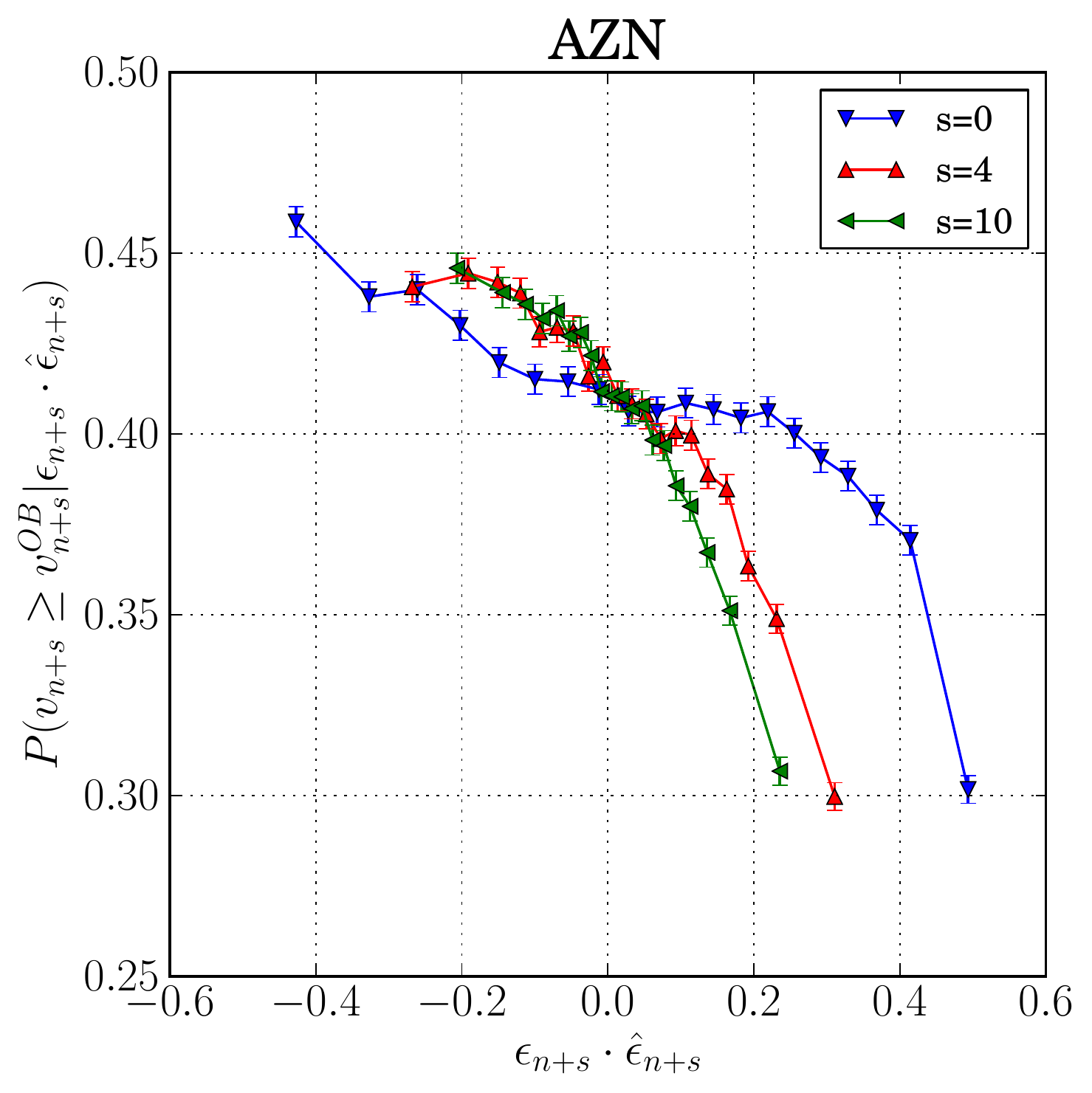}
  \end{minipage}%
  \begin{minipage}[c]{0.5\columnwidth}
    \includegraphics[width=\columnwidth]{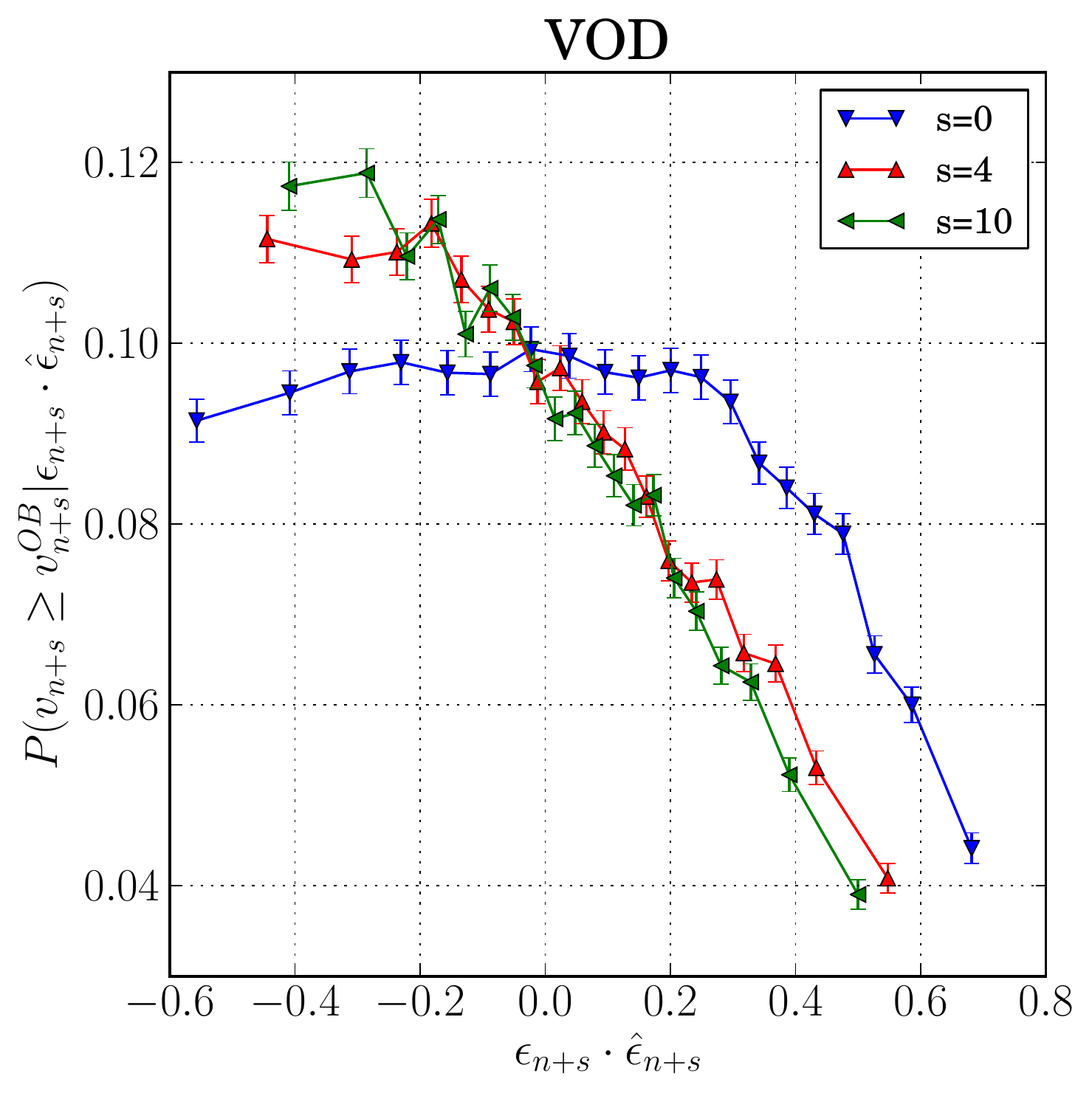}
  \end{minipage} \\
  \begin{minipage}[c]{0.5\columnwidth}
    \includegraphics[width=\columnwidth]{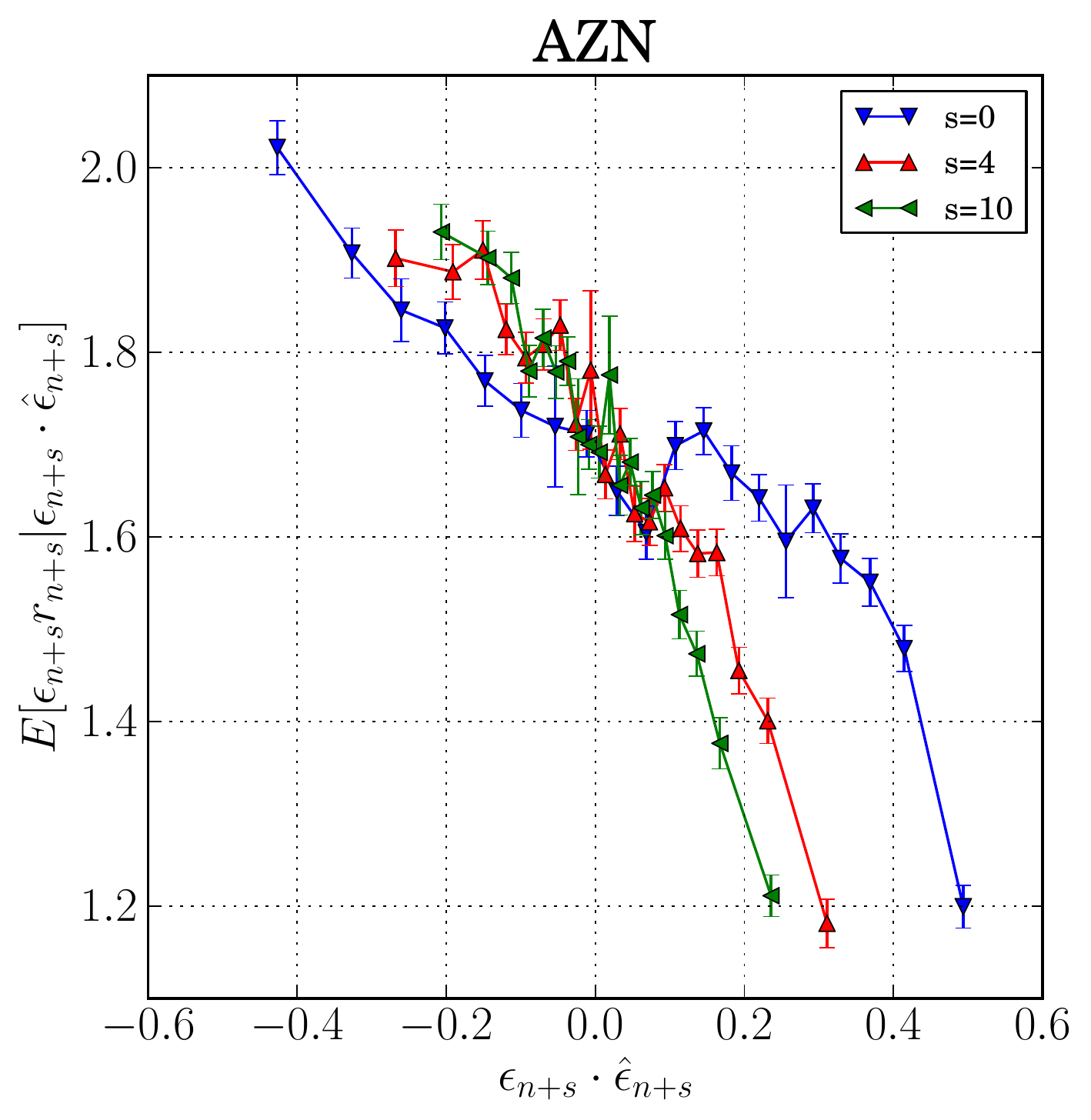}
  \end{minipage}%
  \begin{minipage}[c]{0.5\columnwidth}
    \includegraphics[width=\columnwidth]{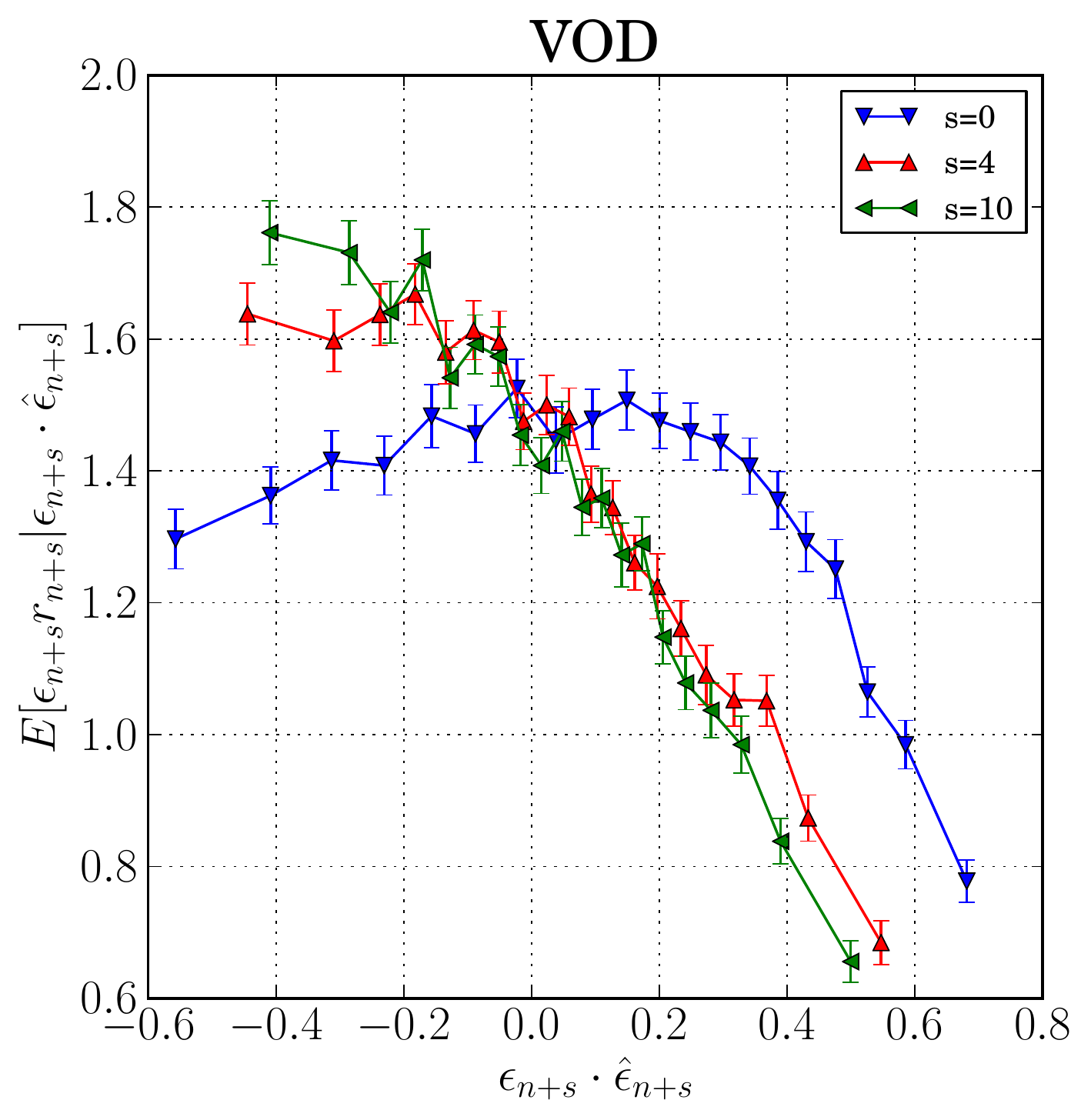}
  \end{minipage}
  \caption{Top row: Lagged probability of penetration $P(v_{n+s} \geqslant v_{n+s}^{OB}|\epsilon_{n+s}\cdot\hat{\epsilon}_{n+s})$. Bottom row: Returns $\mathbb{E}[\epsilon_{n+s}r_{n+s}|\epsilon_{n+s}\cdot\hat{\epsilon}_{n+s}]$ for AZN and VOD and different step values $s$. The error bars are standard errors.}
  \label{fig:lagged_penetration_return2}
\end{figure}

Given that at the individual transaction level clear signs of inefficiency exist, one can ask whether these inefficiencies are removed when one considers the expected signed returns $s$ steps (transactions) ahead, conditional to the present information set. To this end we use the  information set $\Omega_{n-1}$ at time $t_{n-1}$ to build the predictor of market order sign at time $t_{n+s}$. We then compute expectations at time $n+s$ conditional to the variable $\epsilon_{n+s}\cdot\hat{\epsilon}_{n+s}$. In particular we measure
\begin{equation*}
\mathbb{E}[\epsilon_{n+s}r_{n+s}|\epsilon_{n+s}\cdot\hat{\epsilon}_{n+s}], \qquad P(v_{n+s} \geqslant v_{n+s}^{OB}|\epsilon_{n+s}\cdot\hat{\epsilon}_{n+s}).
\end{equation*}

The results of these analysis are shown in Figures~\ref{fig:lagged_penetration_return1}~and~\ref{fig:lagged_penetration_return2}. We consistently observe qualitatively the same behaviour as $s$ increases. The conditional return as a function of $\epsilon_{n+s}\cdot\hat{\epsilon}_{n+s}$ shows evidences of inefficiency for small values of $s$. For larger values of $s$ the curves become closer to a linear relation. A similar transition is observed when one considers the lagged probability of penetration. The linear behaviour is consistent with a linear model of market impact, \emph{i.e.} a model where
\begin{equation*}
r_n=A (\epsilon_n-\hat{\epsilon}_n)+\eta_n\,,
\end{equation*}
where for simplicity we have neglected any dependence on the volume and $\eta_n$ is an idiosyncratic component. In this model we have that
\begin{equation*}
\mathbb{E}[\epsilon_n r_n|\epsilon_n\cdot\hat{\epsilon}_n]=A (1-\epsilon_n\cdot\hat{\epsilon}_n)\,,
\end{equation*}
\emph{i.e.} a linear behaviour. The data shows that this linear behaviour is not observed for $s=0$, but rather for intermediate values of $s$. We postulate therefore that a linear model of market impact could be developed to describe returns on an aggregated time scale.

When $s$ is very large, the conditional return curves become flatter and flatter. The flat behaviour can be understood by considering that when the lag $s$ is very large, the value of the predictor is typically very close to zero and its predictive power is very low. In fact in the limit of no predictability, it is $\mathbb{E}[\epsilon_{n+s}r_{n+s}|\epsilon_{n+s}\cdot\hat{\epsilon}_{n+s}]=\mathbb{E}[\epsilon_{n+s}r_{n+s}]$.

It is important to stress that the number of transactions needed to observe a transition from the behaviour of returns which shows inefficiencies to the linear behaviour is different in the two datasets. Specifically, by observing the figures, we note that the stocks of NASDAQ market reach an approximately linear behaviour of the return for a value of $s$ which is larger than the corresponding $s$ for the LSE stocks. We interpret this fact as a sign that NASDAQ market needs more trade time to process past information than LSE market. This may seem surprising at first view, because modern financial markets are supposed to be more efficient and faster in processing information when compared to several years ago. This is surely true in physical time, but it might be false in trade time. High frequency trading decreases the physical time needed to restore efficiency in the market, but it might increase the trade time.

In order to test this hypothesis more quantitatively, for each stock we estimated the minimal value of $s$ such that the inefficiency condition of Equation~\ref{eqn:cond_inefficiency} is not observed for any value of $\epsilon_n\cdot\hat{\epsilon}_n>0$. We then multiply this value of $s$ by the average time in seconds between trades, already shown in Table~\ref{tab:info_datasets}, in order to get an average minimal time needed to not observe inefficiency as the one of Equation~\ref{eqn:cond_inefficiency}. By spanning  different $s$ values for each stock, we find that this physical time is $5.5\,\mathrm{s}$ ($5$ lags) for AAPL, for MSFT it is $18.7\,\mathrm{s}$ ($11$ lags), for AZN it is $23.1\,\mathrm{s}$ ($1$ lags) and for VOD it is $114.5\,\mathrm{s}$ ($5$ lags). By considering separately large and small tick size stocks, we conclude that recent NASDAQ stocks become efficient in a smaller physical time than older LSE stocks.

\section{Statistical models of order book and predictability of order flow}
\label{sec:Statistical}
Modeling the dynamics of the order book is in general quite complicated and challenging. This is due to its multidimensional nature and to the non trivial coupling between the different components of the process. In recent years there has been a growing interest toward the statistical modeling of the order book~\cite{smith2003statistical}, \textit{i.e.} a modeling approach where the different components of the order flow (limit orders, market orders, and cancellations) are treated as independent Poisson processes and the state of the order book emerges as the result of the interplay between these different components. This type of modeling is sometimes termed “zero-intelligence”, because the flow of each type of orders follows the simplest unconditional process. Despite being conceptually simple, this modeling approach has proven to be surprisingly useful in giving testable predictions of some very short time~\cite{cont2013price} or long time~\cite{farmer2005predictive} properties of the order book. 

From the previous empirical section we have seen that a key element describing the microstructure of financial markets is the fact that the order flow is long range correlated. As we have seen in the Introduction, when a strongly autocorrelated order flow (such as the one of real markets) is used as input of an empirically calibrated statistical model of the order book, an unrealistic price time series emerges. In particular, strong predictability of returns and super-diffusivity of prices are observed. This can be easily understood by considering that we use a strongly correlated input in a Markovian model of the order book. 

In recent years there have been few notable attempts of modeling the limit order book dynamics in presence of a strongly correlated order flow. Mike and Farmer~\cite{mike2008empirical} introduced a model with correlated order flow. However their main task was to reproduce fat tails of returns and not to reproduce diffusive prices and uncorrelated returns, and in fact in their model these last two properties are not verified. 

The attempt of understanding how diffusivity can be recovered in a limit order book model with long memory market order flow has been discussed in~\cite{toth2011anomalous}. As we will detail more below, the proposed model is a variation of the basic zero intelligence model where the market order signs are long memory. In order to limit the effect of the persistence of market orders on prices, authors proposed that market order volume is a random fraction $f$ of the volume at the opposite best price. They claimed that for a fixed level of market order persistence, there is a critical value of the mean value of $f$ such that the price is diffusive (see also~\cite{mastromatteo2013agent}). 

Our model is a variation of the zero intelligence model. More specifically, the order book is modeled as a discrete price grid of constant minimum price increment $w$ (the tick size, that we set to $w=1\,\mathrm{tick}$). In the simulations this grid must be sufficiently large in order to consider it as an infinite support. Each price level is populated by buy limit orders, if the price level is below the current midpoint, or sell limit orders, if the price level is above the current midpoint. This is the instantaneous snapshot of the order book state, whereas its time evolution is dominated by three different types of stochastic processes: limit order placement, market orders arrival, and cancellations of existing limit orders. Limit order placement follows a Poisson process of rate $\lambda$ per tick and unit event time, which for simplicity is uniform across the discrete price grid. For each event time and each price level we draw the number of limit orders of size $s$ (in our simulations $s=100\,\mathrm{shares}$) from a Poisson distribution. Market orders arrival triggers an immediate transaction with limit orders at the opposite side of the book. Market orders arrive at a rate $\mu$ per unit event time, following a Poisson process, which is independent from limit orders and cancellations. Finally, each existing limit order has the same probability $\nu$ per unit event time to be cancelled by liquidity providers.

These features are present also in the zero intelligence model of~\cite{smith2003statistical,cont2013price}. The modification to the zero intelligence model affects mostly the market order stochastic process. We assume that the market order flow sign process has long-range correlations, reflecting the order splitting strategy of large orders used by liquidity takers. In the next section we detail how we model this correlated process. Moreover, as in~\cite{toth2011anomalous} we set the market order volume executed at time $t_n$ as a fraction $f$ of the best opposite volume, $v_n=f\cdot v_n^{OB}$. The scalar $f$ is a random variable drawn from a specific distribution taking values in $f\in[0,1]$, whose shape plays a crucial role in the model. In~\cite{toth2011anomalous} authors proposed that the random scalar $f \in [0,1]$ is drawn from a beta distribution $P_\zeta(f)=\zeta(1-f)^{\zeta-1}$. The parameter $\zeta > 0$ determines the typical relative volume of market orders and the aggressiveness of liquidity takers. In fact, for $\zeta\rightarrow 0$, the distribution peaks around $f=1$; $\zeta=1$ corresponds to a uniform distribution; finally, the limit $\zeta\rightarrow\infty$ corresponds to unit volumes, because we fix a lower bound for market order volumes to $\min(f \cdot v_n^{OB})=1$.

\begin{figure}[p]
\begin{minipage}[c]{0.5\columnwidth}
\includegraphics[width=\columnwidth]{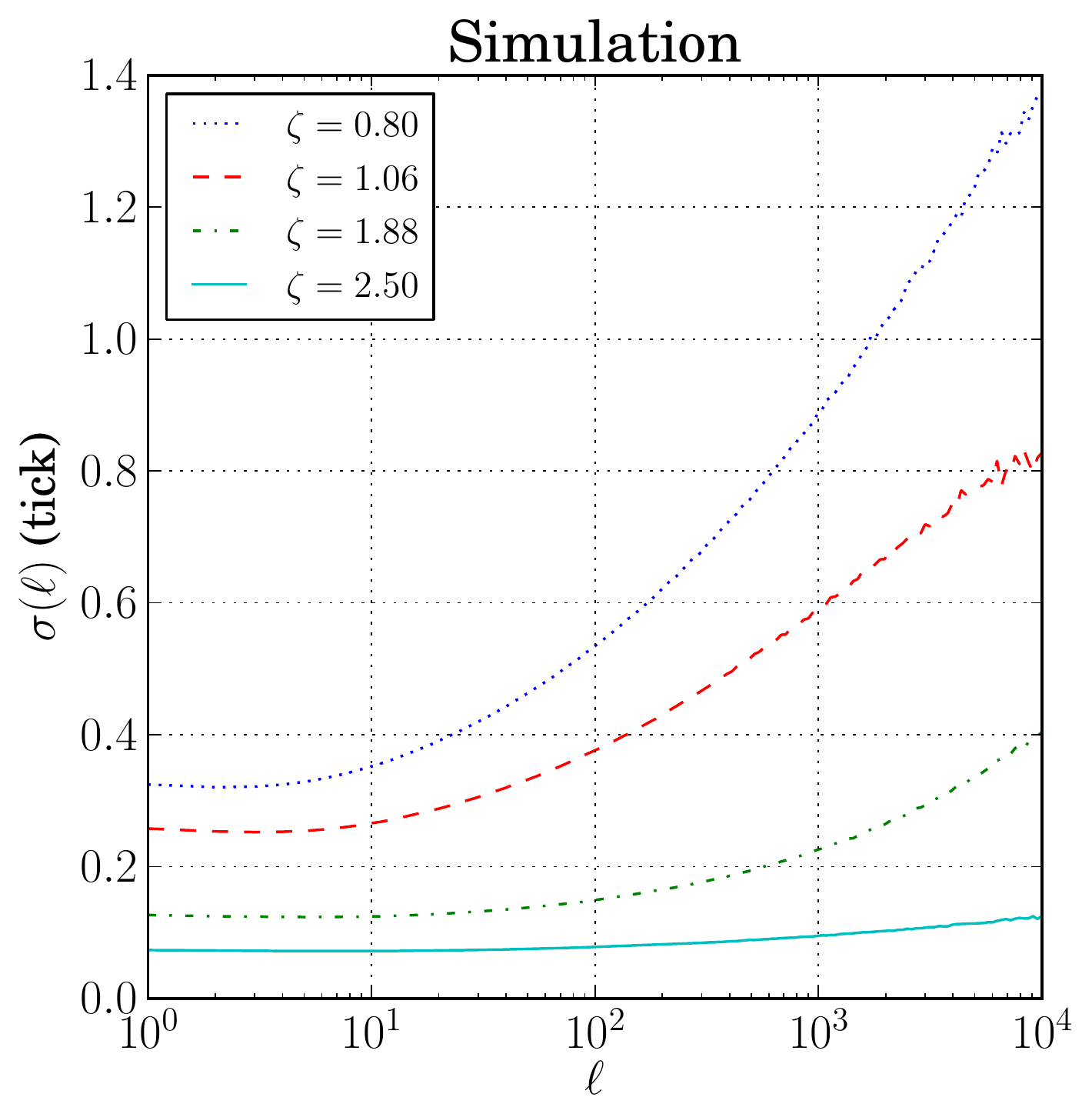}
\end{minipage}%
\begin{minipage}[c]{0.5\columnwidth}
\includegraphics[width=\columnwidth]{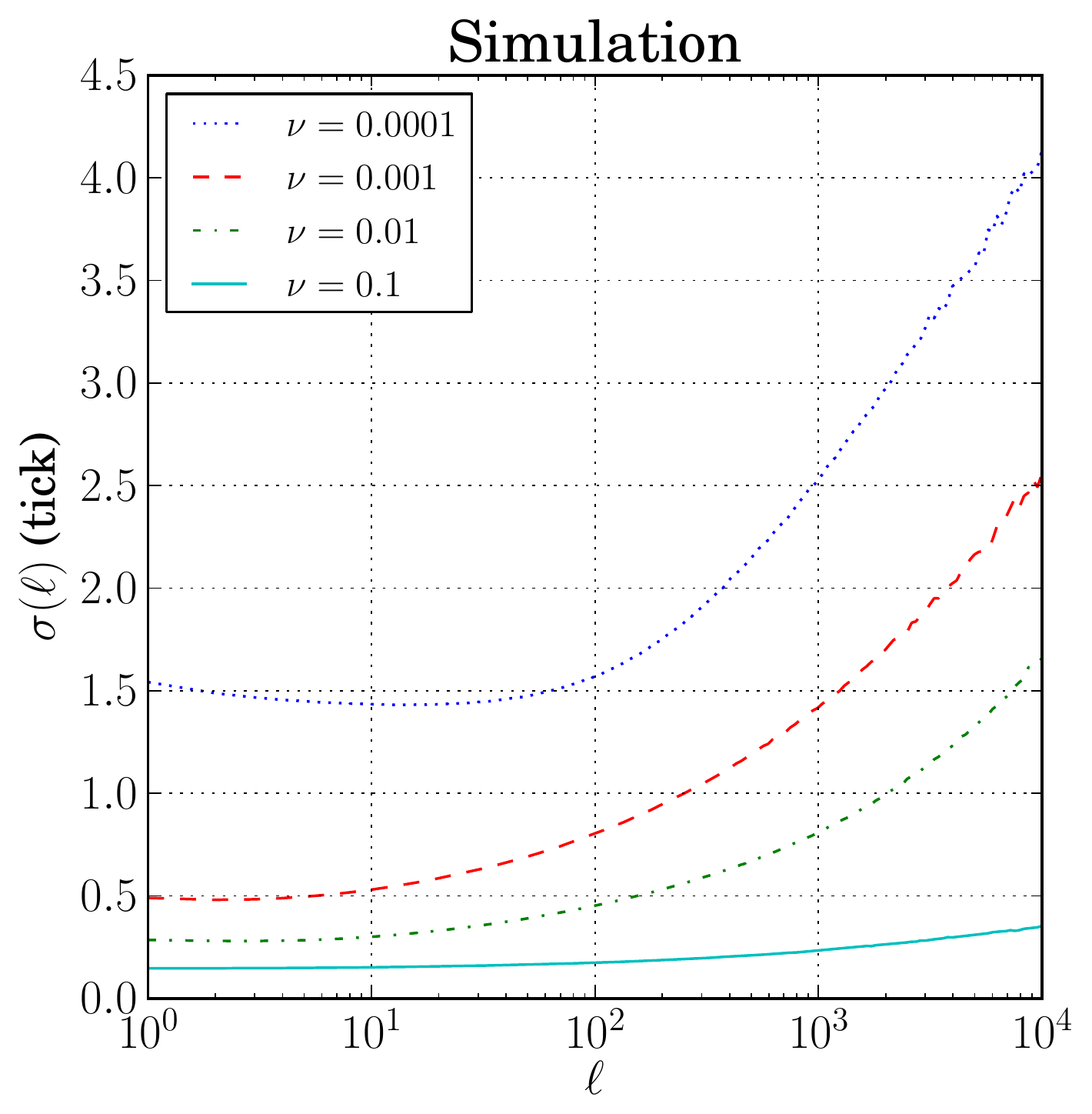}
\end{minipage}
\begin{minipage}[c]{0.5\columnwidth}
\includegraphics[width=\columnwidth]{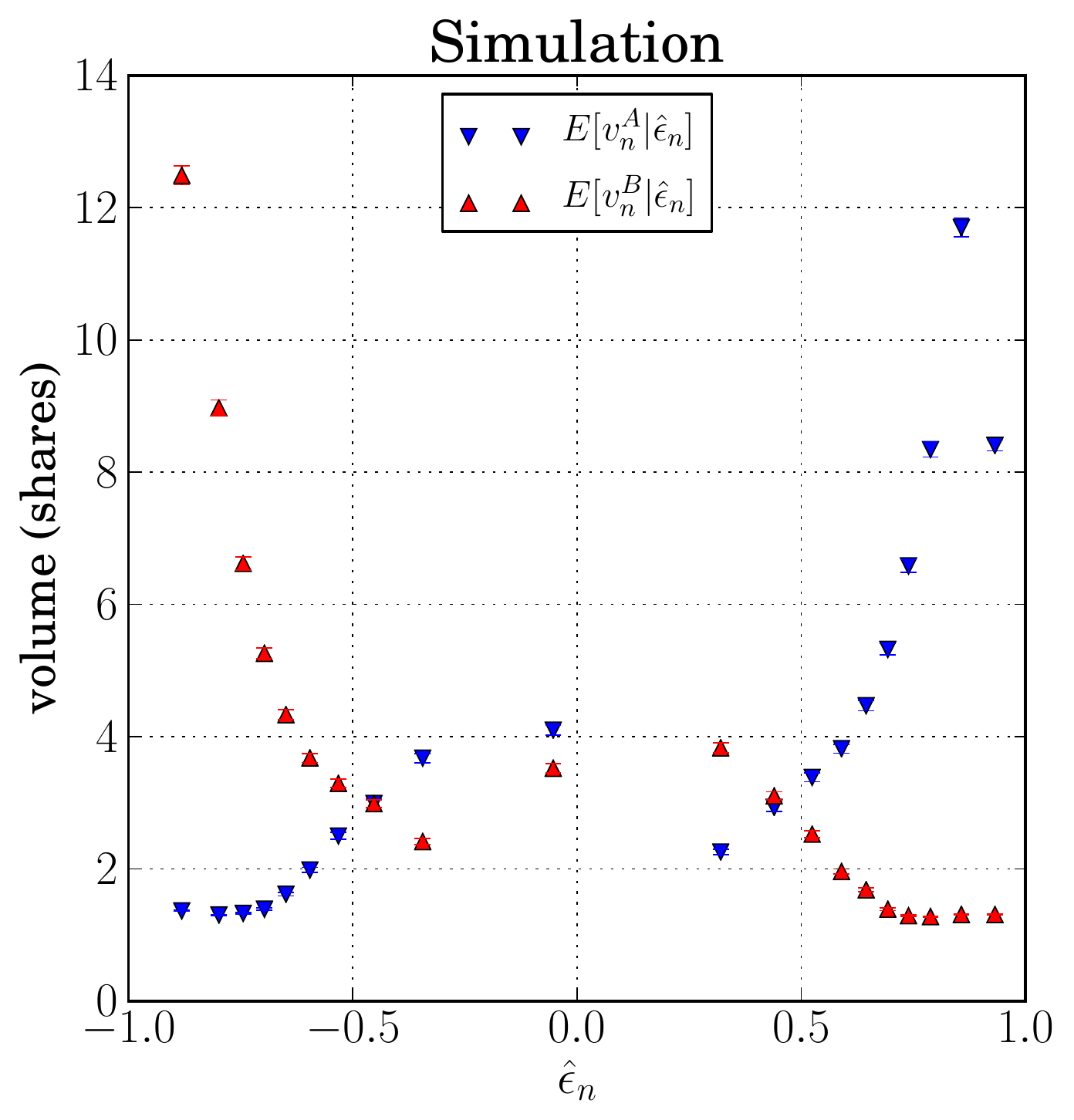}
\end{minipage}%
\begin{minipage}[c]{0.5\columnwidth}
\includegraphics[width=\columnwidth]{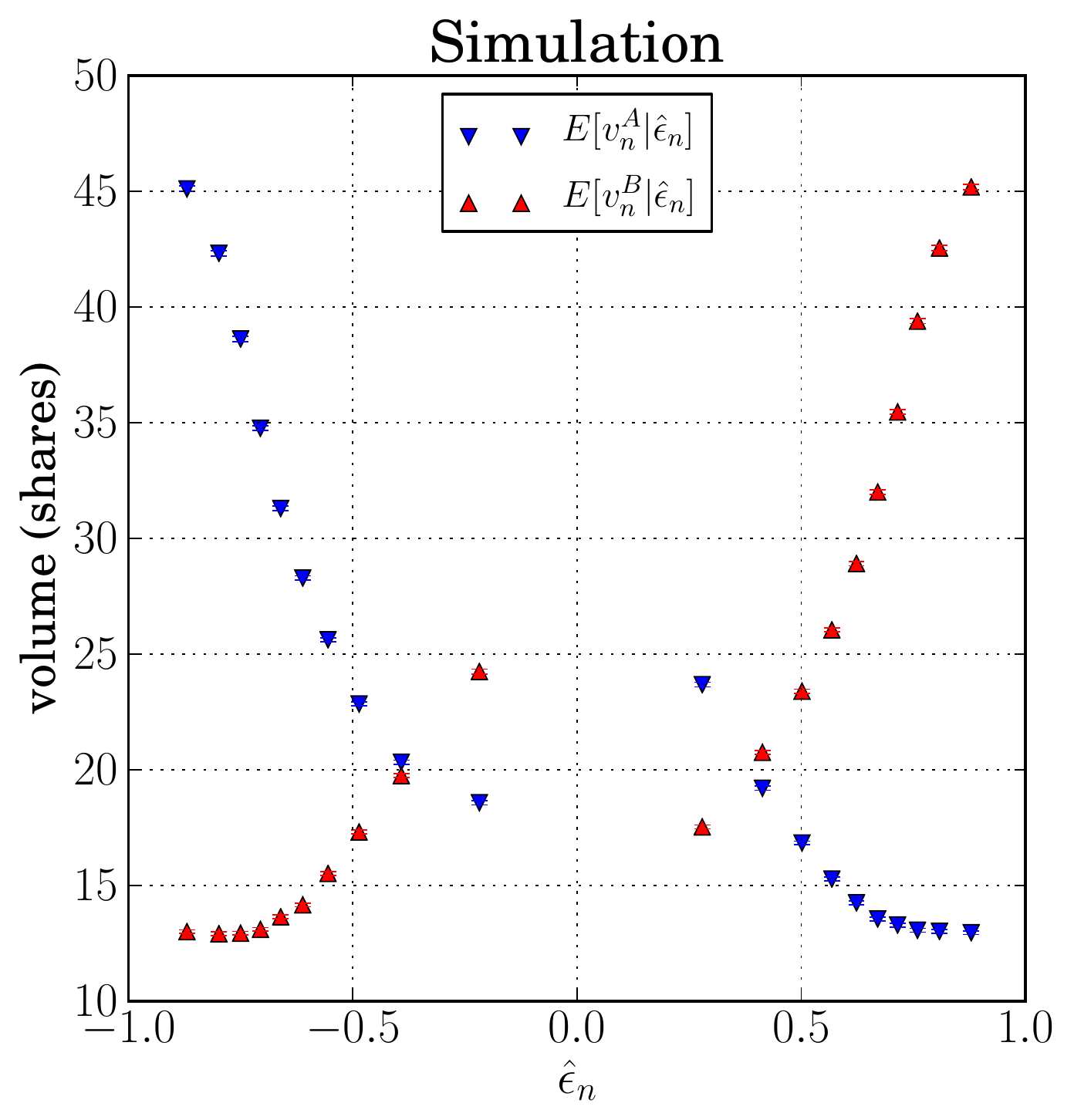}
\end{minipage}
\caption{(Top left) Signature plot $\sigma(\ell)$ for the parameter choice $\mu=0.1\,\mathrm{s^{-1}}$, $\lambda=0.5\,\mathrm{s^{-1}w^{-1}}$, $\nu=0.01\,\mathrm{s^{-1}}$, $\gamma=0.5$ and different values of $\zeta$. The resulting curves show a strong super-diffusive behaviour for large values of $\ell$. When $\zeta\rightarrow\infty$, the volatility converges to zero. (Top right) Signature plot $\sigma(\ell)$ for the parameter choice $\mu=0.1\,\mathrm{s^{-1}}$, $\zeta=0.95$, $\gamma=0.5$, different values of $\nu$ and fixed asymptotic order book depth $\rho_\infty=\lambda w/\nu=50\,\mathrm{shares}$. The resulting curves show a strong super-diffusive behaviour for large values of $\ell$, whereas for low values of $\nu$ the price process has sub-diffusive behaviour for an intermediate time scale region $\ell$. (Bottom) Conditional volumes at the best quotes on different values of the sign predictor, for parameters $\nu=10^{-4}\,\mathrm{s^{-1}}$, $\zeta_c=0.95$ (left) and $\nu=10^{-2}\,\mathrm{s^{-1}}$, $\zeta_c=2.5$ (right). The result is compatible with real markets when the cancellation rate is high, whereas for low values of $\nu$ volumes at the best quotes behave in the opposite way.}
\label{fig:sim_bence}
\end{figure}

In order to test price diffusivity, we investigate the “signature plot” of the model using Equation~\ref{eqn:signplot}. Toth et al.~\cite{toth2011anomalous} have found numerically that there exists a critical value $\zeta_c$ for which the resulting price process of the model is “diffusive” in the intermediate time scale region $\mu^{-1} \ll t \ll \nu^{-1}$, where $t$ is the event time of the model. Thus the lifetime of limit orders $\nu^{-1}$ is a critical ingredient for the diffusivity of the price process. For times longer than this value, the long-range correlation of the order signs dominates and the lagged returns are positively correlated. 

An illustration of this fact is shown in the top panels of Figure~\ref{fig:sim_bence}. The left top panel shows the signature plot of the Toth et al.~\cite{toth2011anomalous} model for different values of the parameter $\zeta$. From the figure, where we set $\nu^{-1}=100\,\mathrm{s}$, it is clear that prices are asymptotically super-diffusive. Moreover when $\zeta\to \infty$ volatility goes to zero. This is due to the fact that in this limit, volume at the best is never eroded by market orders and price remains constant. The top right panel of Figure~\ref{fig:sim_bence} shows the signature plot of the model in~\cite{toth2011anomalous} for different values of cancellation rates, but keeping fixed the asymptotic order book depth $\rho_\infty=\lambda w/\nu=50\,\mathrm{shares}$. The critical value of the cancellation rate is $\nu=10^{-4}$ and we observe an approximately diffusive behaviour for lags $\ell$ between $1$ and $100$ (we convert event time to trade times, like lags $\ell$, by using the relation $\ell \approx t\cdot\mu\,\mathrm{trades}$). As expected, after this value the price becomes highly super-diffusive. By increasing the cancellation rate, the intermediate region for which prices are diffusive shrinks to zero.

In conclusion, the Toth et al.~\cite{toth2011anomalous} model reproduces diffusive prices in a range of lags which strongly depends on the cancellation rate. In order to extend the range of diffusivity one needs to decrease the cancellation rate to very low values. These values are unrealistically small if one wants to consider the model as describing the \textit{real} order book. The authors in~\cite{toth2011anomalous} used this model to describe the latent order book instead. This means that this is a hidden liquidity model, where the values of the rates of cancellation are explicitly chosen to be much smaller than the ones observed in the visible limit order book.

We notice that low cancellation rates lead to wrong predictions of stylized facts of the order book. In the bottom panels of Figure~\ref{fig:sim_bence} we show the volume at the best bid and ask conditional to the value of the predictor $\hat\epsilon_n$. In all cases we have selected the value of $\zeta$ that gives “diffusive” prices, according to the method in~\cite{toth2011anomalous}. The left panel refers to the low cancellation rate regime, while the right panel refers to the high cancellation rate regime (compare with the empirical results of Figure~\ref{fig:volbest}). We observe that in the low cancellation rate regime, the conditional volume at the best is opposite to the one observed in real data, \textit{i.e.} there is more volume at the ask (bid) when it is more likely that the next order is a buy (sell). On the other hand, when the cancellation rate is high (as in real markets) the conditional volume at the best is in agreement with real data. 

Thus if we want diffusive prices on a large range of lags we need low cancellation rates, but in this case the conditional properties of the order book have the wrong sign. If we want to reproduce the latter, we need high and realistic cancellation rates, but in this case the range of diffusivity will be very small. 

We therefore conclude that current statistical models of the order book are unable to reproduce the observed stylized facts when one considers a strongly persistent order flow and is interested in how order book quantities change as a function of order flow predictability as well as efficiency and diffusivity. Even when one uses mechanisms for counterbalancing the persistence of order flow, such as by fine tuning the value of a parameter (e.g. the penetration probability in the~\cite{toth2011anomalous} model), diffusivity is reproduced up to the maximal time scale of the cancellation rate. By decreasing the cancellation rate one obtains a very low volatility and it is not able to reproduce other stylized facts, such as the volume imbalance at bid and ask as a function of order flow predictability.

In the following section we present a statistical order book model with long memory order flow where we are able to simultaneously obtain exact diffusivity of prices and the correct conditional properties of the order book as a function of the order flow predictability. The key intuition behind our modeling scheme is that order book and flow dynamics depend on the predictability of the order flow itself. In other words, instead of fine tuning the value of a parameter (such as the $\zeta$ in~\cite{toth2011anomalous} model), we assume that this “parameter” adapts itself dynamically, depending on the predictability of order flow. This kind of adaptation guarantees diffusivity and the correct dependence of order book quantities on order flow predictability. 

The version of the model we present here aims at modeling how liquidity takers adapt their order flow, keeping the price diffusive and efficient. In this sense our model is close to the one in~\cite{toth2011anomalous}, since the limit order and cancellation processes are totally random. The adaptation occurs inside the market order flow. However we believe that the idea of adaptation could be exported for modeling also the liquidity providers.

\section{Adaptive liquidity model, price diffusivity, and market efficiency}
\label{sec:Adaptive}
The main intuition behind our modeling approach is that the mechanism restoring efficiency (and therefore diffusivity) must depend on the local level of predictability of the order flow. In the previous sections on the empirical analysis we have shown that many quantities of the order flow and of the limit order book depend in fact from the degree of predictability of the order flow, as well as from the fact that the next market order is in agreement or not with the predictor. In particular we have seen that the penetration probability, \textit{i.e.} the probability that the market order volume is larger or equal to the volume at the opposite best, strongly depends on $\epsilon_n\cdot\hat{\epsilon}_n$ (see Figures~\ref{fig:penetration_fraction} and~\ref{fig:vol_mo_ob}). When this quantity is large, \textit{i.e.} the predictability is high and order executed agrees with the predictor, the volume at the opposite best is small, but the penetration probability also declines, suggesting that liquidity takers adjust the volume of their market order to reduce market impact. This behaviour clearly counterbalances the persistence of order flow, making prices more diffusive and efficient. We now show that it is possible to {\it exactly} counterbalance the super-diffusivity of order flow and to give the correct conditional behaviour of limit order book quantity. Before describing the mechanism a caveat is in order. We do not believe that this is the only mechanism responsible for efficiency and diffusivity. We believe that liquidity providers, through the so-called “stimulated refill” mechanism (see~\cite{eisler2012price, mastromatteo2013agent}), are also responsible in part of the restoration of efficiency. However, we think that this mechanism should also be adaptive, depending on the local level of predictability of order flow. 

Our model take as a starting point the model of~\cite{toth2011anomalous}. We assume that the distribution of $f$, the ratio between the volume of the market order and the volume at the opposite best, depends parametrically on  $\epsilon_n\cdot\hat{\epsilon}_n$. In particular, we assume that
\begin{equation*}
P_g(f|\epsilon_n\cdot\hat{\epsilon}_n)=g(\epsilon_n\cdot\hat{\epsilon}_n)(1-f)^{g(\epsilon_n\cdot\hat{\epsilon}_n)-1}\,,
\end{equation*}
where $g(\epsilon_n\cdot\hat{\epsilon}_n)$ is the exponent of the beta distribution, which in~\cite{toth2011anomalous} model is the constant $\zeta$, and fine tuned to recover “diffusivity”. In our model this exponent depends on the predictability of market order flow and the degree of surprise of the market order (\textit{i.e.} if it agrees with the predictor). 

In our model, when $f\in[1-\delta,1]$, where $\delta\in(0,1)$ is a small parameter, the market order volume is equal to the volume at the opposite best and penetrates the book. Thus the conditional probability of penetration is
\begin{eqnarray*}
\fl P\left(v_n = v_n^{OB}\big|\epsilon_n\cdot\hat{\epsilon}_n\right)&=&\int_{1-\delta}^1 P_{g}(f|\epsilon_n\cdot\hat{\epsilon}_n)df \\
&=&\int_{1-\delta}^1 g(\epsilon_n\cdot\hat{\epsilon}_n)(1-f)^{g(\epsilon_n\cdot\hat{\epsilon}_n)-1}df=\delta^{g(\epsilon_n\cdot\hat{\epsilon}_n)}\,.
\end{eqnarray*}

Since $\delta<1$, $P\left(v_n = v_n^{OB}\big|\epsilon_n\cdot\hat{\epsilon}_n\right)$ is a decreasing function of $\epsilon_n\cdot\hat{\epsilon}_n$ if $g$ is an increasing function of its argument. 

This framework reproduces the strategic behaviour of liquidity takers against liquidity providers which operates in a completely random setting. Those who place market orders adjust locally the requirement of liquidity on the level of predictability of the order signs. This mechanism is captured by the model through the adaptive dependence of $P_{g}(f|\epsilon_n\cdot\hat{\epsilon}_n)$ on the sign predictor value. A liquidity taker knows exactly the past history of the market order sign process and the sign of the next order (buy or sell) executed in the market, therefore the choice of the local dependence of $P_g$ appears to us reasonable. We can explain this strategic behaviour of the traders in this way: high predictability of the order flow means that liquidity takers reveal to the market information about their intentions, and in order to control the market impact of their trades, they reduce the volumes of the market orders progressively during the execution of the whole metaorder.

Under some conditions, the penetration probability can be connected to the market impact. In fact, we have
\begin{eqnarray*}
\fl \mathbb{E}[\epsilon_n r_n|\epsilon_n\cdot\hat{\epsilon}_n]&\simeq&\sum_{r_n \neq 0}\epsilon_n r_n P\left(v_n = v_n^{OB},\epsilon_n g_n^{OB} \simeq 2r_n\big|\epsilon_n\cdot\hat{\epsilon}_n\right) \nonumber \\
&\simeq&\sum_{r_n \neq 0}\epsilon_n r_n P\left(v_n = v_n^{OB}\big|\epsilon_n\cdot\hat{\epsilon}_n\right)P\left(\epsilon_n g_n^{OB} \simeq 2r_n\big|\epsilon_n\cdot\hat{\epsilon}_n,v_n = v_n^{OB}\right) \nonumber \\
&\simeq&\frac{1}{2}P\left(v_n = v_n^{OB}\big|\epsilon_n\cdot\hat{\epsilon}_n\right)\mathbb{E}\left[g_n^{OB}\big|\epsilon_n\cdot\hat{\epsilon}_n,v_n = v_n^{OB}\right]\,.
\end{eqnarray*}

We consider a configuration of our model where the price gaps are constant and equal to $w=1\,\mathrm{tick}$ (this is the case of large-tick stocks, where each price level behind the best quotes is populated by limit orders), so that the previous equation reduces to
\begin{equation}
\mathbb{E}[\epsilon_n r_n|\epsilon_n\cdot\hat{\epsilon}_n] \simeq \frac{w}{2}P\left(v_n = v_n^{OB}\big|\epsilon_n\cdot\hat{\epsilon}_n\right) =\frac{w}{2}\delta^{g(\epsilon_n\cdot\hat{\epsilon}_n)}\,.
\label{eqn:return_penetration}
\end{equation}

The expression in the left hand side of Equation~\ref{eqn:return_penetration} is exactly the probability of penetration of market orders, which is the probability that the volume of the market order is equal to the volume at the opposite best. 

It is possible to choose the function $g$ is such a way that
\begin{equation*}
P\left(v_n = v_n^{OB}\big|\epsilon_n\cdot\hat{\epsilon}_n\right)\equiv\alpha+\beta\epsilon_n\cdot\hat{\epsilon}_n\,,
\end{equation*}
\textit{i.e.} the penetration probability is a linear function of $\epsilon_n\cdot\hat{\epsilon}_n$. Taking into account that the last expression is a probability, we fix the constants $\alpha, \beta$ as
\begin{eqnarray*}
P\left(v_n = v_n^{OB}\big|\epsilon_n\cdot\hat{\epsilon}_n=1\right)=0 &\Longrightarrow& \beta=-\alpha \,,\\
P\left(v_n = v_n^{OB}\big|\epsilon_n\cdot\hat{\epsilon}_n=-1\right)\in[0,1] &\Longrightarrow& \alpha\in[0,1/2]\,,
\end{eqnarray*}
and we obtain
\begin{eqnarray*}
g(\epsilon_n\cdot\hat{\epsilon}_n)&=&\frac{\log\alpha+\log(1-\epsilon_n\cdot\hat{\epsilon}_n)}{\log\delta}\,. 
\end{eqnarray*}

Under the large-tick size hypothesis, Equation~\ref{eqn:return_penetration} implies that returns and impact are also linear functions. In fact, we have
\begin{equation*}
\mathbb{E}[\epsilon_n r_n|\epsilon_n\cdot\hat{\epsilon}_n]=A(1-\epsilon_n\cdot\hat{\epsilon}_n)\,,
\end{equation*}
where $A=w \alpha/2$. This model can be rewritten as
\begin{equation}
r_n=p_{n+1}-p_n=A(\epsilon_n-\hat{\epsilon}_n)+\eta_n,\qquad \hat{\epsilon}_n=\mathbb{E}_{n-1}[\epsilon_n|\Omega_{n-1},\mathcal{M}]\,,
\label{eqn:efficient_model}
\end{equation}
where $\eta_n$ is an idiosyncratic IID component of variance $\Sigma^2$. The model is completely defined when we assign the model for the time series of order signs as well as the predictor and the information set used. 

We have thus found a reduced statistical model for price returns starting from a structural model of the order book. As shown in~\cite{bouchaud2009markets}, using as sign predictor an \textit{AR(p)} model when $p\rightarrow\infty$, the statistical model of Equation~\ref{eqn:efficient_model} proposed in~\cite{farmer2006market} is equivalent to the propagator model (see~\cite{bouchaud2004fluctuations,bacry2013hawkes}). We expand this result to the case of the \textit{DAR(p)} model, for which we remind the analytical expression of the sign predictor (see Equation~\ref{eqn:epshat})
\begin{equation*}
\hat{\epsilon}_n=\chi\sum_{i=1}^p \phi_i \epsilon_{n-i},
\end{equation*}
where for simplicity we restrict the model to the case of $\mu_Z=\mathbb{E}[\epsilon_n]=0$. In the propagator model~\cite{bouchaud2004fluctuations}, prices are written as a superposition of past order signs, weighted by a propagator $G_0(\ell)$, and external shocks 
\begin{equation*}
p_n=\sum_{i>0}\left[G_0(i)\epsilon_{n-i}+\eta_{n-i}\right],
\end{equation*}
which leads to the expression of the tick by tick returns of the model,
\begin{equation*}
r_n=p_{n+1}-p_n=G_0(1)\epsilon_n+\sum_{i>0}\left[G_0(i+1)-G_0(i)\right]\epsilon_{n-i}+\eta_n\,.
\end{equation*}

If we impose the equivalence
\begin{equation*}
A \chi \phi_i=G_0(i)-G_0(i+1) \qquad \mbox{or} \qquad G_0(\ell)=A\chi\left[1-\sum_{j=1}^{\ell-1}\phi_j\right]\,,
\end{equation*}
we find the relation between the coefficient of the statistical model of Equation~\ref{eqn:efficient_model} and the functional form of the propagator $G_0(\ell)$ of the model of~\cite{bouchaud2004fluctuations}.

Let us emphasize the statistical properties of returns of the model of Equation~\ref{eqn:efficient_model}. We clearly see that prices are efficient, since
\begin{equation*}
\mathbb{E}_{n-1}[r_n|\Omega_{n-1},\mathcal{M}]=0\,.
\end{equation*}

This means that returns are uncorrelated, $\mathbb{E}[r_n r_{n+\ell}]=0$, for all $\ell>1$. Since $\mathbb{E}[r_n r_{n+\ell}]=\mathbb{E}[\mathbb{E}[r_n r_{n+\ell}|\Omega_{n-1},\mathcal{M}]]$ it is enough to prove that
$\mathbb{E}[r_n r_{n+\ell}|\Omega_{n-1},\mathcal{M}]=0$, which follows making use of the law of iterated expectations.

Furthermore, prices of the model are diffusive for all lags. Let us consider the quantity $p_{n+\ell}-p_n$ and compute its variance, using the last result of uncorrelated returns of the model
\begin{equation*}
\fl \mathbb{E}[(p_{n+\ell}-p_n)^2]=\mathbb{E}\left[\left(\sum_{i=0}^{\ell-1} r_{n+i}\right)^2\right]=\sum_{i=0}^{\ell-1} \mathbb{E}\left[r_{n+i}^2\right] = (\Sigma^2 + A^2)\ell-A^2\sum_{i=0}^{\ell-1} \mathbb{E}\left[\hat{\epsilon}_{n+i}^2\right]\,.
\end{equation*}

The quantity $\mathbb{E}\left[\hat{\epsilon}_{n+i}^2\right]$ depends only on the particular choice of the driving model of the order flow. In the case of the \textit{DAR(p)} model it is constant and independent from $\ell$,
\begin{equation*}
\sum_{i=0}^{\ell-1} \mathbb{E}\left[\hat{\epsilon}_{n+i}^2\right]= \left(\chi^2\sum_{i=1}^p\phi_i^2+2\chi^2\sum_{r>s\geqslant 1}^p\phi_r \phi_s \rho(r-s)\right)\ell\,,
\end{equation*}
where $\rho(\ell)=\mathbb{E}[\epsilon_n\epsilon_{n+\ell}]$ is the empirical autocorrelation of order signs. The unconditional variance finally reads
\begin{equation*}
\mathbb{E}[(p_{n+\ell}-p_n)^2]=\left[\Sigma^2 + A^2\left(1-\chi^2\sum_{i=1}^p\phi_i^2-2\chi^2\sum_{r>s\geqslant 1}^p\phi_r \phi_s \rho(r-s)\right)\right]\ell\,,
\end{equation*}
which scales perfectly as a diffusive process with the lag $\ell$.

\section{Results}
\label{sec:Results}
Since in our model the market order flow plays a crucial role, in this section we present the specific model for the time series describing it. Moreover we shall discuss the different predictors that can be built for this time series. In the final subsection we shall present in detail numerical simulations of the model.

\subsection{Models of the market order flow}
Order flow is strongly autocorrelated in time. As shown in~\cite{toth2011order}, correlation of order flow is mostly due to order splitting, rather than herding. This was originally suggested by Lillo, Mike, and Farmer~\cite{lillo2005theory} on the basis of indirect empirical evidences. In this paper, authors proposed a simple model where the correlation of order flow is a consequence of order splitting and the very heterogeneous distribution of metaorder sizes. Here we use a variation of this model to generate the market order flow which enters the limit order book.

According to the model of~\cite{lillo2005theory}, there are $M$ funds that want to trade one metaorder each of a size $L_i$ ($i=1,..,M$) taken from a distribution $p_L$, where for simplicity $L_i\in \mathbb N^+$. The sign of the each metaorder is taken randomly and at each trade time step, one fund is picked randomly with uniform probability. The selected fund initiates a trade of the sign of its metaorder, and the size of the metaorder is reduced by one unit. When a metaorder is completely traded, a new one is drawn from $p_L$ and assigned a random sign. 

In \cite{lillo2005theory} it is shown how to connect the distribution $p_L$ of metaorder size with the autocorrelation function of trade signs. In particular if the distribution is Pareto
\begin{equation}\label{eqn:Pareto}
p_L=\frac{1}{\zeta(\beta)}\frac{1}{L^{1+\beta}}
\end{equation}
where $\zeta(\beta)$ is the Riemann zeta function, then the autocorrelation function of trade signs decays asymptotically as
\begin{equation}\label{eqn:acfLMF}
\rho_s(\ell)=\mathbb{E}[\epsilon_n \epsilon_{n+\ell}]\sim \frac{M^{\beta-2}}{\ell^{\beta-1}} \sim \frac{1}{\ell^\gamma}
\end{equation}

This model connects the exponent of the autocorrelation function of order signs with the tail exponent of metaorder distribution, since $\gamma=\beta-1$. The market order sign is a long memory process if $\beta<2$, \emph{i.e.} if the variance of the metaorder size diverges. 

There is a growing empirical evidence that the distribution of metaorder size is asymptotically Pareto distributed with a tail exponent close to $\beta=1.5$. Ref.s~\cite{gabaix2006institutional} and~\cite{lillo2005theory} argue that block trades (\emph{i.e.} traded off book) could be used as a proxy of metaorders and find that an exponent very close to $1.5$ describes the tail of the trade size distribution. Ref.~\cite{vaglica2008scaling} use trade data of the Spanish Stock Exchange with an identifier of the broker to statistically reconstruct the metaorders. They find that the size of the metaorder is asymptotically Pareto distributed with an exponent $\beta\approx 1.7$. Finally,~\cite{bershova2013impactlarge} use proprietary data of a set of large institutional metaorders executed at AllianceBernstein’s buy-side trading desk in the US equity market and find that the tail of metaorder size is Pareto with exponent $\beta=1.56$.

In this paper in order to have an analytically tractable expression of the predictor of the order flow, we shall consider a slight modification of the above model. First of all we will consider that only one metaorder is present at each time. This is similar to what is done in~\cite{toth2011anomalous,mastromatteo2013agent}. The second modification is that we will assume that other traders are present and that they contribute with a random background of signs. This can be considered as a large set of metarorders of size $1$. 

More specifically, we introduce the participation ratio $\pi$ of the metaorder, which is the probability that a trade is initiated by the metaorder (of size larger than one), while $1-\pi$ is the probability that the trade is initiated by the noise traders. 

The introduction of the noise traders does not change the long memory properties of the autocorrelation of the order flow. Their only effect is to reduce the global level of the autocorrelation. More specifically, if $\rho_s(\ell)$ is the autocorrelation function of the order flow when one considers only the trades of the metaorder (\emph{i.e.} Equation~\ref{eqn:acfLMF} with $M=1$), one has in presence of noise
\begin{equation*}
\rho(\ell)=\mathbb{E}[\epsilon_n \epsilon_{n+\ell}]\simeq\pi^2 \rho_s (\pi \ell)\,,
\end{equation*}
because the probability that the two trades at time $t_n$ and $t_{n+\ell}$ both come from a metaorder (not necessarily the same) is $\pi^2$ and a time lag of $\ell$ trades corresponds on average to a time lag of $\pi \ell$ trades from the metaorder.

If the metaorder size distribution is Pareto (see Equation~\ref{eqn:Pareto} and~\ref{eqn:acfLMF}), we have
\begin{equation*}
 \rho(\ell)\sim \frac{\pi^2}{(\pi\ell)^{\beta-1}}=\frac{\pi^{3-\beta}}{\ell^{\beta-1}}\,,
\end{equation*}
\emph{i.e.} the autocorrelation function is dampened by a factor $\pi^{3-\beta}$, but it is still long memory with the same Hurst exponent.

\subsection{Predictors of the order flow}
In our model market order volume depends on the predictability of market order flow. Given the time series model described above, we will consider two predictors of the order flow. 

The first predictor is the one associated with the \textit{DAR(p)} model discussed in the empirical section and reviewed in the appendix. The $p$ signs of past market orders are used to build the expected value of the next sign. Clearly this predictor does not have any direct information on how many metaorders were present in the estimation window, thus we call it the “public” sign predictor. Given the fact that our order flow model is composed by one metaorder at a time (plus the noise background), it is likely that the estimation window of $p$ past signs includes orders that are coming from past (\emph{i.e.} not anymore active) metaorders. This adds of course noise and decreases the forecasting ability of the predictor. On the opposite side, if the metaorder is longer than $\pi p$, from a certain point on, the predictor is using information of the most recent part of the metaorder and it is discarding information of the first part of the metaorder. As we will see below, this will have an effect on the diffusivity of price at time scales longer than $p$ trades.

The second predictor is the one which makes use of the information allowing to discriminate the orders due to the active metaorder to those due to the noisy background. This information is not typically of public domain and therefore cannot be used by the liquidity providers, therefore we call it the “private” sign predictor. In our model the liquidity taker adjusts the volume of their market orders to the degree of predictability of the order flow. Given their active role, they are able to use a predictor that takes into account the history of the recent order flow and the information on the current length of the metaorder.

The key point is that the correlation of the order flow comes from the presence of the metaorder. If $m$ trades of the current metaorder has been already traded, the probability that the metaorder continues is~\cite{farmer2013efficiency}
\begin{equation*}
{\cal P}_m=\frac{\sum_{i=m+1}^\infty p_i} {\sum_{i=m}^\infty p_i}\,. 
\end{equation*}
For example, if the metaorder size distribution is Pareto (see Equation~\ref{eqn:Pareto}) this continuation probability is
\begin{equation*}
{\cal P}_m=\frac{\zeta(1+\beta,1+m)}{\zeta(1+\beta,m)}\simeq \left(\frac{m}{m+1}\right)^\beta\sim 1-\frac{\beta}{m}\,,
\end{equation*}
where $\zeta(s,a)$ is the generalized Riemann zeta function (also called the Hurwitz zeta function). The approximations are valid in the large $m$ limit. This means that the longer the metaorder has been active, the more likely is that it continues.

Let us suppose that the active metaorder is a buy and the participation rate is $\pi$. The probability that the next order is a buy is
\begin{equation*}
p^+_m=\frac{1-\pi}{2}+\pi\left({\cal P}_m+\frac{1-{\cal P}_m}{2}\right)=\frac{1+\pi{\cal P}_m}{2}\,.
\end{equation*}
The first term describes the event in which the next order is from a noise trader, which with probability $1/2$ will place a buy. The second term describes the event in which the next order comes from a metaorder. Moreover, the first term in parenthesis gives the probability that the active metaorder is not finished (and one trade from it will be surely a buy), while the second term in brackets describes the possibility that the active metaorder is finished and that the order comes from a new metaorder, which with $1/2$ probability is a buy. Similarly if the active metaorder is a sell, rather than a buy, then $p^+_m=(1-\pi{\cal P}_m)/2$. If we indicate with $s_n$ the sign of the active metaorder at time $t_n$, we can rewrite
\begin{equation*}
p^+_n=\frac{1+s_n\pi{\cal P}_m}{2}\,.
\end{equation*}

In general, since the sign predictor $\hat \epsilon_n \equiv p^+_n-p^-_n$ and obviously $p^+_n+p^-_n=1$, we have
\begin{equation*}
p^+_n=\frac{1+\hat\epsilon_n}{2}\,,\qquad p^-_n= \frac{1-\hat\epsilon_n}{2}\,,
\end{equation*}
or, in other words, it is 
\begin{equation*}
\hat{\epsilon}_n^\mathrm{LMF}=\mathbb{E}[\epsilon_n|\Omega_{n-1},\mathrm{LMF}]=2p^+_n-1\,.
\end{equation*}
where LMF refers to the model developed by Lillo, Mike, and Farmer in~\cite{lillo2005theory} and described above. This means that the predictor which allows to discriminate the trades from the metaorder is
\begin{equation}
\hat{\epsilon}_n^\mathrm{LMF}=s_n\pi {\cal P}_m\,. \label{eqn:exact_pred}
\end{equation}

\begin{figure}[t]
\centering
\includegraphics[width=0.75\columnwidth]{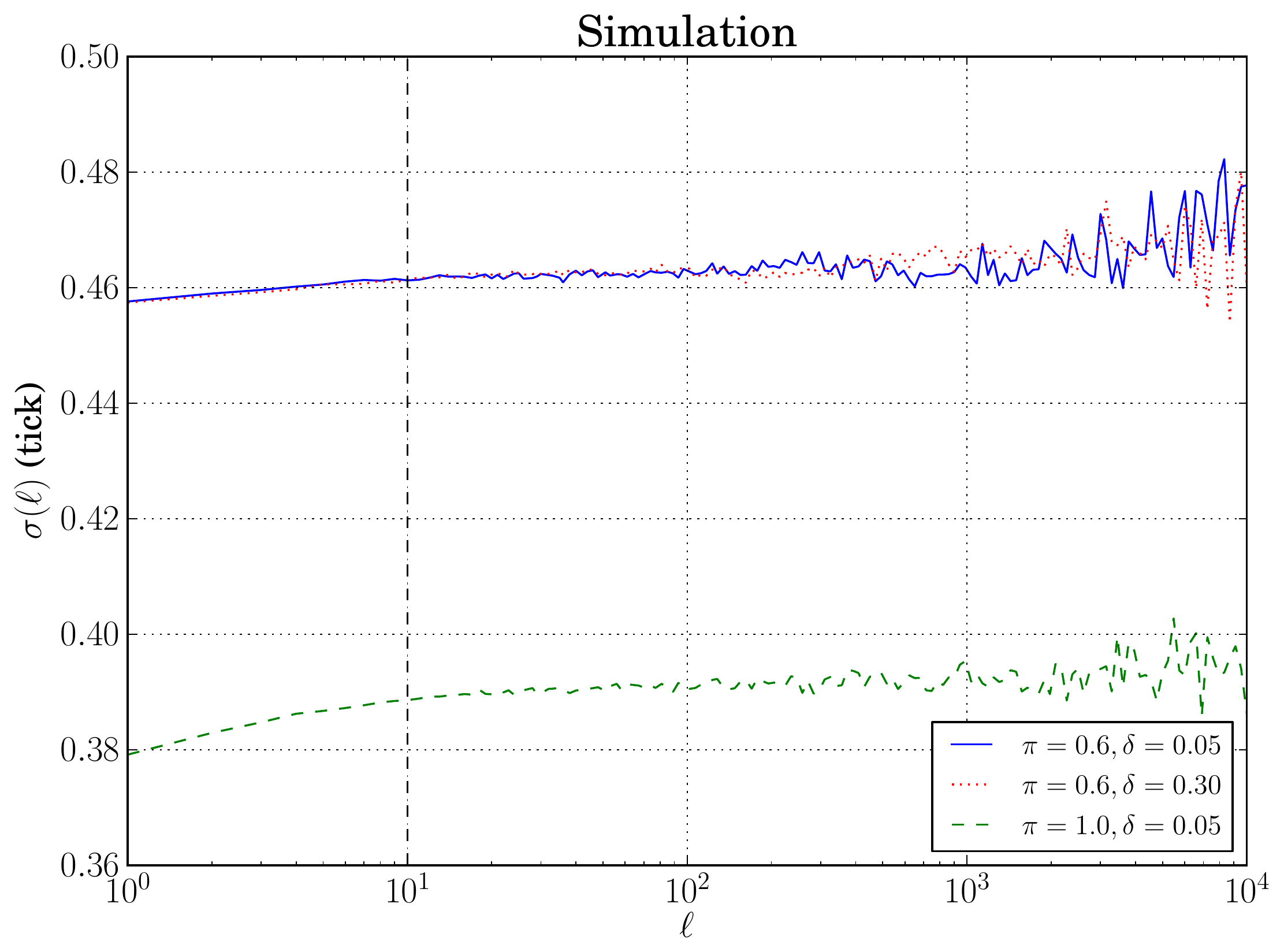}
\caption{Signature plot of the model as a function of the (tick) time lag for different values of participation rate $\pi$ and $\delta$, and for the parameter choice $\mu=0.1\,\mathrm{s^{-1}}$, $\lambda=0.5\,\mathrm{s^{-1}w^{-1}}$, $\nu=0.01\,\mathrm{s^{-1}}$, $\gamma=0.5$, $\alpha=0.5$. The vertical line is the lifetime of limit orders $\nu^{-1}\mu\,\mathrm{trades}$.}
\label{fig:vol_pareto}
\end{figure}

\subsection{Numerical results}
\label{sec:Numerical}
Numerical simulations of the model confirm the theoretical prediction explained above. We have measured the signature plot (see Equation~\ref{eqn:signplot}) of the price process as result of the interaction of market orders, limit orders, and cancellation of the model. We have used the “private” sign predictor of Equation~\ref{eqn:exact_pred}. Figure~\ref{fig:vol_pareto} shows the signature plot of the model. As one can observe, the volatility as a function of the lag $\ell$ is almost constant, $\sigma(\ell)=D$, and it is compatible with a diffusive process. The vertical line is the lifetime of limit orders $\nu^{-1}\mu=10\,\mathrm{trades}$, which is in the model of Ref.~\cite{toth2011anomalous} the maximum time scale for which prices are still diffusive. Furthermore, by construction the resulting prices of the model are informationally efficient. This characteristic does not depend on the particular choice of the participation $\pi$ and of the parameter $\delta$. In fact we observe that, as expected from the theoretical analysis, volatility does not depend on $\delta$ for a fixed value of $\pi$. On the other hand, volatility is lower for higher values of $\pi$, which is not surprising because high levels of participation rate lead to lower uncertainty in the order flow and lower volatility.

\begin{figure}[t]
\centering
\includegraphics[width=0.75\columnwidth]{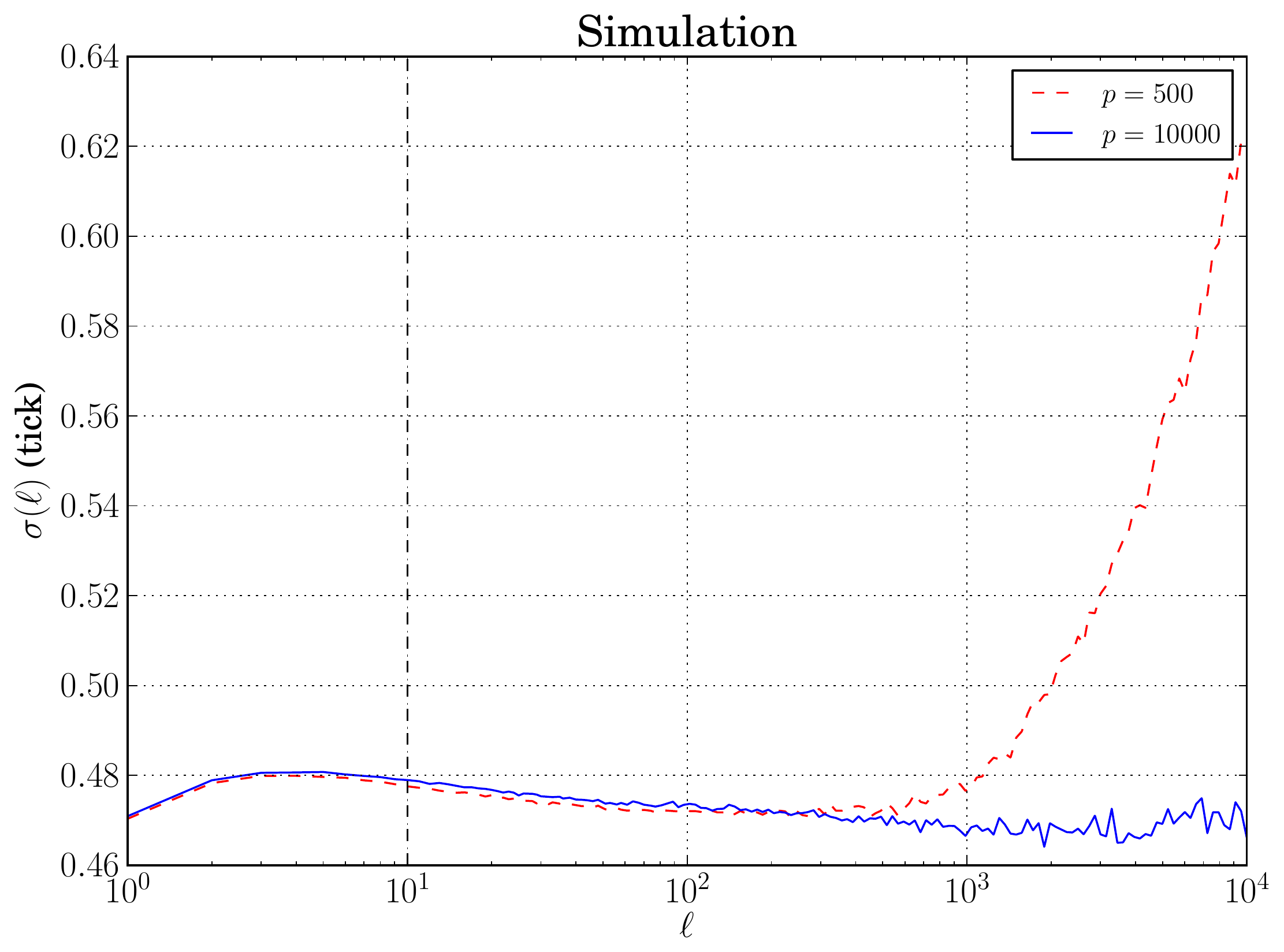}
\caption{Signature plot of the model as a function of the (tick) time lag for different values of order $p$ of the \textit{DAR(p)} process used for the computation of the sign predictor, using $\mu=0.1\,\mathrm{s^{-1}}$, $\lambda=0.5\,\mathrm{s^{-1}w^{-1}}$, $\nu=0.01\,\mathrm{s^{-1}}$, $\gamma=0.5$, $\pi=0.6$, $\delta=0.05$, $\alpha=0.5$. The vertical line is the lifetime of limit orders $\nu^{-1}\mu\,\mathrm{trades}$.}
\label{fig:vol_dar}
\end{figure}

We have repeated the same simulations by using the sign predictor of the \textit{DAR(p)} process, which uses only the signs of past order flow, and no information on metaorders. Figure~\ref{fig:vol_dar} shows the signature plot of the resulting price process in such a setting. The time scale for which the price process is diffusive depends on the chosen order $p$ of the \textit{DAR(p)} process, and $\sigma(\ell)$ is constant for $\ell<p$. This is not surprising because if one considers a time window of length $\ell>p$, there might exist non vanishing positive correlations of order signs due to metaorders longer than $\pi p$, but the predictor considers only the past $p$ trades. By taking longer windows for the \textit{DAR(p)} predictor (or by considering models with shorter metorders) one recovers diffusivity at all scales. An interesting result of the simulations is that the volatility using the two different sign predictor has approximately the same value, \emph{i.e.} it is almost independent from the choice of the particular set of information used for the sign predictor.

\begin{figure}[t]
\begin{minipage}[c]{0.5\columnwidth}
\includegraphics[width=\columnwidth]{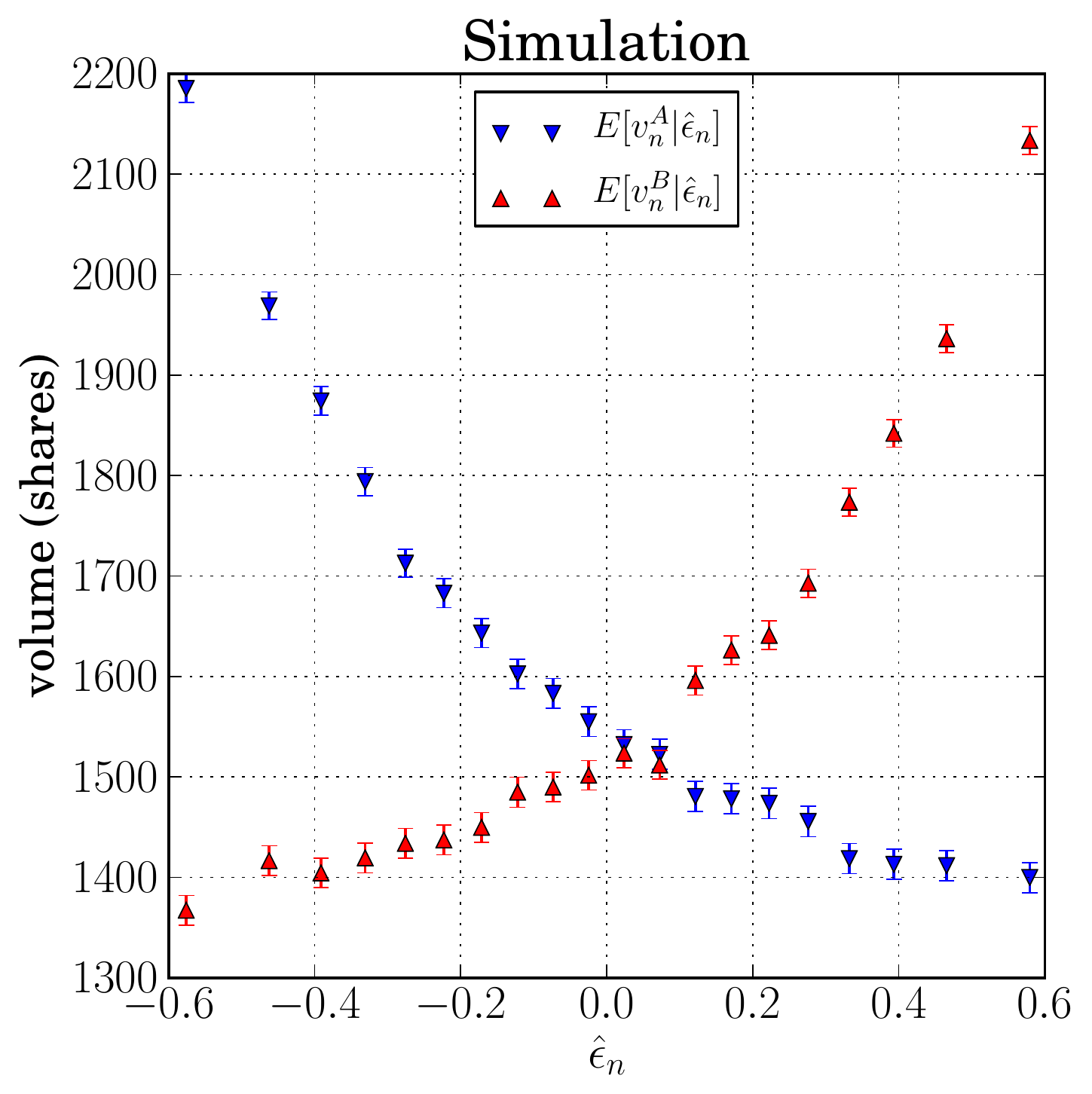}
\end{minipage}%
\begin{minipage}[c]{0.5\columnwidth}
\includegraphics[width=\columnwidth]{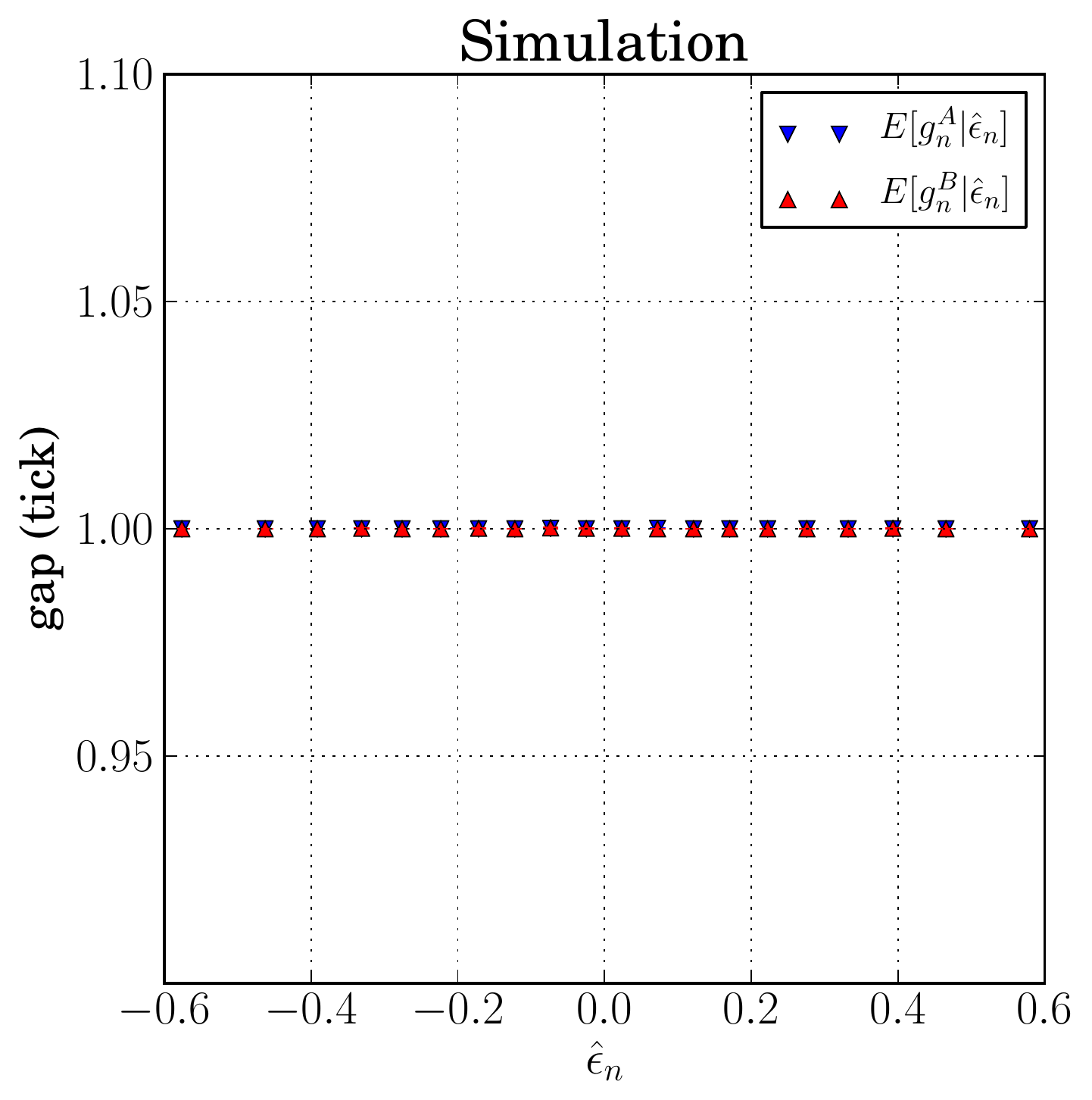}
\end{minipage}
\begin{minipage}[c]{0.5\columnwidth}
\includegraphics[width=\columnwidth]{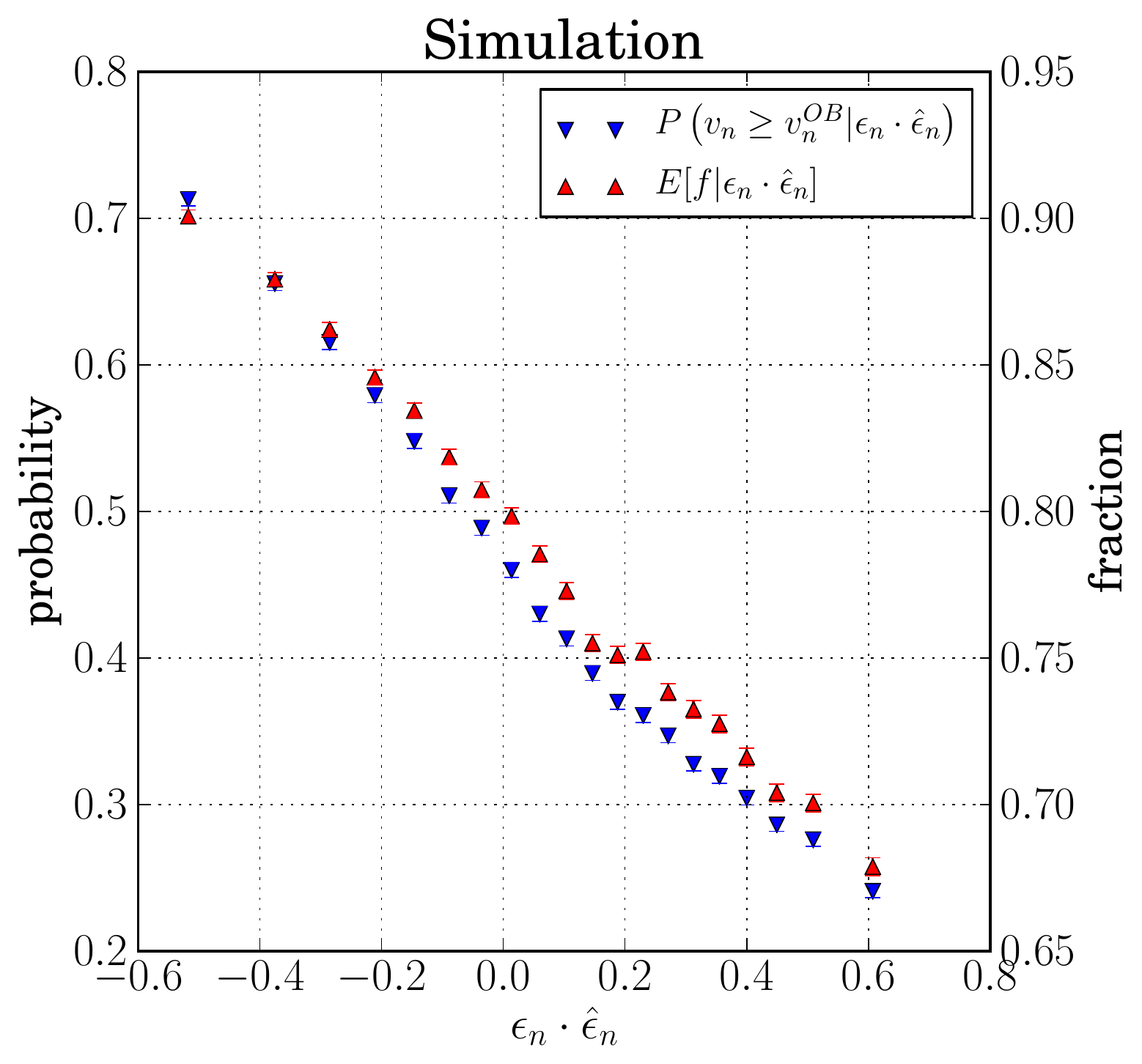}
\end{minipage}%
\begin{minipage}[c]{0.5\columnwidth}
\includegraphics[width=\columnwidth]{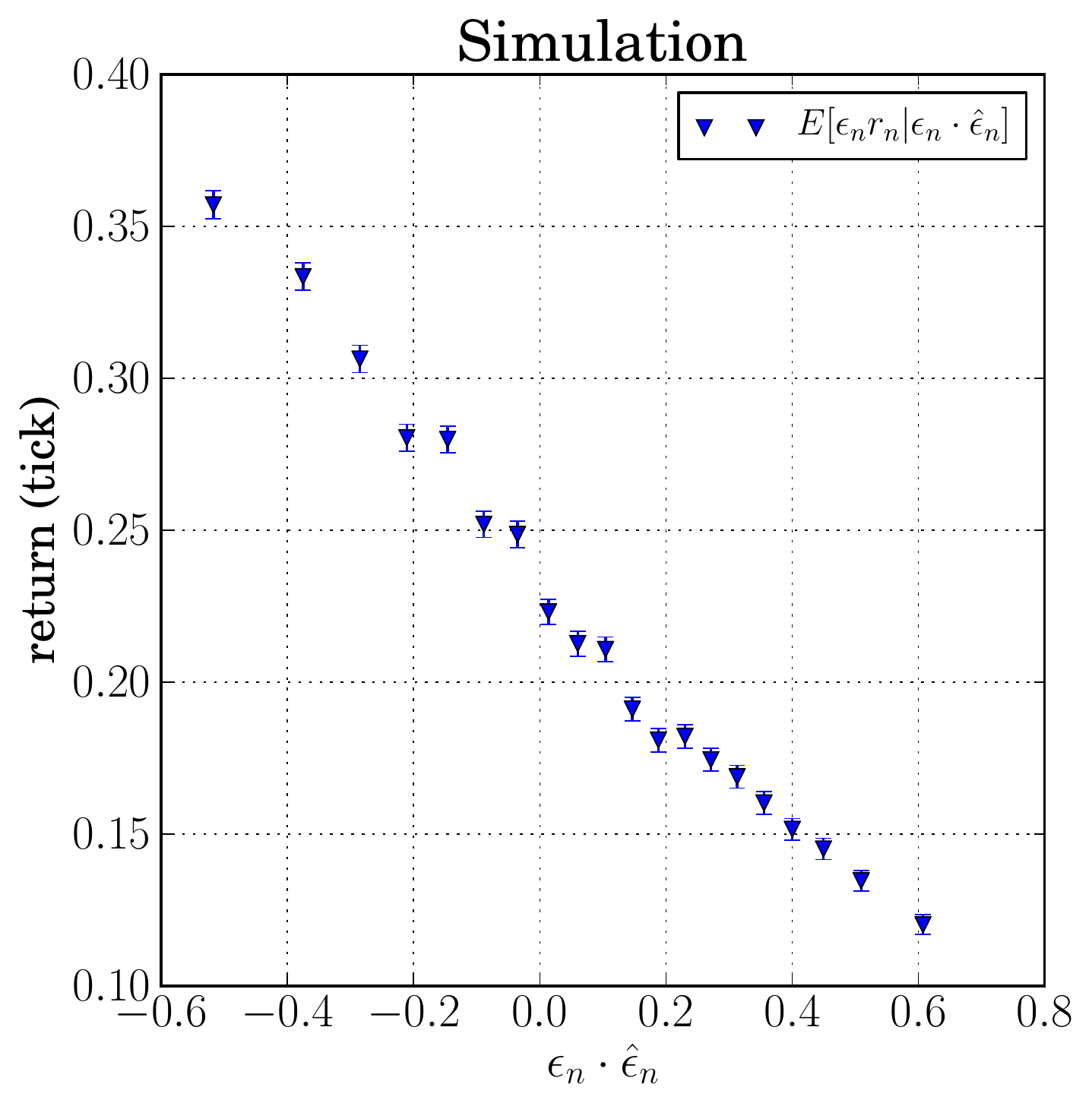}
\end{minipage}
\caption{(Top left) Conditional volumes at the best, (top right) conditional price gaps on different sign predictor values, (bottom left) probability of penetration and conditional fraction, and (bottom right) conditional return on $\epsilon_n\cdot\hat{\epsilon}_n$ of the model. The parameters of the model were $\mu=0.1\,\mathrm{s^{-1}}$, $\lambda=0.5\,\mathrm{s^{-1}w^{-1}}$, $\nu=0.01\,\mathrm{s^{-1}}$, $\gamma=0.5$, $\pi=0.6$, $\delta=0.05$, $\alpha=0.5$.}
\label{fig:sim}
\end{figure}

We therefore conclude that our model is able to give exactly diffusive prices. Our model is also able to reproduce the empirically observed dependencies of order book quantities from the predictability of the order flow. Specifically, we have measured the same order book quantities of section~\ref{sec:Empirical} in our synthetic market, using in $P_g$ the “private” sign predictor. However, the conditional expected values of volume at best quotes, price gaps, probability of penetration, fraction, and returns are computed conditioning them on the sign predictor of the \textit{DAR(p)} process. In this way, we can compare the results from our model with the evidences from real markets.

The results are shown in Figure~\ref{fig:sim}. The model reproduces quite well the behaviour of real order books. In particular, volume at the bid is higher (smaller) than the volume at the ask when the most likely next market order is a buy (sell), as observed in Figure~\ref{fig:volbest}. Gaps are constant because we are working in the large tick approximation, where all the level of the order book are occupied (see the right panels of Figure~\ref{fig:gap}). The penetration probability and the average fraction $f$ both decline with $\epsilon_n\cdot\hat{\epsilon}_n$, as seen in Figure~\ref{fig:penetration_fraction}\footnote{In the case of NASDAQ stocks, one can observe this behaviour in the region of high predictability where the majority of the mass of the distribution of predictors is concentrated.}. Finally, the impact $\epsilon_n r_n$ declines with $\epsilon_n\cdot\hat{\epsilon}_n$, as postulated by the asymmetric liquidity mechanism, and as observed in real data in the region where the mass of the distribution of sign predictors is concentrated (see Figure~\ref{fig:return}).

Let us comment the property of efficiency of our model. The synthetic market simulated by the model is more efficient than the real ones. In fact, the signature plot is almost constant for every time scale $\ell$ and conditional returns are linear in $\epsilon_n\cdot\hat{\epsilon}_n$ like the efficient model in Equation~\ref{eqn:efficient_model}. In modern markets linearity is recovered after few trades (see Figures~\ref{fig:lagged_penetration_return1} and~\ref{fig:lagged_penetration_return2}). We are confident that introducing some inefficiency in the model, it can reproduce some effects measured during our empirical analysis of real stocks.

\section{Conclusions}
\label{sec:Conclusions}
In this paper we have considered the subtle issue of reconciling the persistence of order flow with price efficiency and return diffusivity. Since on average a buyer initiated trade pushes the price up while a seller initiated trade pushes it down, in a naive view the strong positive correlation between trades measured empirically would lead to strongly correlated returns. However, the empirical evidence of price efficiency clashes with this view. 

We have investigated the microstructural mechanisms able to reconcile both evidences. In the first part of our analysis we have performed an empirical study of the behaviour of four stocks, Astrazeneca, Vodafone, Apple, and Microsoft, which have been selected in light of their different features. While the order book data for the former two stocks were recorded at the London Stock Exchange in 2004, for the latter two stocks the data sample is relatively more recent and was recorded at NASDAQ in July/August 2009. Moreover, while Vodafone and Microsoft are stocks whose tick-to-price ratio is on average large, for Astrazeneca and Apple the ratio is very small. Our choice should guarantee independence of our results on the specificity of the stocks and of the market place. Nonetheless, we have planned a future extension of our analysis to a wider data sample.

A possible mechanism able to reconcile the persistence of trade signs and price efficiency is the asymmetric liquidity mechanism proposed by~\cite{lillo2004long}: The price impact of an order is inversely related to the probability of its occurrence. This means that if at a certain point in time it is more likely that the next trade is a buy rather than a sell, a buyer initiated trade will have a smaller impact than a seller initiated trade. There is therefore a compensation between the probability of an event and its effect on the price. In spite of its conceptual simplicity there are many possible microstructural mechanisms responsible for it. Among the several explanations of the drop of the impact one could consider the case where efficiency is guaranteed by the agents initiating the trade and adjusting the volume of their trades to the volume outstanding on the opposite side of the order book. A second explanation would focus on the leading role of liquidity providers revising their quotes after a trade in order to compensate for the impact due to liquidity takers. Our empirical analysis evidences that when the order flow predictability increases in one direction (buy or sell) the volume outstanding at the opposite best decreases, the opposite side of the book becomes more and more sparse, but the probability that a trade moves the price decreases significantly. While the last mechanism is able to counterbalance the persistence of order flow and restore efficiency and diffusivity, the first two act in the opposite direction. Moreover, disentangling each return in a component due to a mechanical impact and in a second aggregated component due to the revision of liquidity providers, we have measured a positive correlation between impact and quote revision. However, this effect tends to disappear when the order sign predictability increases.  

The above empirical evidences lead to significant challenges in the modeling of the order book dynamics. A growing strand of literature is dealing with this issue, and in the second part of our paper we have introduced a statistical model designed for large tick stocks which is able to successfully recover the empirical findings in the presence of a strongly persistent order flow. The main intuition behind our approach is that the mechanism restoring efficiency must depend on the local level of predictability of the order flow. More precisely, the agent placing a market order knows exactly the past history of the market order sign process and the sign of the next order (buy or sell) she is going to execute and adapts her order volume to the level of predictability of the order sign. We explain this strategic behaviour in this way: High predictability of the order flow means that liquidity takers reveal to the market information about their intentions, and in order to control the market impact of their trades, they reduce the volumes of the market orders progressively during the execution of the whole metaorder. We have supported our conclusions with extensive Monte Carlo simulations.

The adaptive liquidity taking mechanism described above is, however, only part of the story. In spite of its effectiveness, it is indeed evident that a determinant role has to be also played by liquidity providers. While in the current paper we have focused on modeling the strategic behaviour of liquidity takers, we are currently working on the extension of the statistical model in order to include the strategic behaviour of market makers.

\ack
The authors acknowledge partial support by the grant, SNS13LILLB “Systemic risk in financial markets across time scales”. We also thank Jean-Philippe Bouchaud, Iacopo Mastromatteo, and Bence Toth for inspiring discussions.

\appendix
\section{Building a predictor for the order flow sign: the \textit{DAR(p)} model}
\label{app:DAR}
The \textit{DAR(p)} model defines a general class of simple models for discrete variate time series. It generates a sequence of stationary discrete random variables with the properties of a $p$-th order Markov process. These properties are reflected by the fact that the distribution of $X_n$ only depends on $\Omega_{n-1}=\lbrace X_{n-1},\ldots,X_{n-p} \rbrace$. The process is specified by the stationary marginal distribution of $X_n$ and by the correlation structure of the sequence.

{\it Definition.}
The $p$-th order discrete autoregressive model \textit{DAR(p)} is given by
\begin{equation}
X_n=V_nX_{n-A_n}+(1-V_n)Z_n \label{eqn:darp}\,.
\end{equation}

The sequence $\lbrace Z_n \rbrace_{\mathbb{N}_\lambda}$ is composed by IID random values drawn by a marginal distribution $\Xi$, whose sample space is a subset of the integers $\mathbb{N}_\lambda$, where $\lambda$ is the cardinality of the sample space. Furthermore, $\lbrace V_n \rbrace$ is a sequence of IID random values following a Bernoulli distribution $\mathcal{B}(1,\chi)$. Therefore we have
\begin{equation*}
P(V_n=1)=1-P(V_n=0)=\chi, \qquad \mbox{with } 0 \leqslant \chi < 1\,.
\end{equation*}

Finally, $\lbrace A_n \rbrace$ is a sequence of IID random values drawn from a multinomial distribution $\mathcal{M}(1,\vec{\phi})$, with states $\lbrace 1,2, \ldots p \rbrace$ and probabilities
\begin{equation*}
P(A_n=i)=\phi_i\geqslant0, \qquad i \in \lbrace 1,2, \ldots p \rbrace\,,
\end{equation*}
where the parameter vector $\vec{\phi}=(\phi_1,\ldots,\phi_p)$ is normalised to unity, $\sum_{i=1}^p \phi_i=1$.

Let us explain in a less formal way the \textit{DAR(p)} process of Equation~\ref{eqn:darp}: The value $X_n$ is either taken from the history of $\lbrace X_n \rbrace$ (with probability $\chi$) or drawn randomly from the $\Xi$ distribution (with probability $1-\chi$). The random values $V_n$ have the function of a switch between the two cases. In the case of a positive Bernoulli trial ($V_n=1$), $X_n$ is determined by moving $A_n$ steps back in the past observations of $\lbrace X_n \rbrace$, with $A_n$ assuming values in $\lbrace 1,2, \ldots p \rbrace$ with probability given by the parameter vector $\vec{\phi}=(\phi_1,\ldots,\phi_p)$. Therefore, with probability $\chi \phi_i$, $X_n=X_{n-i}$, for $i=1,2,\ldots,p$. In the second case, when $V_n=0$, $X_n=Z_n$ is drawn randomly from the specific marginal distribution $\Xi$, which has a discrete state space.

It is possible to select an initial distribution of $X_0$ which yields a stationary sequence $\lbrace X_n \rbrace_{\mathbb{N}_\lambda}$ with marginal distribution $\Xi$, and it is possible to prove that this initial distribution coincides with the marginal distribution $\Xi$.

This procedure differs essentially from the representation of a Markov process as a matrix of transition probabilities, where the number of parameters to estimate is $\lambda^{p+1}-\lambda$. Here a smaller number $p+1$ of effective parameters allows a better control of the statistical properties of the sequence, while in the case of a transition probability matrix one has many independent parameters, each of which regulates only a minor aspect of the process.

{\it Autocovariance structure.}
Let $\lbrace X_n \rbrace_{\mathbb{N}_\lambda}$ be a stationary \textit{DAR(p)} process with marginal distribution $\Xi$, parameter $\chi$ and parameter vector $\vec{\phi}=(\phi_1,\ldots,\phi_p)$. From Equation~\ref{eqn:darp}, we immediately find that $\mu_X=\mathbb{E}[X_n]=\mathbb{E}[Z_n]=\mu_Z$.

We center the $X_n$'s with the unconditional mean, $\widetilde{X}_n = X_n-\mu_X$, multiply Equation~\ref{eqn:darp} by $\widetilde{X}_{n-k}$, $k>0$ and we take the expectation of both sides
\begin{equation*}
  \gamma_k=\mathbb{E}[\widetilde{X}_n \widetilde{X}_{n-k}]=\chi\sum_{i=1}^p \phi_i \mathbb{E}[\widetilde{X}_{n-i}\widetilde{X}_{n-k}]+(1-\chi)\mathbb{E}[(Z_n-\mu_X)\widetilde{X}_{n-k}]\,.
\end{equation*}

Dividing both sides by the variance of $\widetilde{X}_n$, we obtain the corresponding relation for the autocorrelations
\begin{equation}
  \rho_k=\chi \sum_{i=1}^p \phi_i \rho_{k-i}, \qquad k\geqslant 1\,,
  \label{eqn:YuleWalker}
\end{equation}
which are the usual Yule-Walker equations~\cite{hamilton1994time}. This linear system can be solved recursively after the computation of the sample autocorrelations from the time series. Given $\rho_1,\rho_2,\ldots,\rho_p$, the first $p$ equations can be solved for the $p$ parameters $\phi_1,\ldots,\phi_{p-1}$ and $\chi$. The parameter $\phi_p$ is given by $(1-\phi_1-\ldots-\phi_{p-1})$. The estimations of the components of the parameter vector $\vec{\phi}$ can lead to negative values, but probabilities must be always greater than or equal to zero, $\phi_i \geq 0$. This problem is important when we perform simulations of the process, in this case we smooth the empirical coefficients performing a moving average which spans ten points and finally we set the negative elements to zero.

The advantages of the \textit{DAR(p)} model is that it is intrinsically autoregressive, and its parameters can be easily computed by means of the sample autocorrelation.

{\it Forecasting.}
We can now construct the best predictor of variable $X_n$ within this model. We recall the fact that all sequences $\lbrace V_n \rbrace$, $\lbrace A_n \rbrace$ and $\lbrace Z_n \rbrace_{\mathbb{N}_\lambda}$ are independent one from each other. We take the expected values both unconditional and conditional on $\Omega_{n-1}$ and we calculate the second conditional moment,
\begin{eqnarray}
\mathbb{E}[X_n]&=&\mu_Z\,, \nonumber \\
\mathbb{E}[X_n|\Omega_{n-1}]&=&\chi \sum_{i=1}^p \phi_i X_{n-i} +\mu_Z(1-\chi)\equiv\hat{X}^{DAR}_n, \nonumber \\
\mathbb{E}[X^2_n|\Omega_{n-1}]&=&\chi \mathbb{E}[X^2_{n-A_n}|\Omega_{n-1}]+\mu_Z(1-\chi) \nonumber \\ &=&\chi \sum_{i=1}^p\phi_i \mathbb{E}[X^2_{n-i}|\Omega_{n-1}]+\mu_Z(1-\chi) \nonumber \\
&=&\chi \sum_{i=1}^p\phi_i X^2_{n-i}+\mu_Z(1-\chi)\,. \label{eqn:preddar}
\end{eqnarray}

The expression of the predictor can be extended by computing the conditional expected value of $X_{n+s}$, for $s=1,2,\ldots$. After simple calculations, we find that
\begin{equation}
  \hat{X}_{n+s}^{DAR}=\chi\sum_{i=1}^p \phi_i Y_{n+s-i}+\mu_Z(1-\chi)\,,
  \label{eqn:epshat}
\end{equation}
where
\begin{equation*}
Y_{n+s-i} =
\left\{
\begin{array}{rl}
\hat{X}_{n+s-i}^{DAR} & \mbox{for } i \leqslant s \\
X_{n+s-i} & \mbox{for } i > s
\end{array}
\right.\,.
\end{equation*}

\end{document}